\documentclass[fleqn,
11pt, 
english, 
singlespacing, 
headsepline, 
]{MastersDoctoralThesis} 

\usepackage[utf8]{inputenc} 
\usepackage[T1]{fontenc} 

\usepackage{mathpazo} 

\usepackage[backend=bibtex,style=numeric-comp,natbib=true,sorting=none,maxbibnames=99]{biblatex} 

\AtEveryBibitem{
    \clearfield{urlyear}
    \clearfield{urlmonth}
}

\usepackage[autostyle=true]{csquotes} 

\usepackage{comment}

\usepackage{xcolor}
\usepackage[linesnumbered,ruled,vlined]{algorithm2e}

\usepackage{amsmath}

\usepackage{bbding}
\usepackage{pifont}
\usepackage{wasysym}
\usepackage{amssymb}

\usepackage{multirow}
\usepackage{longtable}
\usepackage{hhline}

\usepackage{bm} 
\usepackage{booktabs}
\usepackage{soul} 
\usepackage{graphicx} 
\usepackage{float}
\newcommand{\tabitem}{~~\llap{\textbullet}~~}

\SetKwInput{KwInput}{Input}
\SetKwInput{KwOutput}{Output}

\SetCommentSty{mycommfont}

\usepackage{pifont}
\newcommand{\xmark}{\ding{55}}%

\usepackage{chngpage}
\usepackage[export]{adjustbox}
\usepackage{textgreek} 

\usepackage{CJKutf8}

\usepackage{pdflscape}
\usepackage{afterpage}

\usepackage{makecell}

\usepackage{svg}
\svgsetup{inkscapelatex=false}

\thesistitle{Distinguishability Investigation on Longa’s Atomic Patterns when used as a Basis for Implementing Elliptic Curve Scalar Multiplication Algorithms} 
\supervisor{Hon. Prof. Dr.-Ing. Zoya \textsc{Dyka}} 
\examiner{} 
\degree{Master of Science in Cyber Security} 
\author{Sze Hei \textsc{Li}} 
\addresses{} 

\subject{Cyber Security} 
\keywords{} 
\university{{Brandenburg University of Technology Cottbus-Senftenberg}} 
\department{{Chair of Wireless Systems}} 
\group{{Research Group Name}} 
\faculty{{Faculty 1 - MINT - Mathematics, Computer Science, Physics, Electrical Engineering and Information Technology}} 

\begin{document}

\begin{CJK*}{UTF8}{gbsn}

\frontmatter 

\pagestyle{plain} 


\begin{titlepage}

\includegraphics[width=0.5\textwidth]{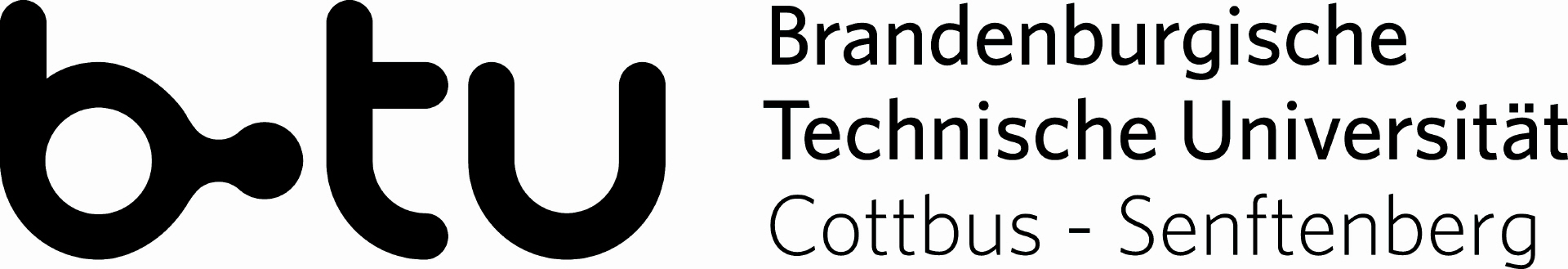}\hfill
\includegraphics[width=0.15\textwidth]{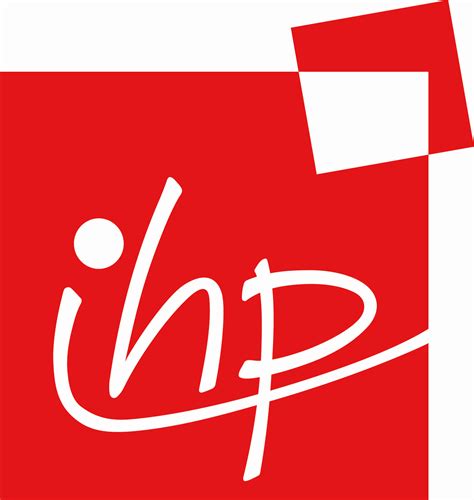}
\begin{center}

\vspace*{.06\textheight}
\univname\\ \vspace{0.1cm} Faculty 1 - MINT - Mathematics, Computer Science, Physics, \\Electrical Engineering and Information Technology\\ \vspace{0.1cm} \deptname \\
\vspace{1cm}

\vspace{0.3cm}
{\huge \bfseries \ttitle\par}\vspace{0.5cm} 

\textsc{\Large Master's Thesis\\ \vspace{0.25cm}}
\vspace{0.5cm}
\textbf{\authorname}\\ \vspace{0.1cm} 
This thesis is submitted for the degree of Master of Science in Cyber Security
\vspace{1cm}

{\large August 5, 2024}\\[4cm] 

\vspace{1cm}
Supervisor: \supname \\  
Second Examiner: Dr.-Ing. Ievgen \textsc{Kabin} \\


 

 
\end{center}
\end{titlepage}

\begin{acknowledgements}
\noindent I am deeply grateful to Hon. Prof. Dr.-Ing. Zoya Dyka and Dr.-Ing. Ievgen Kabin for their invaluable guidance and support throughout my research. Their expertise, encouragement, and constructive feedback have been instrumental in shaping this thesis. I sincerely appreciate their time and dedication, which have significantly contributed to the success of this work.
\end{acknowledgements}


\begin{abstract}
\noindent In the evolving landscape of cryptographic security, the robustness of Elliptic Curve Cryptography (ECC) against side-channel analysis (SCA) attacks is of paramount importance due to the widespread use of ECC and the growing sophistication of SCAs. This thesis delves into the investigation of Longa's atomic patterns applied within Elliptic Curve scalar multiplication algorithms, assessing their resistance to horizontal SCAs. The research employs these atomic patterns in practical implementation on a microcontroller (Texas Instruments Launchpad F28379 board) using the open-source cryptographic library FLECC in C.

In our analysis, we only focused on the distinguishability of the first atomic block in the Elliptic Curve point doubling and point addition patterns. Due to various technical limitations, we were unable to determine significant differences in the execution time and the shapes of the atomic blocks. Further investigations of the SCA-resistance can be performed based on this work.

A significant contribution of this work is the identification and correction of several discrepancies in Longa's original atomic patterns. This thesis marks the first practical implementation of Longa’s patterns, extending the theoretical research into empirical analysis.

\end{abstract}


\tableofcontents 

\listoffigures 
\addchaptertocentry{List of Figures}
\listoftables 
\addchaptertocentry{List of Tables}
\listofalgorithms
\addchaptertocentry{List of Algorithms}

\begin{abbreviations}{ll} 
\addchaptertocentry{List of Abbreviations}
\textbf{A} & \textbf{A}ddition\\
\textbf{CCS} & \textbf{C}ode \textbf{C}omposer \textbf{S}tudio\\
\textbf{DPA} & \textbf{D}ifferential \textbf{P}ower \textbf{A}nalysis\\
\textbf{DSCA} & \textbf{D}ifferential \textbf{S}ide-\textbf{C}hannel \textbf{A}nalysis\\
\textbf{EC} & \textbf{E}lliptic \textbf{C}urve\\
\textbf{ECC} & \textbf{E}lliptic \textbf{C}urve \textbf{C}ryptosystem\\
\textbf{ECDH} & \textbf{E}lliptic \textbf{C}urve \textbf{D}iffie-\textbf{H}ellman\\
\textbf{ECDSA} & \textbf{E}lliptic \textbf{C}urve \textbf{D}igital \textbf{S}ignature \textbf{A}lgorithm\\
\textbf{EM} & \textbf{E}lectro\textbf{m}agnetic\\
\textbf{FLECC} & \textbf{FLECC} in C\\
\textbf{Frac-wmbNAF} & \textbf{Frac}tional Window-\textbf{w} \textbf{m}ulti\textbf{b}ase \textbf{N}on \textbf{A}djacent \textbf{F}orm\\
\textbf{HCCA} & \textbf{H}orizontal \textbf{C}ollision \textbf{C}orrelation \textbf{A}nalysis\\
\textbf{ICS} & \textbf{I}ntegrated \textbf{C}ircuit \textbf{S}canner\\
\textbf{IDE} & \textbf{I}ntegrated \textbf{D}evelopment \textbf{E}nvironment\\
\textbf{LM} & \textbf{L}onga-\textbf{M}iri precomputation\\
\textbf{M} & \textbf{M}ultiplication\\
\textbf{mbNAF} & \textbf{m}ulti\textbf{b}ase \textbf{N}on \textbf{A}djacent \textbf{F}orm\\
\textbf{N} & \textbf{N}egation\\
\textbf{NOP} & \textbf{N}o \textbf{Op}eration\\
\textbf{PA} & \textbf{P}oint \textbf{A}ddition\\
\textbf{PD} & \textbf{P}oint \textbf{D}oubling\\
\textbf{RAM} & \textbf{R}andom \textbf{A}ccess \textbf{M}emory\\
\textbf{RSA} & \textbf{R}ivest-\textbf{S}hamir-\textbf{A}dleman cryptosystem\\
\textbf{SCA} & \textbf{S}ide-\textbf{C}hannel \textbf{A}nalysis\\
\textbf{SOS} & \textbf{S}eparated \textbf{O}perand \textbf{S}canning\\
\textbf{SPA} & \textbf{S}imple \textbf{P}ower \textbf{A}nalysis\\
\textbf{SSCA} & \textbf{S}imple \textbf{S}ide-\textbf{C}hannel \textbf{A}nalysis\\
\textbf{wmbNAF} & Window-\textbf{w} \textbf{m}ulti\textbf{b}ase \textbf{N}on \textbf{A}djacent \textbf{F}orm\\

\end{abbreviations}















\mainmatter 

\pagestyle{thesis} 



\chapter{Introduction} 

\label{Chapter1} 


\newcommand{\keyword}[1]{\textbf{#1}}
\newcommand{\tabhead}[1]{\textbf{#1}}
\newcommand{\code}[1]{\texttt{#1}}
\newcommand{\file}[1]{\texttt{\bfseries#1}}
\newcommand{\option}[1]{\texttt{\itshape#1}}


\section{Motivation}
Based on the side-channel atomicity principle introduced by Chevallier-Mames et al. \nocite{chevallier-mames_low-cost_2004}[1], Longa developed Simple Side-Channel Analysis (SSCA)-resistant Elliptic Curve (EC) scalar multiplication atomic formulae in his work \nocite{longa_accelerating_2008}[2], which have up to 22\% faster speed than using traditional atomic structures and improved the security of Elliptic Curve scalar multiplication ($k\bm{P}$) against side-channel analysis (SCA) attacks. Building resistance for $k\bm{P}$ algorithms against simple SCA attacks is the primary goal of the atomicity principle.

Recent work by Kabin \nocite{kabin_horizontal_2023}[3] has shown that the address-bit vulnerability can be exploited using horizontal, i.e., single trace, SCA attacks. Key-dependent addressing of registers is an inherent part of regular as well as atomic patterns-based $k\bm{P}$ algorithms.
Although many studies in SCA have focused on improving the efficiency and security of EC scalar multiplication theoretically, there are very few studies focusing on the practical implementation and analysis of the existing approaches for hardware or embedded devices. 
Thus, Longa's atomic patterns can be an insufficient SCA countermeasure.
This work fills that research gap by performing horizontal SCA attacks against an implementation of a binary scalar multiplication $k\bm{P}$ algorithm using Longa's MNAMNAA atomic pattern [2]. 

\section{Contributions}
The contributions of this thesis are:
\begin{itemize}
    \item An overview of the literature that referenced Longa’s master’s thesis [2]
    \item The correction of Longa’s atomic patterns
    \item The implementation of a $k\bm{P}$ algorithm using Longa's MNAMNAA atomic patterns on a Texas Instruments Launchpad F28379 evaluation board TMS320F28379D with a 32-bit microcontroller using the open-source cryptographic library FLECC in C
    \item The implementation of SSCA attacks analysing an electromagnetic trace of a $k\bm{P}$ execution
\end{itemize}

To the best of our knowledge, we are the first to implement Longa’s point doubling and mixed point addition atomic patterns in an embedded device. 

\section{Structure of This Thesis}
Following the introduction, this thesis is organized as follows: Chapter \ref{Chapter2} explains the basic concepts regarding Elliptic Curve Cryptosystem, SCA, $k\bm{P}$ algorithm vulnerabilities and atomicity principle. In Chapter \ref{Chapter3}, we review some state-of-the-art literature related to Longa’s atomic pattern algorithms. Chapter \ref{Chapter4} describes our experimental setup regarding the implementation of Longa’s atomic pattern algorithms. In Chapter \ref{Chapter5}, we look into a selected number of open-source cryptographic libraries to choose a suitable one for our constant-time atomic $k\bm{P}$ algorithm implementation on our target device. Chapter \ref{Chapter6} describes the practical implementation of Longa’s atomic pattern algorithms and our SCA methodology. Chapter \ref{Chapter8} summarizes this thesis. 
Finally, Chapter \ref{Chapter9} serves as the appendix, where the detailed descriptions of the literature overview, corrections of Longa's atomic patterns, the distinguishability between Doubling 1 and Doubling 2, the execution time of "no operations", and an attack scenario are presented. These contents are placed in the final chapter to improve the overall readability of the thesis. The appendix represents a significant part of the investigations performed in this work.


\chapter{Background} 

\label{Chapter2} 

\section{Elliptic Curve Cryptosystem}\label{ecc}
Elliptic Curve Cryptosystem (ECC) was introduced independently by Miller \nocite{williams_use_1986}[4] and Koblitz \nocite{koblitz_elliptic_1987}[5] in 1985. It is a public-key cryptosystem used to perform critical security functions including encryption, authentication and digital signatures. Not only can ECC achieve higher security per key bit compared to RSA, its other benefits such as reduced memory requirements, lower energy consumption and faster execution time have made it suitable for implementation of cryptographic protocols for resource-constrained embedded devices.

In this thesis, we discuss an implementation of elliptic curve point multiplication using an elliptic curve $E$ over a prime finite field $\mathbb F_p$, where $p$ is a large prime and represents the number of elements of the finite field. The EC over $\mathbb F_p$ is denoted by $E(\mathbb F_p)$ and defined as follows \nocite{hankerson_guide_2004}[6]:
\begin{equation} \label{ec_eqn}
E(\mathbb F_p): y^2=x^3+ax+b
\end{equation}
where $x,y,a,b \in \mathbb F_{p}, p > 3$ and $\Delta=4a^{3}+27b^{2}\neq 0$. \\

A point $\bm{P}$ on an EC is a solution $\bm{P}=(x,y)$ to the equation $E(\mathbb F_p)$. The set of points satisfying equation (\ref{ec_eqn}), along with the so-called point at infinity $\bm{O}$, form the EC over the finite field $\mathbb F_p$. The total number of elements in this set, known as the cardinality, is referred to as the order of the elliptic curve $E$ and is usually denoted by \textepsilon. 
For the points on the EC, two operations are defined: point doubling and point addition.
In cryptographic protocols, the primary operation involving EC points is scalar multiplication, denoted as $k\bm{P}$ (also known as the $k\bm{P}$-operation, EC point multiplication, or EC scalar multiplication). This operation can be computed through a series of point doubling ($\bm{Q}'$= 2$\bm{P}$) and point addition ($\bm{Q}'$= $\bm{P}$+$\bm{Q}$) operations.
All points on an EC over a finite field form an additive group. Therefore, the result $\bm{Q}'$ is also a point on $E(\mathbb F_p)$ of order \textepsilon.

An important parameter of an EC point is its order. The order $q$\footnote{For NIST standardized elliptic curves over prime finite field, $q$=\textepsilon.} of an EC point $\bm{P}$ is an integer $q$ defined as the following: if $q\bm{G} = \bm{P} + … + \bm{P} = \bm{O}$ but $k\bm{P} \neq \bm{O}$ for all integers 1 $\leq k < q$, where $\bm{O}$ is the point at infinity. If such a $q$ exists, then $\bm{P}$ has finite order. In EC scalar multiplication, we make use of this property to generate a unique resulting point on the EC for each secret scalar $k$ used. The EC scalar multiplication is executed to generate a private/public key pair as follows:
\begin{equation} \label{kp_eqn}
\bm{P}=k\bm{G}
\end{equation}
where $\bm{P}$ and $\bm{G}$ are points on $E(\mathbb F_p)$ of order \textepsilon, and $k$ is a randomly generated big binary secret scalar smaller than order $q$ of the point $\bm{G}$. For standardized ECs, point $\bm{G}$ is the publicly known base point with its coordinates given \nocite{chen_digital_2023}[7]. The private key is the secret scalar $k$; the public key is the point $\bm{P}$.

The security of ECCs, particularly the key pair generation protocol, is based on the complex mathematical problem known as the Elliptic Curve Discrete Logarithm Problem (ECDLP), that claims: for $\bm{Q}=k\bm{P}$, given $\bm{P}$ and $\bm{Q}$, finding $k$ is computationally infeasible. Thus, for the key pair generation protocol (\ref{kp_eqn}), given the public key $\bm{P}$ and the base point $\bm{G}$, finding the private key $k$ is computationally infeasible. The same is true for the $r\bm{G}$ operation in Elliptic Curve Digital Signature Algorithm (ECDSA)\nocite{johnson_elliptic_2001}[8]:
\begin{equation} \label{RrG_eqn}
\bm{R}=r\bm{G}
\end{equation}
with publicly known EC points $\bm{R}$ and $\bm{G}$, and a random integer $r$.

Both operations (\ref{kp_eqn}) and (\ref{RrG_eqn}) are critical from the security point of view: The scalars $k$ from (\ref{kp_eqn}) and $r$ from (\ref{RrG_eqn}) have to be kept secret. Note that the knowledge of $r$ allows to calculate the private key used in the EC signature generation process. 

Thus, $k\bm{P}$ operation, which is the main resource-consuming operation in elliptic-curve-based cryptosystems, is also security-critical. Secure $k\bm{P}$ algorithms have to be resistant against different attacks, particularly SCA attacks, which can reveal the scalar $k$ through statistical or machine learning analysis of side-channel information obtained during the execution of $k\bm{P}$ algorithms. More detailed SCA attacks are described in Chapter \ref{sca}.

\subsection{Binary Algorithms for EC Scalar Multiplication}\label{binary-method}
The binary method is a traditional scalar multiplication method based on the binary expansion of the $l$-bit long scalar ${k=k_{l-1}k_{l-2}…k_2k_1k_0}$, with $k_i\in\{0,1\}$. Given the binary representation of the scalar $k$, the $k\bm{P}$ operation can be computed by processing the scalar $k$ bit-by-bit, for example, from left to right using the traditional double-and-add algorithm, as shown in Algorithm \ref{alg:binary-kp}, taken from \nocite{hankerson_guide_2004}[6].

\begin{algorithm}[!ht]
\DontPrintSemicolon
  
  \KwInput{$k=(k_{l-1},...,k_0)_{2},\bm{P}\in E(\mathbb F_p)$}
  \KwOutput{$k\bm{P}$}
  $\bm{Q}=\infty$\tcp*{Point at infinity $\bm{O}$}
  \For{$i=(l-1)$ downto 0}
  {
    $\bm{Q} = 2\bm{Q}$\tcp*{Point doubling}
    \If{$k_i=1$}
        {$\bm{Q}=\bm{Q}+\bm{P}$\tcp*{Point addition}}
    }
    return $\bm{Q}$

\caption{Left-to-right binary double-and-add $k\bm{P}$ algorithm}
\label{alg:binary-kp}
\end{algorithm}

Numerous other binary $k\bm{P}$ algorithms exist, as well as $k\bm{P}$ algorithms that utilize different scalar coding methods, such as the Window-$w$ Multibase Non-Adjacent Form ($wmb$NAF) proposed in [2]. In this work, we focus on the binary left-to-right $k\bm{P}$ algorithm implemented for the P-256 curve, which is an EC over prime finite field and currently recommended for use for ECDSA and EC key establishment by NIST \nocite{chen_digital_2023, chen_recommendations_2023}[7, 9]. For this curve, all mathematical operations involving point arithmetic (i.e., point doublings and point additions in Algorithm \ref{alg:binary-kp}) use elements of the prime finite field $\mathbb F_p$, with the Mersenne prime $p=2^{256}$- $2^{224} + 2^{192} + 2^{96}$- 1 \nocite{national_bureau_of_standards_federal_2000}[10].

Point doublings and point additions can be represented as sequences of mathematical operations, whereby the sequence of the operations for a point doubling differs from the one for a point addition. In addition to mathematical operations, each point doubling and point addition operation also requires storing and reading data to and from registers. The sequence of these operations depends on the processed scalar $k$ (further also denoted as key), whereby the sequence corresponding to the processing of a key bit value ‘1’ differs from the sequence corresponding to the processing of a key bit value ‘0’. This can be used to reveal the processed scalar $k$ by performing various attacks.

\section{Side-Channel Analysis Attacks}\label{sca}
Side-Channel Analysis (SCA) is a security exploit which aims to extract secrets by analysing physical parameters (e.g., power consumption, electromagnetic emanation and timing, etc.) measured during the execution of a cryptographic algorithm running on a physical device. A trace refers to a set of side-channel information (e.g., power consumption, electromagnetic emanation or timing) measurements taken during a cryptographic operation. Depending on the number of traces needed for the analysis, the analysis can be classified into two categories: horizontal SCA and vertical SCA. Horizontal SCA utilizes only one single trace on a single algorithm execution to extract sensitive information. Whereas vertical SCA requires several\footnote{At least two.} traces to perform the statistical analysis. Key randomization, EC point blinding and randomization of EC point coordinates proposed by Coron \nocite{koc_resistance_1999}[11] are some effective countermeasures against vertical SCA. However, these countermeasures are ineffective against horizontal SCA \nocite{bauer_horizontal_2015, kabin_evaluation_2017}[12, 13].

Depending on the analysis method, SCA attacks can be classified into two categories: Simple Side-Channel Analysis (SSCA) and Differential Side-Channel Analysis (DSCA). SSCA is a technique that uses a trace for direct interpretation of the behavior to infer the secret key. In contrast, DSCA exploits side-channel information leakages by using statistical or clustering methods across many hundreds or even millions of traces to deduce the processed key bits.  It is important to note that various statistical and machine learning methods can also be applied to analyse a single $k\bm{P}$ trace, for example \nocite{bauer_horizontal_2015, batina_balanced_2005, aftowicz_horizontal_2020, kabin_vulnerability_2023}[12, 14-16]. The differences in the shape of the measured trace(s) are searched and analysed using statistical or machine learning methods. The similarities in small parts of traces are usually searched using correlation power analysis, e.g., Pearson's correlation coefficient.

\section{Vulnerabilities of \textit{k\textbf{P}} Implementations to SCA}\label{kp-vulnerabilities}
The shape of each measured side-channel trace of a $k\bm{P}$ execution depends on the processed input data, the scalar $k$, the $k\bm{P}$ algorithm implementation being attacked, target frequency, and the target platform or device, i.e., the technology used for the circuit implementation and the optimization options applied for the design synthesis \nocite{avakian_optimizing_2005}[17]. 

Simple SCA attacks on a $k\bm{P}$ implementation exploit differences in $k\bm{P}$ trace parts corresponding to the processing of `0' and `1' values of $k$, allowing attackers to reveal the key $k$. 
Primarily, binary $k\bm{P}$ algorithms are vulnerable due to the distinguishability of sequences of mathematical operations for point doubling and point addition. If these point operations’ power shapes are distinguishable, attackers can infer the processed bits of $k$ from the sequence of these operations, revealing the private key.

To counteract SCA, the power shapes of the processing of key bit value `0' should be indistinguishable from that of `1' through regularization or randomization. In another word, an SCA-resistant algorithm aims to generate either very similar shapes for the processing of both `0' and `1', or totally random-looking shapes. The double-and-add-always algorithm proposed in 1999 by Coron \nocite{koc_resistance_1999}[11] is a regular $k\bm{P}$ algorithm performing a $2\bm{P}$ and a $\bm{P}+\bm{Q}$ for each bit, but it is inefficient due to dummy operations \nocite{hankerson_guide_2004}[6]. Atomic pattern algorithms, introduced in \nocite{chevallier-mames_low-cost_2004}[1], are a faster and more energy-efficient kind of $k\bm{P}$ algorithms with a reduced number of dummy operations. Different algorithmic techniques can further increase the resistance of $k\bm{P}$ implementations by removing, reducing or hiding the algorithm’s dependency on $k$. Examples include key masking or blinding, and randomizing steps within the $k\bm{P}$ algorithm. 

In summary, the distinguishability of processed key bits arises from:
\begin{enumerate}
    \item Differences in sequences of mathematical operations ($2\bm{P}$ vs. $\bm{P}+\bm{Q}$), resulting in different shapes of profiles for the computations. \\Existing countermeasures:
    \begin{itemize}
        \item Dummy EC point addition operations: double-and-add-always \nocite{hankerson_guide_2004}[6]
        \item Universal formulae for $2\bm{P}$ and $\bm{P}+\bm{Q}$ calculations using the same sequence of field operations \nocite{goos_montgomery_2003}[18]
        \item Atomicity, e.g., \nocite{chevallier-mames_low-cost_2004}[1]
    \end{itemize}
    \item Key-dependent data processing differences for $2\bm{P}$ vs. $\bm{P}+\bm{Q}$ (data-bit vulnerability), usually exploited by vertical SCA. \\Algorithmic data randomization techniques \nocite{koc_resistance_1999}[11] are well-known effective countermeasures:
    \begin{itemize}
        \item Key randomization
        \item Projective coordinates randomization
        \item Point blinding
    \end{itemize}
    \item Key-dependent differences in the addressing of different blocks/registers in $2\bm{P}$ vs. $\bm{P}+\bm{Q}$ (address-bit vulnerability), usually exploited by vertical SCA \nocite{goos_address-bit_2003}[19]. Algorithmic address randomization \nocite{goos_address-bit_2003, izumi_improved_2010}[19, 20] is a known effective countermeasure, but it is not effective against horizontal attacks according to \nocite{kabin_randomized_2023}[21].
\end{enumerate}

Binary $k\bm{P}$ algorithms, such as regular Montgomery ladder using Lopez-Dahab projective coordinates \nocite{goos_software_2000}[22], atomic pattern algorithms \nocite{francillon_revisiting_2014}[23] or even Montgomery ladder with randomized processing of point operations utilizing a universal point addition formula \nocite{batina_side-channel_2010}[24], are vulnerable to horizontal address-bit SCA \nocite{kabin_breaking_2020, kabin_horizontal_2017, sigourou_successful_2023}[25-27], at least when implemented in hardware. 

This work only focuses on atomic pattern algorithms [2] as SSCA countermeasures for embedded devices. The goal is to investigate if the atomic patterns are vulnerable to SSCA and which kind of vulnerability - data-bit or address-bit - is dominant.

\section{Atomicity as a Countermeasure Against SSCA}\label{atomicity}
Numerous proposals have been put forward to prevent SSCA. The side-channel atomicity principle introduced in 2004 by Chevallier-Mames et al. \nocite{chevallier-mames_low-cost_2004}[1] aims to counteract SSCA and is especially suitable for embedded devices, due to the significantly reduced computation burden in comparison to traditional double-and-add-always $k\bm{P}$ algorithms.  Extensive research have been dedicated to enhancing this countermeasure by Longa [2] and by Giraud and Verneuil \nocite{hutchison_atomicity_2010}[28], who proposed new – optimized – atomic pattern algorithms to improve the computation efficiency of scalar multiplication.

In the atomicity principle, each EC point operation is viewed as a process, whereby a process (e.g., EC point doubling) is viewed as a set of blocks consisting of the same sequence of instructions. Two instruction sequences are said to be side-channel equivalent if they are indistinguishable through side-channel analysis. Each of these side-channel equivalent instruction sequences is called a common side-channel atomic block, and a series of these atomic blocks forms an atomic pattern for an EC operation. Each complex calculation process (e.g., $2\bm{P}$ or $\bm{P}+\bm{Q}$ operation) can then be expressed as a repetition of the side-channel atomic blocks represented as a repetition of the same atomic patterns, which carries a regular sequence of basic field operations. 

Side-channel atomicity involves careful design of cryptographic operations implementation such that individual operations (e.g., point doubling and point addition) are side-channel equivalent, which means upon execution, it is impossible to distinguish one operation from the other or even to identify the start and end of these operations. Note that this principle is based on the assumption that the loading/storing of data from/to different registers are equivalent operations \nocite{goos_montgomery_2003}[18]. 

Since this work focuses on investigating the potential vulnerabilities of the atomic pattern algorithms proposed by Longa [2], we specifically selected the MNAMNAA-based atomic pattern algorithms to discuss in greater detail in this thesis.

Table \ref{point-doubling} shows the atomic point doubling algorithm, which uses standard Jacobian coordinates. Table \ref{point-addition} presents the atomic mixed point addition algorithm
, which uses a mix of Jacobian and affine coordinates. 

\begin{table}[H]
\centering
\begin{tabular}{|l|l|l|l|l|}
\multicolumn{5}{l}{Input: $\bm{P}=(X_1,Y_1,Z_1)$}\\
\multicolumn{5}{l}{Output: $2\bm{P} = (X_3,Y_3,Z_3) = (T_1,T_2,T_3)$}\\
\multicolumn{5}{l}{$T_1\gets X_1,T_2 \gets Y_1, T_3 \gets Z_1$}\\ \hline
&\multicolumn{1}{|c|}{$\Delta1$} & \multicolumn{1}{c|}{$\Delta2$} & \multicolumn{1}{c|}{$\Delta3$} & \multicolumn{1}{c|}{$\Delta4$} \\ \hline  
M& $T_4=T_3^2$  & $T_5=T_4\times T_5$  & $T_5=T_4^2$ & $T_2=T_2^2$ \\ 
N& * & * & * & $T_5=-T_1$ \\
A& $T_5=T_1+T_4$  & $T_4=T_5+T_5$  & $T_6=T_2+T_2$ & $T_5=T_5+T_6$ \\
M& $T_6=T_2^2$ & $T_3=T_2\times T_3$ & $T_6=T_1\times T_6$ & $T_5=T_4\times T_5$ \\
N& $T_4=-T_4$ & * & $T_1=-T_6$ & $T_2=-T_2$ \\
A& $T_2=T_2+T_2$ &  $T_4=T_4+T_5$ & $T_1=T_1+T_1$ & $T_2=T_2+T_2$ \\
A& $T_4=T_1+T_4$ &  $T_2=T_6+T_6$ & $T_1=T_1+T_5$ & $T_2=T_2+T_4$ \\ \hline
\end{tabular}
\parbox{14.5cm}{\caption{\label{point-doubling}MNAMNAA-based atomic point doubling, taken from [2].}}
\end{table}

\begin{table}[H]
\centering
\begin{tabular}{|l|l|l|l|}
\multicolumn{4}{l}{Input: $\bm{P}=(X_1,Y_1,Z_1)$ and $\bm{Q}=(X_2,Y_2)$}\\
\multicolumn{4}{l}{Output: $\bm{P}+\bm{Q} = (X_3,Y_3,Z_3,X_1',Y_1') = (T_1,T_2,T_3,T_4,T_5)$}\\
\multicolumn{4}{l}{$T_1\gets X_1,T_2 \gets Y_1, T_3 \gets Z_1, T_x\gets X_2, T_y \gets Y_2$}\\ \hline
& \multicolumn{1}{|c|}{$\Delta1$} & \multicolumn{1}{c|}{$\Delta2$} & \multicolumn{1}{c|}{$\Delta3$}  \\ \hline

M& $T_4=T_3^2$ & $T_6=T_5^2$ & $T_9=T_5\times T_6$ \\
N& * & * & *   \\
A& * & * & $T_8=T_8+T_9$ \\
M& $T_5=T_x\times T_4$ & $T_7=T_1\times T_6$ & $T_4=T_3\times T_4$  \\
N& $T_6=-T_1$ & * & *  \\
A& $T_5=T_5+T_6$ & $T_8=T_1+T_1$ & *  \\
A& * & * & * \\ \hline

& \multicolumn{1}{|c|}{$\Delta4$} & \multicolumn{1}{c|}{$\Delta5$} & \multicolumn{1}{c|}{$\Delta6$} \\ \hline

M& $T_4=T_y\times T_4$ & $T_8=T_2\times T_9$ & $T_3=T_3\times T_5$ \\
N& $T_{10}=-T_2$ & $T_6=-T_1$ & $T_4=-T_7$\\
A& $T_4=T_4+T_{10}$ & $T_6=T_6+T_7$ & $T_4=T_1+T_4$ \\
M&$T_{10}=T_4^2$ &  $T_6=T_6\times T_{10}$ & $T_5=T_4^2$\\
N&$T_8=-T_8$ & $T_9=-T_8$ & $T_6=-T_8$\\
A&$T_1=T_6+T_8$ & $T_2=T_6+T_9$ & $T_6=T_2+T_6$\\
A&* & * & *  \\

\hline
\end{tabular}
\caption{\label{point-addition}MNAMNAA-based atomic point addition, taken from [2].}
\end{table}

Each $\Delta i$ with $i\in\{1,2,3,4,5,6\}$  in Table \ref{point-doubling} and Table \ref{point-addition} represents an atomic block. Within an atomic block, each line contains one field operation. In this case, it is either a multiplication (M), a negation (N) or an addition (A). Each * represents a dummy field operation conforming to the atomic pattern as a disguise, but the dummy operations are unnecessary for the overall result calculation. Following the atomic pattern MNAMNAA, point doubling can be computed in 4 atomic blocks, and point addition in 6 atomic blocks. Attackers can then only observe a sequence of side-channel atomic block executions and are unable to distinguish the underlying operations using SSCA, since side-channel information like timing, power consumption and memory access patterns are theoretically equivalent for every atomic block.

\section{Vulnerabilities of the Atomic Patterns}\label{atomic-patterns-vulnerabilities}
\subsection{Key-dependent Data Processing}\label{data-bit}
Key-dependent differences in the shape of measured traces can arise not only from different sequences of instructions but also from processing different data within the same instruction or operation. These differences are typically small and dependent on the Hamming distance of the data processed. Various statistical methods, such as correlation analysis, can exploit these differences to reveal the key. This is often referred to as exploiting data-dependent vulnerabilities through data-bit SCA. While data-dependent vulnerabilities are usually exploited in vertical attacks, there are known successful horizontal attacks against $k\bm{P}$ implementations that also leverage data-dependent vulnerabilities.

Horizontal Collision Correlation Analysis (HCCA) \nocite{bauer_horizontal_2015}[12] is an attack that exploits data-bit vulnerabilities on elliptic curves atomic implementations with input randomization analysing a single $k\bm{P}$ execution trace. The main assumption is that the adversary can detect when two field multiplications have at least one operand in common or no shared operand at all. The HCCA attack can be used to detect distinguishing property of sharing of operands among EC point doubling and point addition operations, even if they are SCA-protected by applying atomicity principle. It can thus reveal the underlying secret key bit successively and expose the entire secret key.
Bauer et al.'s work \nocite{bauer_horizontal_2015}[12] demonstrated through experiments that the atomic pattern proposed by Chevallier et al. in \nocite{chevallier-mames_low-cost_2004}[1] is vulnerable to HCCA. In addition, they pointed out that Longa's MANA atomic pattern [2] and Giraud and Verneuil's atomic pattern \nocite{hutchison_atomicity_2010}[28] are theoretically vulnerable to HCCA too, due to their similarities to Chevallier et al.'s atomic pattern.

\subsection{Key-dependent Addressing of Registers}\label{address-bit}
An address-bit vulnerability in devices performing cryptographic operations arises from the unintended leakage of information through observable memory access patterns. For instance, an attacker can analyse the power consumption associated with key-dependent addressing registers to gain insights into cryptographic keys or other confidential data. This is possible because the addresses of different registers induce varying bus activity for the connection of the addressed blocks, particularly in hardware implementations. Consequently, this can serve as a source of side-channel leakage.

The power consumption or electromagnetic emanation during a $k\bm{P}$ execution fluctuates based on the executed operations and the Hamming distances of data transmitted within the device. Additionally, this also relates to the Hamming distances of the registers' or blocks' addresses used during the $k\bm{P}$ execution. The address-bit vulnerability stems from the requirement that a register must have an address assigned to it and has to be "called" before data can be loaded into or from it. If different register addresses are used within the same instruction sequence, the power consumption or electromagnetic emanation of the device will vary according to the differences in the Hamming weights of these register addresses. This dependency is indirect and often challenging to measure.

In 1999, the side-channel analysis attack - address-bit Differential Power Analysis (DPA) was first introduced by Messerges et al. \nocite{messerges_investigations_1999}[29]. The correlation between the addresses of registers and the secret information was analysed, using hundreds of power traces of a single encryption operation\footnote{The same inputs were used every time, without input randomization.}. They claimed that the horizontal address-bit SCA attack is powerful in revealing bit transition information. However, due to the lack of concrete results in their experiments, the address-bit DPA did not attract significant attention. Then, Itoh et al. \nocite{goos_address-bit_2003}[19] extended address-bit DPA to ECC design implementing Montgomery ladder, showing that the multiple-trace attack is successful when a small number of registers are used in the algorithm. Further research by Kabin et al. \nocite{kabin_horizontal_2017}[26] proposed that the addressing of source and destination registers or blocks going through the bus in a $k\bm{P}$ hardware implementation can be distinguishable using horizontal SCA attacks, i.e., analysing only a single $k\bm{P}$ execution trace. In 2023, Kabin \nocite{kabin_horizontal_2023}[3] described successful horizontal address-bit SCA attacks against hardware accelerators of $k\bm{P}$ algorithms implementing highly regular Montgomery ladder as well as an atomic pattern $k\bm{P}$ algorithm. 

\chapter{State-of-the-art Literature Overview} 

\label{Chapter3} 
We carried out the literature overview to understand the progress made by researchers over the years regarding the atomic pattern algorithms proposed in [2], to validate our initial idea that no one before us has ever done a SCA on an implementation of Longa’s MNAMNAA-based atomic pattern algorithm on an embedded device. Also, we used the literature overview to assess if there was any previous work we could compare our thesis’s analysis results with, to help us draw more comprehensive conclusions.

\begin{figure}[H]
\centering
\includegraphics[width=0.75\textwidth]{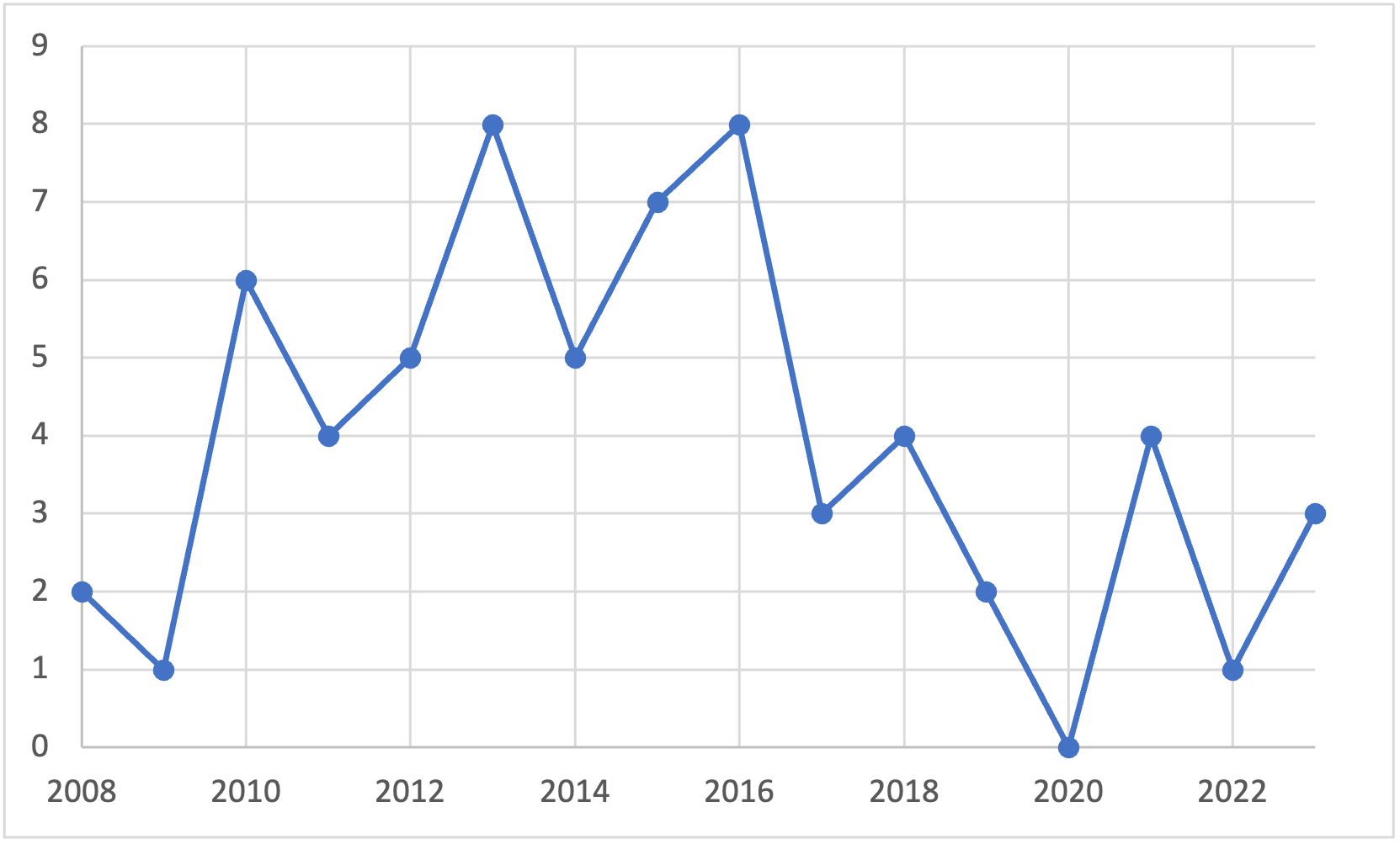}
\caption{Number of published papers that referenced [2] over the years.}\label{fig:papers-years}
\end{figure}

Using Google Scholar and Semantic Scholar as search engines at the end of 2023, we reviewed all of the known literature which cited Longa’s master's thesis [2]. There are 75 literature cited [2] reported by Google Scholar and 64 reported by Semantic Scholar, summing to 84 distinct records in 4 languages altogether. However, only 63 out of the 84 records are nonduplicate, available to view and have truly cited [2]. From Figure \ref{fig:papers-years}, we can see that [2] demonstrated certain influence on others’ research activities over the past 15 years continually.

We have studied all these 63 reference papers \nocite{kabin_horizontal_2023, bauer_horizontal_2015, kabin_randomized_2023, francillon_revisiting_2014, hutchison_atomicity_2010, sigourou_successful_2023, lu_general_2013, verneuil_elliptic_2012, das_improved_2016, kabin_ec_2021-1, kabin_ec_2021, longa_high-speed_2011, faye_algorithmes_2014, venelli_contribution_2011, houssain_elliptic_2012, houssain_power_2012, tawalbeh_towards_2016, gebotys_elliptic_2010, purohit_fast_2011, hutchison_recoding_2012,purohit_elliptic_2012, chabrier_arithmetic_2013, chabrier_--fly_2013, reyes_performance_2013, dygin_efficient_2013, hutchison_joint_2013, hutchison_expansion_2013, li_fast_2014, al_saffar_improved_2014, ajeena_point_2014, ahmad_x-tract_2015, lopez_analysis_2015, dou_fast_2015, _jacobian_2015, __2015,  das_secure_2016, romaniuk_usage_2016, quetny__2016, guerrini_randomized_2018, khleborodov_fast_2018, ho_kim_speeding_2020, cai_handshake_2018, wojcik_partially_2018, khleborodov_fast_2018-1, yq_dou_revisiting_2018, ilyenko_perspectives_2019, andres_lara-nino_comparison_2021, shuang-gen_liu_fast_2021, xie_secure_2022, longa_setting_2008, longa_analysis_2010, chung_fast_2011, adikari_hybrid_2011, almohaimeed_increasing_2013, meier_side-channel_2014, das_exploiting_2015, sako_side-channel_2016, das_inner_2016, das_automatic_2019, ryan_safe-errors_2016, cramer_new_2008, jarecki_fast_2009, hutchison_efficient_2010}[3, 12, 21, 23, 27, 28, 30–86] and classified them corresponding to the context, in which the atomic pattern algorithm [2] was referred. 

\begin{figure}[H]
\centering
\includegraphics[width=0.8\textwidth]{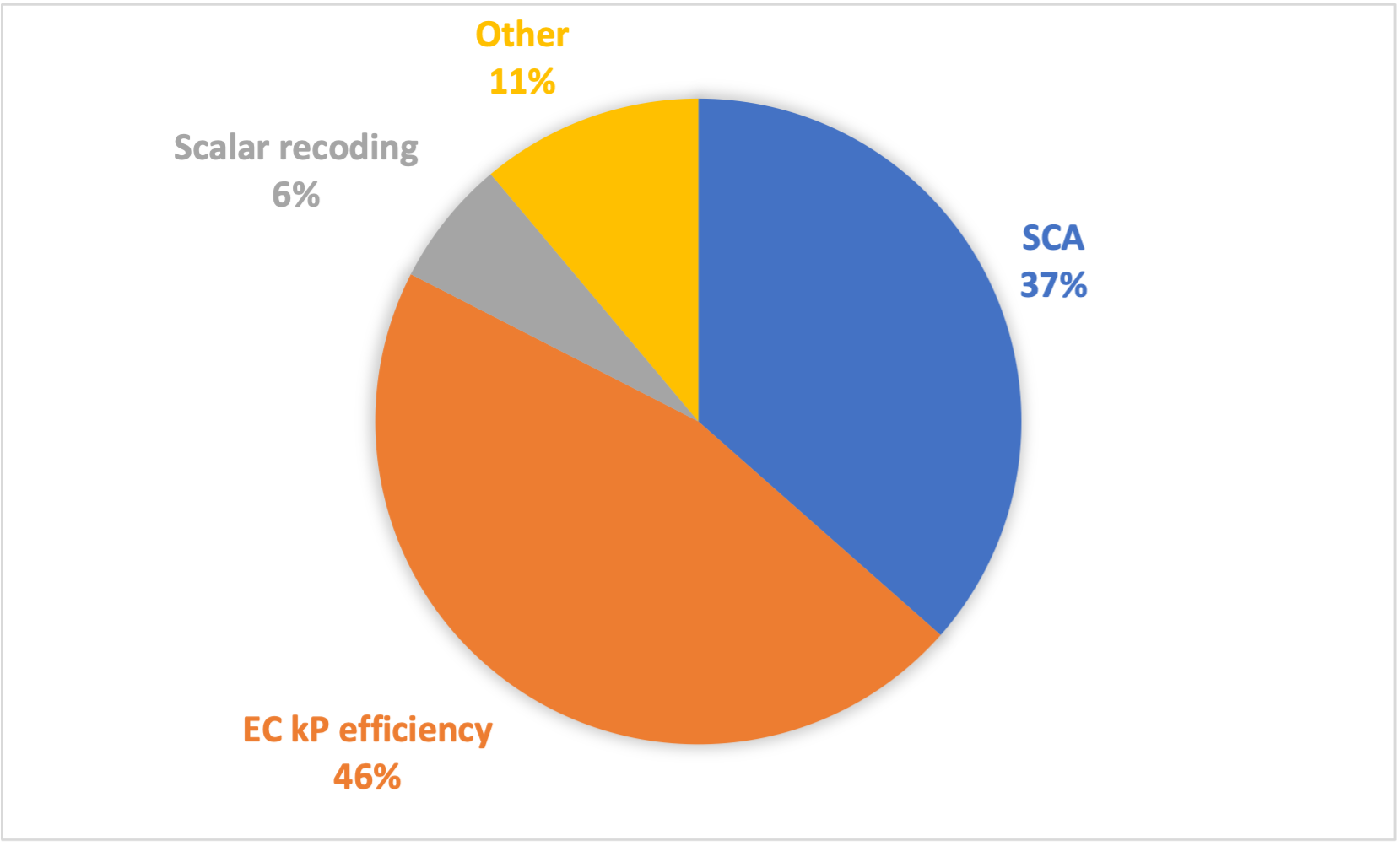}
\caption{Distribution of themes of literature referencing [2].}
\label{fig:papers-themes}
\end{figure}

\noindent The themes of reference papers citing [2] include:
\begin{itemize}
    \item SCA and/or SCA countermeasures,
    \item EC $k\bm{P}$ algorithm efficiency optimization or analysis without any modifications of the atomic patterns,
    \item recoding algorithms, and
    \item other miscellaneous topics
\end{itemize}
The results of the literature investigation are represented in Chapter \ref{Appendix1} (see Table \ref{tab:lit-overview}). Figure \ref{fig:papers-themes} represents the results of the literature investigation graphically.

As shown in Figure \ref{fig:papers-themes}, of all the reference papers citing [2], 23 papers, i.e., 37\%, are about SCA and/or SCA countermeasure \nocite{hutchison_atomicity_2010, lu_general_2013, verneuil_elliptic_2012, francillon_revisiting_2014, bauer_horizontal_2015, das_improved_2016, venelli_contribution_2011, houssain_elliptic_2012, houssain_power_2012, tawalbeh_towards_2016, kabin_ec_2021, kabin_ec_2021-1, kabin_horizontal_2023, sigourou_successful_2023, kabin_randomized_2023, almohaimeed_increasing_2013, meier_side-channel_2014, das_exploiting_2015, sako_side-channel_2016, das_inner_2016, ryan_safe-errors_2016, das_automatic_2019}[3, 12, 21, 23, 27, 28, 30–34, 37–40, 77–
83]. It prompted us to further examine the improvements, implementations and analysis made based on Longa’s atomic pattern algorithms. 15 papers \nocite{hutchison_atomicity_2010, lu_general_2013, verneuil_elliptic_2012, francillon_revisiting_2014, bauer_horizontal_2015, das_improved_2016, venelli_contribution_2011, houssain_elliptic_2012, houssain_power_2012, tawalbeh_towards_2016, kabin_ec_2021, kabin_ec_2021-1, kabin_horizontal_2023, sigourou_successful_2023, kabin_randomized_2023}[3, 12, 21,
23, 27, 28, 30–34, 37–40] out of 23 mentioned [2] as an example of the state-of-the-art countermeasure against SSCA. In the following section of this thesis, we will delve into the enhancement of SCA resistance and the implementation of atomicity as a SCA countermeasure, specifically focusing on Longa’s [2] atomic patterns. We selected 15 out of 23 papers from the category "SCA" (see Figure \ref{fig:papers-themes}) for a detailed overview as they are highly relevant to this work.

\section{Analysis of MNAMNAA-based Algorithms}\label{mnamnaa-analysis}
In this subsection, we summarize the reference papers \nocite{hutchison_atomicity_2010, lu_general_2013, verneuil_elliptic_2012, francillon_revisiting_2014, bauer_horizontal_2015, das_improved_2016, faye_algorithmes_2014, kabin_ec_2021, kabin_ec_2021-1, kabin_horizontal_2023, kabin_randomized_2023, sigourou_successful_2023, venelli_contribution_2011, houssain_elliptic_2012, houssain_power_2012, tawalbeh_towards_2016, longa_high-speed_2011}[3, 12, 21, 23, 27, 28, 30–40] that cited Longa’s atomic pattern algorithms as state-of-the-art, used it as a basis for improvement, and/or those which analysed an atomic patterns algorithm.

In 2010, Giraud and Verneuil \nocite{hutchison_atomicity_2010}[28] claimed that Longa's atomic patterns cannot defeat DSCA, as the DSCA countermeasures - projective coordinates randomization or random curve cannot be applied on his atomic pattern algorithms. They improved Longa’s algorithms by introducing new fast doubling, general doubling and readdition algorithms using Longa’s atomic patterns MNAMNAA, SNAMNAA and MNAMNAA, respectively. By maximizing the use of field squarings to replace multiplications, and minimizing the use of field additions and negations, they proposed a right-to-left mixed scalar multiplication algorithm. It was proven to be 10\% more efficient than any previous methods at the time. Later in 2012, Verneuil published his PhD thesis \nocite{verneuil_elliptic_2012}[31] covering the same contributions he made in \nocite{hutchison_atomicity_2010}[28].

Lu et al. \nocite{lu_general_2013}[30] introduced a general framework of side-channel atomicity to support the proposed $\tau$-scalar, $\xi$-base representation for scalar multiplication, which has shown improved resistance to simple power analysis and supports Longa’s single scalar multiplication algorithms. They used AMD Athlon X2 245-based hardware to test SPA-resistance on a 2-scalar multiplication algorithm. It shows that the implementation is SPA-resistant but with overheads. However, Lu et al. only implemented Chevallier-Mames et al.'s atomic pattern “MNAA” \nocite{chevallier-mames_low-cost_2004}[1] due to hardware space limitation. They claimed that their algorithm hypothetically should be able to accommodate Longa's atomic pattern without changing its effectiveness.

In 2013, Rondepierre \nocite{francillon_revisiting_2014}[23] proposed a new atomic pattern for EC point doubling and point addition operations for double scalar multiplications, which can be adapted to process single scalar multiplication too. He used the Straus-Shamir trick and scalar recoding to optimize the efficiency of his atomic pattern algorithms. He showed that the implementation of his single scalar multiplication algorithm on a smart card has an efficiency improvement of 40\% when compared to Giraud and Verneuil’s  algorithm implementation. Although he only showed memory cost reduction by optimizing intermediate registers used in his pattern and Giraud and Verneuil’s pattern, he expected that this optimization can also be applied on Longa’s pattern.

Bauer et al. \nocite{bauer_horizontal_2015}[12] introduced HCCA, and performed a theoretical analysis on Longa’s atomic pattern point doubling and point addition algorithms. They pointed out that Longa’s algorithms should be susceptible to HCCA. However, they only implemented Chevallier-Mames et al.’s atomic pattern scheme \nocite{chevallier-mames_low-cost_2004}[1] but not Longa’s in their experiments. They demonstrated that HCCA against Chevallier-Mames et al.’s atomic pattern algorithm works for 8- and 32-bit microprocessors and elliptic curves of size 160, 250 and 384 bits.

Das et al. \nocite{das_improved_2016}[32] recalled the proposition made by Bauer et al. in \nocite{bauer_horizontal_2015}[12] that Longa’s atomic pattern $k\bm{P}$ algorithm is vulnerable to HCCA, since the operand-sharing property among field operations can be detected. They observed that side-channel leakage between two field multiplications can be quantified by computing the distance between two leakages using Pearson correlation coefficient. In addition, they proposed a new protected atomic pattern to demonstrate how to make Giraud and Verneuil’s right-to-left atomic scheme safe against HCCA and the improved Big Mac attack \nocite{goos_sliding_2001}[87], with a small overhead. 

Kabin et al. \nocite{kabin_ec_2021, kabin_ec_2021-1}[33, 34] implemented the atomic pattern from \nocite{francillon_revisiting_2014}[23] in hardware, performed a horizontal SCA attack (adapting the technique \textit{comparison to the mean} to atomic patterns) and mentioned that all atomic pattern algorithms may be vulnerable to horizontal address-bit SCAs, at least when implemented in hardware.

Longa \nocite{longa_high-speed_2011}[35] evaluated the costs of point doubling and doubling-addition from [2] in three different processors, and optimized the $k\bm{P}$ computing costs in the proposed Longa-Miri precomputation (LM) scheme.

Faye \nocite{faye_algorithmes_2014}[36] evaluated Longa's special addition and point tripling to provide a summary of state-of-the-art algorithms for EC scalar multiplication in terms of efficiency. Also, he studied parallel calculation of $k\bm{P}$ using Longa's point doubling algorithm on a 3-processor architecture theoretically.

Venelli \nocite{venelli_contribution_2011}[37], Houssain \nocite{houssain_elliptic_2012}[38], Houssain et al. \nocite{houssain_power_2012}[39] and Lo'Ai et al. \nocite{tawalbeh_towards_2016}[40] all mentioned Longa's atomic patterns as state-of-the-art SCA countermeasure, but their works only serve as review papers without any implementation or improvement on the atomic pattern algorithms.

Kabin \nocite{kabin_horizontal_2023}[3], Kabin et al. \nocite{kabin_randomized_2023}[21] and Sigourou et al. \nocite{sigourou_successful_2023}[27] mentioned Longa's atomic patterns as state-of-the-art SCA countermeasure, performed successful SSCA against an atomic pattern $k\bm{P}$ algorithm (see \nocite{kabin_horizontal_2023, kabin_randomized_2023}[3, 21]) or Montgomery ladder (see \nocite{sigourou_successful_2023}[27]) in their works, but did not implement Longa's atomic pattern algorithms.

\begin{longtable}[H]{|p{22mm}|l|p{97mm}|}
\hline
Author(s) & Year & Contributions \\ \hline
                                        
\endfirsthead

\hline
Author(s) & Year & Contributions \\ \hline

\endhead


Giraud and Verneuil \nocite{hutchison_atomicity_2010}[28]
& 2010 & 
\begin{tabular}{l@{}l@{}l@{}}
\tabitem Claimed that Longa’s atomic patterns cannot defeat\\ DSCA.\\
\tabitem Proposed new composite operations using Longa’s\\ atomic patterns.\\
\tabitem Proposed right-to-left mixed scalar multiplication\\ algorithm.
\end{tabular}
\\ \hline
Verneuil \nocite{verneuil_elliptic_2012}[31] & 2012 & 
\begin{tabular}{l@{}l@{}l@{}}
\tabitem Covered the same contributions he made in \nocite{hutchison_atomicity_2010}[28].
\end{tabular}
\\ \hline
Lu et al. \nocite{lu_general_2013}[30] &  2013 & 
\begin{tabular}{l@{}l@{}l@{}l@{}}
\tabitem Introduced a general framework of $k\bm{P}$ side-channel \\atomicity.\\
\tabitem Experimented with AMD Athlon X2 245-based\\ hardware. \\
\tabitem Only implemented the atomic pattern “MNAA”. \\
\tabitem Their algorithm hypothetically can use Longa's\\ atomic pattern.
\end{tabular}
\\ \hline
Rondepierre \nocite{francillon_revisiting_2014}[23] & 2014 & 
\begin{tabular}{l@{}l@{}l@{}l@{}}
\tabitem Proposed a new atomic pattern for scalar\\ multiplications.\\
\tabitem Used Straus-Shamir trick and scalar recoding to \\optimize algorithms.\\
\tabitem Implemented single scalar multiplication algorithm\\ on a smart card.\\
\tabitem Showed memory cost reduction by optimizing \\intermediate registers. 
\end{tabular}
\\ \hline
Bauer et al. \nocite{bauer_horizontal_2015}[12] & 2015 &  
\begin{tabular}{l@{}l@{}l@{}l@{}}
\tabitem Introduced HCCA.\\
\tabitem Performed theoretical analysis on Longa’s atomic \\pattern algorithms.\\
\tabitem Claimed Longa’s algorithms should be susceptible\\ to SCCA.\\
\tabitem Only implemented Chevallier-Mames et al.’s atomic \\pattern scheme\nocite{chevallier-mames_low-cost_2004}[1].\\
\tabitem Demonstrated that HCCA against atomic pattern \\algorithm\nocite{chevallier-mames_low-cost_2004}[1] works.
\end{tabular}
\\ \hline
Das et al. \nocite{das_improved_2016}[32] & 2016 & 
\begin{tabular}{l@{}l@{}l@{}l@{}}
\tabitem Recalled that Longa’s atomic pattern algorithm is \\vulnerable to HCCA.\\
\tabitem Quantified side-channel leakage between two field \\multiplications.\\
\tabitem Proposed a new atomic pattern to safeguard the \\atomic scheme\nocite{hutchison_atomicity_2010}[28].
\end{tabular}
\\ \hline
Kabin et al. \nocite{kabin_ec_2021, kabin_ec_2021-1}[33, 34] & 2021 &  
\begin{tabular}{l@{}l@{}l@{}l@{}}
\tabitem Implemented the atomic pattern from \nocite{francillon_revisiting_2014}[23] in\\ hardware.\\
\tabitem Performed horizontal SCA attack.\\
\tabitem All atomic pattern algorithms may be vulnerable to \\their attack.\\
\end{tabular}
\\ \hline

Longa \nocite{longa_high-speed_2011}[35] & 2011 &  
\begin{tabular}{l@{}l@{}l@{}l@{}}
\tabitem Evaluated the costs of $2\bm{P}$ and $2\bm{P}+\bm{Q}$ from [2] in 3 \\different processors.\\
\tabitem Optimized the $k\bm{P}$ computing costs in the proposed\\ LM scheme.\\
\end{tabular}
\\ \hline

\begin{tabular}{l@{}l@{}l@{}l@{}}Faye \nocite{faye_algorithmes_2014}[36]\end{tabular} & 2014 &  
\begin{tabular}{l@{}l@{}l@{}l@{}}
\tabitem Evaluated Longa's special addition and point tripling.\\
\tabitem Summarized state-of-the-art $k\bm{P}$ algorithms in terms of \\efficiency.\\
\tabitem Studied parallel calculation of $k\bm{P}$ using Longa's $2\bm{P}$ \\algorithm.\\
\end{tabular}
\\ \hline

\begin{tabular}{l@{}l@{}}Venelli \nocite{venelli_contribution_2011}[37], \\Houssain \\ \nocite{houssain_elliptic_2012}[38], \\Houssain \\et al. \nocite{houssain_power_2012}[39] \\and Lo'Ai \\et al. \nocite{tawalbeh_towards_2016}[40]\end{tabular} & \begin{tabular}[c]{@{}l@{}}2011-\\2016\end{tabular} &  
\begin{tabular}{l@{}l@{}l@{}l@{}}
\tabitem Reviewed numerous papers related to SCA.\\
\tabitem Cited Longa's atomic patterns as state-of-the-art SCA \\countermeasure.\\
\end{tabular}
\\ \hline

\begin{tabular}{l@{}l@{}l@{}l@{}}Kabin \nocite{kabin_horizontal_2023}[3], \\Kabin et al. \\\nocite{kabin_randomized_2023}[21] and \\Sigourou\\ et al. \nocite{sigourou_successful_2023}[27]\end{tabular} & 2023 &  
\begin{tabular}{l@{}l@{}l@{}l@{}}
\tabitem Cited Longa's atomic patterns as state-of-the-art SCA \\countermeasure.\\
\tabitem Performed successful SSCA against atomic pattern $k\bm{P}$ \\algorithm \nocite{kabin_horizontal_2023, sigourou_successful_2023}[3, 27] or against Montgomery ladder \nocite{kabin_randomized_2023}[21].\\
\end{tabular}
\\ \hline

\caption{\label{tab:lit}Summary of literature citing Longa's atomic patterns as state-of-the-art, analysing or improving Longa's atomic patterns.}
\end{longtable}

We found that none of the papers listed in Table \ref{tab:lit} mentioned or reported about the implementation of Longa’s atomic pattern algorithms. Thus, the indistinguishability of Longa's atomic pattern was never investigated practically. The paper which is the most suitable for the comparison to our experiments is \nocite{sigourou_successful_2023}[27], in which the authors implemented an atomic pattern $k\bm{P}$ algorithm for an embedded device and performed SSCA analysing a measured electromagnetic trace.

\chapter{Board for Experiments} 
\label{Chapter4} 

\begin{figure}[H]
\centering
\includegraphics[width=0.7\textwidth]{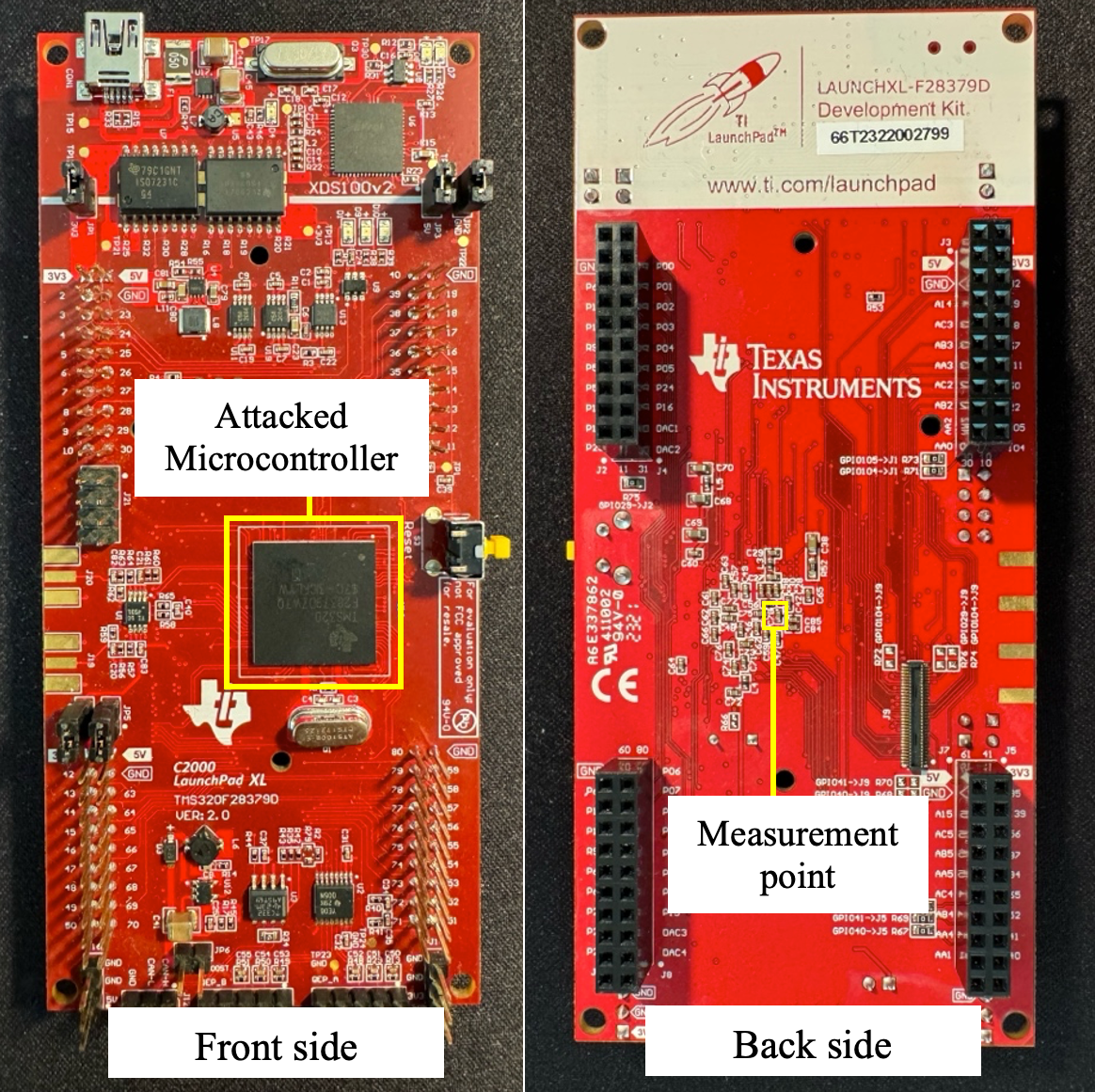}
\caption{Front side and back side of the board attacked.}\label{fig:board}
\end{figure}

As the target platform for our experiments, we used a Texas Instruments LAUNCHXL-F28379D LaunchPad. It is a development kit that provides on-board emulation for programming and debugging, along with a TMS320F28379D C2000 32-bit dual-core microcontroller as our target device. The microcontroller has 1 MB flash memory, 204 kB random-access memory (RAM) and operates real-time maximum at 200 MHz. Figure \ref{fig:board} shows the board attacked.

In Figure \ref{fig:board}, we can see the attacked microcontroller on the front side of the board, and our measurement point at the back of the microcontroller as indicated. More details for the measurement setup will be given in Chapter \ref{measurements}.

We used Code Composer Studio (CCS) version 12.4.0.00007 as our integrated development environment (IDE) for the microcontroller. It supports compilation of C and C++.

Since the target device is resource constrained, we need to find a suitable cryptographic library to implement the atomic pattern $k\bm{P}$ algorithm [2] such that the program is portable to the target device, among other criteria.


\chapter{Selection of a Suitable Open-source Cryptographic Library} 
\label{Chapter5} 

A recent work by Sigourou et al. \nocite{sigourou_successful_2023}[27] describes the implementation of the atomic pattern $k\bm{P}$ algorithm from Rondepierre \nocite{francillon_revisiting_2014}[23]. They have explored the five most popular open-source cryptographic libraries at the time, namely OpenSSL \nocite{openssl_software_foundation_openssl_2023}[88], MIRACL \nocite{mccusker_miracl_2019}[89], Crypto++ \nocite{dai_crypto_2023}[90], Cryptlib \nocite{gutmann_cryptlib_2021}[91] and FLECC in C (FLECC) \nocite{werner_flecc_2017}[92], regarding their suitability for implementation of ECC protocols (e.g., ECDSA, ECDH) on resource-constrained devices. Their study shows that FLECC is the best suited library among others. They considered each library’s suitability for embedded devices in terms of source code language used, portability to the attacked device in terms of memory size, access to modular arithmetic functions, and presence of constant-runtime functions. Their study used FLECC for the EC $k\bm{P}$ implementation of Rondepierre’s atomic pattern \nocite{francillon_revisiting_2014}[23] on a LAUNCHXL-F280025C 32-bit low-cost real-time microcontroller board with 128 kB flash memory and 24 kB RAM. However, even though some modular arithmetic functions from FLECC library are claimed to be constant-runtime, the authors of \nocite{sigourou_successful_2023}[27] observed that these functions are not truly time-constant at hardware level. 

Our work focuses on examining open-source cryptographic libraries to find the libraries which might truly have constant-runtime functions we need at hardware level, for the implementation of our chosen atomic pattern algorithms as a countermeasure against timing attacks on our target device, which is similar to but not the same as the one used in \nocite{sigourou_successful_2023}[27]. Since the study results on cryptographic libraries in \nocite{sigourou_successful_2023}[27] state that FLECC fulfilled all the criteria for the implementation of atomic $k\bm{P}$ algorithm on their resource-constrained device, we examined this library to see if our $k\bm{P}$ implementation developed using it can be run in our environment and our target device. 
Additionally, we examined the suitability of using two other libraries – Botan \nocite{botan_contributors_botan_2023}[93] and Mbed TLS \nocite{mbed_tls_contributors_mbed_2023}[94] – which also have constant-runtime functions according to a study by Jancar et al. \nocite{jancar_theyre_2022}[95]. Table \ref{libraries} shows the investigation results of the cryptographic libraries we looked into. Eventually, we selected FLECC for the algorithm implementation on our board. The reasons behind will be elaborated in the subsections.

\begin{table}[H]
\centering
\begin{tabular}{|c|c|c|c|c|}
\hline
 & Modular   & {Constant-runtime} & Suitable for  & {Portable to }\\ 
Library &  arithmetic & {modular arithmetic} &  our device  & {our board}\\
 &  functions & functions &  & \\ \hline
FLECC in C  & \checkmark & \checkmark  & \checkmark & \checkmark \\ \hline
Mbed TLS & \checkmark & {None} & \checkmark & Not checked \\ \hline
Botan & \checkmark & {\xmark Negation}  & \checkmark & Not checked       \\ 
 & & {\xmark Multiplication} & & \\ \hline
\end{tabular}
\caption{\label{libraries}Summary of cryptographic libraries studied.}
\end{table}

\section{FLECC in C}
FLECC in C (FLECC) \nocite{werner_flecc_2017}[92] is a C library supporting constant-runtime modular arithmetic functions. However, it does not have a single function supporting the modular multiplication of two numbers (i.e., $a \cdot b$ mod $p$) in constant time, but the available constant-time Montgomery modular multiplication can be applied twice for a single field product calculation.  Efficiency is not our concern in this thesis, as we only focus on studying the SCA-resistance provided by atomicity. Thus, we selected FLECC for our implementation on hardware, as according to the library description \footnote{https://github.com/IAIK/flecc\_in\_c/blob/develop/src/gfp/gfp\_const\_runtime.c}, it provides all the constant-time functions we needed for the algorithm implementation. We used the first version of FLECC and CCS version 12.4.0.00007 for the implementation.

\section{Mbed TLS}
Mbed TLS \nocite{mbed_tls_contributors_mbed_2023}[94] is a C library with modular arithmetic functions available and with some constant-runtime functions. In this thesis, we looked into Mbed TLS version 3.5.0. However, none of its modular arithmetic functions is declared as constant-runtime. Hence, this library was not selected for our implementation.

\section{Botan}
Botan \nocite{botan_contributors_botan_2023}[93] is a C++ library with modular arithmetic functions available and with a range of constant-runtime functions. We looked into Botan version 3.1.1 and found that the suitable constant-time functions provided are big integers modulo operation, big integers addition and Montgomery modular reduction. But it does not offer any constant-time field element multiplication or negation operation which are necessary for the implementation of our chosen atomic pattern [2]. Furthermore, it requires C++ 11 standard for compilation, which our CCS version does not support. Therefore, it was not selected for our implementation.
 

\chapter{Investigations} 

\label{Chapter6} 

\section{Implemented Atomic Patterns \textit{k\textbf{P}} Algorithm}\label{impl-atomic-kp}
We implemented Longa’s [2] MNAMNAA atomic patterns for EC point doubling (PD) and EC point addition (PA) computations required in binary $k\bm{P}$ algorithms, as mentioned in Chapter \ref{Chapter2}. 
To investigate the distinguishability of the atomic patterns, we used them in the binary left-to-right $k\bm{P}$ algorithm corresponding to Algorithm \ref{modified-kp}, which differs from Algorithm \ref{alg:binary-kp} in its initialization phase (see line 1). 
\\ \vspace{0.3cm}

\begin{algorithm}[H]
\DontPrintSemicolon
  
  \KwInput{$k=(k_{l-1},...,k_0)_{2},\bm{P}\in E(\mathbb F_p)$}
  \KwOutput{ $k\bm{P}$ }
  $\bm{Q}=\bm{P}$\\
  \For{$i=(l-2)$ downto 0}
  {
    $\bm{Q} = 2\bm{Q}$\tcp*{Point doubling atomic pattern (4 atomic blocks)}
    \If{$k_i=1$}
        {$\bm{Q}=\bm{Q}+\bm{P}$\tcp*{Point addition atomic pattern (6 atomic blocks)}}
    }
    return $\bm{Q}$

\caption{Modified left-to-right binary double-and-add $k\bm{P}$ algorithm}
\label{modified-kp}
\end{algorithm}
\vspace{0.3cm}

\noindent Our point doubling implementation follows formula (2.13) from [2]:
\begin{eqnarray} X_3=\alpha^2-2\beta,  Y_{3}=\alpha(\beta-X_3)-8Y_{1}^{4},  Z_3=2Y_{1}Z_{1} \end{eqnarray}
where $\alpha=3X_1^2+aZ_1^4, \beta = 4X_1Y_1^2$
\vspace{0.5cm} \\
During the implementation and functionality tests, we found four erroneous registers used in Longa’s MNAMNAA atomic patterns point doubling and point addition [2]. We examined the correspondence of the atomic patterns to formulae proposed in [2] and corrected the atomic patterns correspondingly. The corrected atomic patterns are represented in Table \ref{impl-point-doubling} and Table \ref{impl-point-addition}. 

\begin{table}[H]
\centering
\begin{tabular}{|l|l|l|l|l|}
\multicolumn{4}{l}{Input: $\bm{P}=(X_1,Y_1,Z_1)$}\\
\multicolumn{4}{l}{Output: $2\bm{P} = (X_3,Y_3,Z_3) = (T_1,T_2,T_3)$}\\
\multicolumn{4}{l}{$T_1\gets X_1,T_2 \gets Y_1, T_3 \gets Z_1$}\\ \hline
\multicolumn{1}{|c|}{}&\multicolumn{1}{c|}{$\Delta1$} & \multicolumn{1}{c|}{$\Delta2$} & \multicolumn{1}{c|}{$\Delta3$} & \multicolumn{1}{c|}{$\Delta4$} \\ \hline
M& $T_4=T_3^2$  & $T_5=T_4\cdot T_5$  & $T_5=T_4^2$ & $T_2=T_2^2$ \\ 
N&* & * & * & $T_5=-T_1$ \\
A&$T_5=T_1+T_4$  & $T_4=T_5+T_5$  & $T_6=T_2+T_2$ & $T_5=T_5+T_6$ \\
M&$T_6=T_2^2$ & $T_3=T_2\cdot T_3$ & $T_6=T_1\cdot T_6$ & $T_5=T_4\cdot T_5$ \\
N&$T_4=-T_4$ & * & $T_1=-T_6$ & $T_2=-T_2$ \\
A&$T_2=T_2+T_2$ &  $T_4=T_4+T_5$ & $T_1=T_1+T_1$ & $T_2=T_2+T_2$ \\
A&$T_4=T_1+T_4$ &  $T_2=T_6+T_6$ & $T_1=T_1+T_5$ & $T_2=T_2+$\hl{$\pmb{T_5}$} \\ \hline
\end{tabular}
\caption{\label{impl-point-doubling}MNAMNAA atomic patterns sequence for EC point doubling implemented in this work.}
\end{table}

\begin{table}[H]
\centering
\begin{tabular}{|l|l|l|l|}
\multicolumn{4}{l}{Input: $\bm{P}=(X_1,Y_1,Z_1)$ and $\bm{Q}=(X_2,Y_2)$}\\
\multicolumn{4}{l}{Output: $\bm{P}+\bm{Q} = (X_3,Y_3,Z_3,X_1',Y_1') = (T_1,T_2,T_3,T_4,T_5)$}\\
\multicolumn{4}{l}{$T_1\gets X_1,T_2 \gets Y_1, T_3 \gets Z_1, T_x\gets X_2, T_y \gets Y_2$}\\ \hline
\multicolumn{1}{|c|}{}&\multicolumn{1}{c|}{$\Delta1$} & \multicolumn{1}{c|}{$\Delta2$} & \multicolumn{1}{c|}{$\Delta3$} \\ \hline

M& $T_4=T_3^2$ & $T_6=T_5^2$ & $T_9=T_5\cdot T_6$ \\
N&* & * & *  \\
A&* & * & $T_8=T_8+T_9$   \\
M&$T_5=T_x\cdot T_4$ & $T_7=T_1\cdot T_6$ & $T_4=T_3\cdot T_4$    \\
N&$T_6=-T_1$ & * & *   \\
A&$T_5=T_5+T_6$ & $T_8=$\hl{$\pmb{T_7}$}+\hl{$\pmb{T_7}$} & *    \\
A&* & * & *  \\ \hline

\multicolumn{1}{|c|}{} & \multicolumn{1}{c|}{$\Delta4$} & \multicolumn{1}{|c|}{$\Delta5$} & \multicolumn{1}{c|}{$\Delta6$} \\ \hline
M& $T_4=T_y\cdot T_4$ & $T_8=T_2\cdot T_9$ & $T_3=T_3\cdot T_5$  \\
N& $T_{10}=-T_2$ & $T_6=-T_1$ & $T_4=-T_7$ \\
A& $T_4=T_4+T_{10}$ & $T_6=T_6+T_7$ & $T_4=T_1+T_4$ \\
M& $T_{10}=T_4^2$ &$T_6=T_6\cdot$ \hl{$\pmb{T_{4}}$} & $T_5=T_4^2$ \\
N& $T_8=-T_8$ & $T_9=-T_8$ & $T_6=-T_8$ \\
A& $T_1=$\hl{$\pmb{T_{10}}$}$+T_8$ & $T_2=T_6+T_9$ & $T_6=T_2+T_6$ \\
A& * & * & * \\ \hline
\end{tabular}
\caption{\label{impl-point-addition}MNAMNAA atomic patterns sequence for EC point addition implemented in this work.}
\end{table}

\noindent The corrections are shown in bold with yellow highlight. The corrected patterns for EC operations were then verified to be able to produce correct results in point doubling and point addition by being implemented additionally as software code.

In this work, we examined the correctness of all the other atomic patterns proposed in [2]. The corrected patterns are presented in Chapter \ref{Appendix2}.

Accordingly, we implemented point doubling and point addition operations corresponding to Longa’s MNAMNAA atomic patterns as shown in Table \ref{impl-point-doubling} and Table \ref{impl-point-addition} using the open-source library FLECC to form the core part of $k\bm{P}$ computation of Algorithm \ref{modified-kp}. The implementation was done for the secp256r1 elliptic curve in FLECC, which is the EC denoted as P-256 by NIST \nocite{chen_digital_2023}[7]. For the calculation of a single field product, we always used the constant-runtime Montgomery modular multiplication function available in FLECC twice, similar to the implementation in \nocite{sigourou_successful_2023}[27]. Montgomery modular multiplication \nocite{montgomery_modular_1985}[96] works by first transforming the multipliers into Montgomery space, where modular multiplication can be performed inexpensively, and then transforming them back when their actual values are needed. 
The Montgomery space is defined by the modulo $n$ and a positive integer $R \geq n$ coprime to $n$. The Montgomery multiplication based on Separated Operand Scanning (SOS) method \nocite{kaya_koc_analyzing_1996}[97] provided by FLECC library that we use involves modulo and division by $R$, where $R=2^{sw}$, $w$ is the computer’s word size in bits and $s$ is the number of words required to represent a large prime. In our case, we use the default values from the library  $R=2^{(256)(8)}=2^{2048}$ and $n$\footnote{$n$= 0xFFFFFFFF 00000000 FFFFFFFF FFFFFFFF BCE6FAAD A7179E84 F3B9CAC2 FC632551, the same number as modulo $p$.}. In order to implement the field product calculation as a constant-runtime function, we have to execute two Montgomery modular multiplications for each finite field multiplication in Longa’s atomic patterns. We denote here a single Montgomery modular multiplication function as "X". This approach highly deteriorates the performance of the original design of the atomic pattern MNAMNAA, as the atomic pattern now becomes XXNAXXNAA, with doubled the number of multiplications, which are the most computationally expensive operations among other operations used. Using the FLECC library, a multiplication ($c=a\cdot b$ mod $p$) will be implemented as the following two steps: $c=a\cdot b\cdot R^{-1}$ and $c=c\cdot R^2\cdot R^{-1}$. All implementation steps are shown in Table \ref{op-table} in details. The implemented $k\bm{P}$ algorithm can be found in \nocite{li_elliptic_2024}[98].



\clearpage
\begin{landscape}
\begin{longtable}[H]{|l|l|ll|ll|}

\hline

\multirow{6}{*}{\rotatebox{90}{Atomic blocks}}                       & \multirow{5}{*}{}      & \multicolumn{2}{l|}{EC Point Doubling}                                      & \multicolumn{2}{l|}{EC Point Addition}  \\
                                        &                        & \multicolumn{2}{l|}{Input: $\bm{P}=(X_1,Y_1,Z_1)$}                                                  & \multicolumn{2}{l|}{Input: $\bm{P}=(X_1,Y_1,Z_1), \bm{Q}=(X_2,Y_2)$}   \\
                                        &                        & \multicolumn{2}{l|}{Output: $2\bm{P}=(X_3,Y_3,Z_3)=(T_1,T_2,T_3)$}                                             & \multicolumn{2}{l|}{Output: $\bm{P}+\bm{Q}=(X_3,Y_3,Z_3,X_1',Y_1')=(T_1,T_2,T_3,T_4,T_5)$}  \\
                                        &                        & \multicolumn{2}{l|}{$T_1\gets X_1,T_2 \gets Y_1, T_3 \gets Z_1$}                                        & \multicolumn{2}{l|}{$T_1\gets X_1,T_2 \gets Y_1, T_3 \gets Z_1, T_x\gets X_2,T_y \gets Y_2$} \\ \cline{3-6}
                 & Operations   & \multicolumn{1}{l|}{Original atomic pattern}                   & Our implementation using& \multicolumn{1}{l|}{Original atomic pattern}                   & Our implementation using  \\
                                        &                        & \multicolumn{1}{l|}{corresponding to Table \ref{impl-point-doubling}}                    & constant-time FLECC functions & \multicolumn{1}{l|}{corresponding to Table \ref{impl-point-addition}}                    & constant-time FLECC functions \\
\hline

\endfirsthead

\hline
& \multirow{3}{*}{}    & \multicolumn{2}{l|}{EC Point Doubling}                      & \multicolumn{2}{l|}{EC Point Addition}   \\ \cline{3-6}
                 & Operations   & \multicolumn{1}{l|}{Original atomic pattern}                   & Our implementation using& \multicolumn{1}{l|}{Original atomic pattern}                   & Our implementation using  \\
                                        &                        & \multicolumn{1}{l|}{corresponding to Table \ref{impl-point-doubling}}                    & constant-time FLECC functions & \multicolumn{1}{l|}{corresponding to Table \ref{impl-point-addition}}                    & constant-time FLECC functions \\ 
                                        \hline
\endhead

\multirow{9}{*}{$\Delta1$} & \multirow{2}{*}{OP1:M} & \multicolumn{1}{l|}{\multirow{2}{*}{$T_4 \gets T_3 \cdot T_3$}}                                & $T_4 \gets T_3 \cdot T_3 \cdot R^{-1}$                   & \multicolumn{1}{l|}{\multirow{2}{*}{$T_4 \gets T_3 \cdot T_3$}}  & $T_4 \gets T_3 \cdot T_3 \cdot R^{-1}$ \\
                                        &    & \multicolumn{1}{l|}{}    &  $T_4 \gets T_4 \cdot R^2 \cdot R^{-1}$  & \multicolumn{1}{l|}{}  &  $T_4 \gets T_4 \cdot R^2 \cdot R^{-1}$  \\ \cline{2-6} 
                                        & OP2:N                  & \multicolumn{1}{l|}{*} & $T_0 \gets -T_1$ & \multicolumn{1}{l|}{*} & $T_0 \gets -T_1$ \\ \cline{2-6} 
                                        & OP3:A                  & \multicolumn{1}{l|}{$T_5 \gets T_1 + T_4$} & $T_5 \gets T_1 + T_4$ & \multicolumn{1}{l|}{*} & $T_5 \gets T_1 + T_4$ \\ \cline{2-6} 
                                        & \multirow{2}{*}{OP4:M} & \multicolumn{1}{l|}{\multirow{2}{*}{$T_6 \gets T_2 \cdot T_2$}} & $T_6 \gets T_2 \cdot T_2 \cdot R^{-1}$ & \multicolumn{1}{l|}{\multirow{2}{*}{$T_5 \gets T_x \cdot T_4$}}  & $T_5 \gets T_x \cdot T_4 \cdot R^{-1}$ \\
                                        &                        & \multicolumn{1}{l|}{} & $T_6 \gets T_6 \cdot R^2 \cdot R^{-1}$ & \multicolumn{1}{l|}{} &  $T_5 \gets T_5 \cdot R^2 \cdot R^{-1}$ \\ \cline{2-6} 
                                        & OP5:N                  & \multicolumn{1}{l|}{$T_4 \gets -T_4$} & $T_4 \gets -T_4$ & \multicolumn{1}{l|}{$T_6 \gets -T_1$} & $T_6 \gets -T_1$ \\ \cline{2-6} 
                                        & OP6:A                  & \multicolumn{1}{l|}{$T_2 \gets T_2 + T_2$} & $T_2 \gets T_2 + T_2$ & \multicolumn{1}{l|}{$T_5 \gets T_5 + T_6$} & $T_5 \gets T_5 + T_6$ \\ \cline{2-6} 
                                        & OP7:A                  & \multicolumn{1}{l|}{$T_4 \gets T_1 + T_4$} & $T_4 \gets T_1 + T_4$ & \multicolumn{1}{l|}{*} & $T_0 \gets T_1 + T_4$ \\ \hline \hhline{|=|=|=|=|=|=|}
\multirow{9}{*}{$\Delta2$} & \multirow{2}{*}{OP8:M} & \multicolumn{1}{l|}{\multirow{2}{*}{$T_5 \gets T_4 \cdot T_5$}} & $T_5 \gets T_4 \cdot T_5 \cdot R^{-1}$  & \multicolumn{1}{l|}{\multirow{2}{*}{$T_6 \gets T_5 \cdot T_5$}}  & $T_6 \gets T_5 \cdot T_5 \cdot R^{-1}$   \\
                        &    & \multicolumn{1}{l|}{}  & $T_5 \gets T_5 \cdot R^2 \cdot R^{-1}$ & \multicolumn{1}{l|}{}  & $T_6 \gets T_6 \cdot R^2 \cdot R^{-1}$ \\ \cline{2-6} 
                        & OP9:N                  & \multicolumn{1}{l|}{*}  &  $T_0 \gets -T_2$ & \multicolumn{1}{l|}{*} &  $T_0 \gets -T_2$  \\ \cline{2-6} 
                        & OP10:A                  & \multicolumn{1}{l|}{$T_4 \gets T_5 + T_5$}  & $T_4 \gets T_5 + T_5$ & \multicolumn{1}{l|}{*}                                 &  $T_0 \gets T_5 + T_5$ \\ \cline{2-6} 
                        & \multirow{2}{*}{OP11:M} & \multicolumn{1}{l|}{\multirow{2}{*}{$T_3 \gets T_2 \cdot T_3$}} & $T_3 \gets T_2 \cdot T_3 \cdot R^{-1}$ & \multicolumn{1}{l|}{\multirow{2}{*}{$T_7 \gets T_1 \cdot T_6$}} & $T_7 \gets T_1 \cdot T_6 \cdot R^{-1}$ \\
                        &     & \multicolumn{1}{l|}{} &  $T_3 \gets T_3 \cdot R^2 \cdot R^{-1}$ & \multicolumn{1}{l|}{} & $T_7 \gets T_7 \cdot R^2 \cdot R^{-1}$ \\ \cline{2-6} 
                        & OP12:N                  & \multicolumn{1}{l|}{*} & $T_0 \gets -T_4$ & \multicolumn{1}{l|}{*} & $T_0 \gets -T_4$ \\ \cline{2-6} 
                        & OP13:A                  & \multicolumn{1}{l|}{$T_4 \gets T_4 + T_5$} & ${T_4 \gets T_4 + T_5}$ & \multicolumn{1}{l|}{$T_8 \gets$ \hl{$\pmb{T_7}$} $+$ \hl{$\pmb{T_7}$}}  & {$T_8 \gets T_7 + T_7$} \\ \cline{2-6} 
                        & OP14:A                 & \multicolumn{1}{l|}{$T_2 \gets T_6 + T_6$}  & {$T_2 \gets T_6 + T_6$} & \multicolumn{1}{l|}{*} & {$T_0 \gets T_0 + T_0$}\\ \hline \hhline{|=|=|=|=|=|=|}
\clearpage
\multirow{9}{*}{$\Delta3$} & \multirow{2}{*}{OP15:M} & \multicolumn{1}{l|}{\multirow{2}{*}{$T_5 \gets T_4 \cdot T_4$}} & {$T_5 \gets T_4 \cdot T_4 \cdot R^{-1}$} & \multicolumn{1}{l|}{\multirow{2}{*}{$T_9 \gets T_5 \cdot T_6$}} & {$T_9 \gets T_5 \cdot T_6 \cdot R^{-1}$} \\
                        &                        & \multicolumn{1}{l|}{} & {$T_5 \gets T_5 \cdot R^2 \cdot R^{-1}$} & \multicolumn{1}{l|}{} & {$T_9 \gets T_9 \cdot R^2 \cdot R^{-1}$}  \\ \cline{2-6} 
                        & OP16:N                  & \multicolumn{1}{l|}{*}  & {$T_0 \gets -T_1$} & \multicolumn{1}{l|}{*} & {$T_0 \gets -T_1$} \\ \cline{2-6} 
                        & OP17:A                  & \multicolumn{1}{l|}{$T_6 \gets T_2 + T_2$}  & {$T_6 \gets T_2 + T_2$} & \multicolumn{1}{l|}{$T_8 \gets T_8 + T_9$} & {$T_8 \gets T_8 + T_9$}  \\ \cline{2-6} 
                        & \multirow{2}{*}{OP18:M} & \multicolumn{1}{l|}{\multirow{2}{*}{$T_6 \gets T_1 \cdot T_6$}} & {$T_6 \gets T_1 \cdot T_6 \cdot R^{-1}$} & \multicolumn{1}{l|}{\multirow{2}{*}{$T_4 \gets T_3 \cdot T_4$}} & {$T_4 \gets T_3 \cdot T_4 \cdot R^{-1}$} \\
                        &                        & \multicolumn{1}{l|}{} & {$T_6 \gets T_6 \cdot R^2 \cdot R^{-1}$} & \multicolumn{1}{l|}{} & {$T_4 \gets T_4 \cdot R^2 \cdot R^{-1}$} \\ \cline{2-6} 
                        & OP19:N                  & \multicolumn{1}{l|}{$T_1 \gets -T_6$}  & {$T_1 \gets -T_6$} & \multicolumn{1}{l|}{*} & {$T_0 \gets -T_6$} \\ \cline{2-6} 
                        & OP20:A                  & \multicolumn{1}{l|}{$T_1 \gets T_1 + T_1$}  & {$T_1 \gets T_1 + T_1$} & \multicolumn{1}{l|}{*} & {$T_0 \gets T_1 + T_1$} \\ \cline{2-6} 
                        & OP21:A                  & \multicolumn{1}{l|}{$T_1 \gets T_1 + T_5$}  & {$T_1 \gets T_1 + T_5$} & \multicolumn{1}{l|}{*} & {$T_0 \gets T_1 + T_5$} \\ \hline \hhline{|=|=|=|=|=|=|}

\multirow{9}{*}{$\Delta4$} & \multirow{2}{*}{OP22:M} & \multicolumn{1}{l|}{\multirow{2}{*}{$T_2 \gets T_2 \cdot T_2$}} & {$T_2 \gets T_2 \cdot T_2 \cdot R^{-1}$} & \multicolumn{1}{l|}{\multirow{2}{*}{$T_4 \gets T_y \cdot T_4$}} & {$T_4 \gets T_y \cdot T_4 \cdot R^{-1}$} \\
                        &                        & \multicolumn{1}{l|}{} & {$T_2 \gets T_2 \cdot R^2 \cdot R^{-1}$} & \multicolumn{1}{l|}{} & {$T_4 \gets T_4 \cdot R^2 \cdot R^{-1}$}  \\ \cline{2-6} 
                        & OP23:N                  & \multicolumn{1}{l|}{$T_5 \gets -T_1$}  & {$T_5 \gets -T_1$} & \multicolumn{1}{l|}{$T_{10} \gets -T_2$} & {$T_{10} \gets -T_2$}  \\ \cline{2-6} 
                        & OP24:A                  & \multicolumn{1}{l|}{$T_5 \gets T_5 + T_6$}  & {$T_5 \gets T_5 + T_6$} & \multicolumn{1}{l|}{$T_4 \gets T_4 + T_{10}$} & {$T_4 \gets T_4 + T_{10}$} \\ \cline{2-6} 
                        & \multirow{2}{*}{OP25:M} & \multicolumn{1}{l|}{\multirow{2}{*}{$T_5 \gets T_4 \cdot T_5$}} & {$T_5 \gets T_4 \cdot T_5 \cdot R^{-1}$} & \multicolumn{1}{l|}{\multirow{2}{*}{$T_{10} \gets T_4 \cdot T_4$}} & {$T_{10} \gets T_4 \cdot T_4 \cdot R^{-1}$}  \\
                        &                        & \multicolumn{1}{l|}{} & {$T_5 \gets T_5 \cdot R^2 \cdot R^{-1}$} & \multicolumn{1}{l|}{} & {$T_{10} \gets T_{10} \cdot R^2 \cdot R^{-1}$}  \\ \cline{2-6} 
                        & OP26:N                  & \multicolumn{1}{l|}{$T_2 \gets -T_2$}  & {$T_2 \gets -T_2$} & \multicolumn{1}{l|}{$T_8 \gets -T_8$} & {$T_8 \gets -T_8$} \\ \cline{2-6} 
                        & OP27:A                  & \multicolumn{1}{l|}{$T_2 \gets T_2 + T_2$}  & {$T_2 \gets T_2 + T_2$} & \multicolumn{1}{l|}{$T_1 \gets $ \hl{$\pmb{T_{10}}$} $+ T_8$} & {$T_1 \gets T_{10} + T_8$} \\ \cline{2-6} 
                        & OP28:A                  & \multicolumn{1}{l|}{$T_2 \gets T_2$ + \hl{$\pmb{T_5}$}}  & {$T_2 \gets T_2 + T_5$} & \multicolumn{1}{l|}{*} & {$T_0 \gets T_2 + T_5$} \\ \hline \hhline{|=|=|=|=|=|=|}
\clearpage
\multirow{9}{*}{$\Delta5$} & \multirow{2}{*}{OP29:M} & \multicolumn{1}{l|}{\multirow{2}{*}{}} & {} & \multicolumn{1}{l|}{\multirow{2}{*}{$T_8 \gets T_2 \cdot T_9$}} & {$T_8 \gets T_2 \cdot T_9 \cdot R^{-1}$} \\
                        &                        & \multicolumn{1}{l|}{} & {} & \multicolumn{1}{l|}{} & {$T_8 \gets T_8 \cdot R^2 \cdot R^{-1}$} \\ \cline{2-6} 
                        & OP30:N                  & \multicolumn{1}{l|}{}  & {} & \multicolumn{1}{l|}{$T_6 \gets -T_1$} & {$T_6 \gets -T_1$} \\ \cline{2-6} 
                        & OP31:A                  & \multicolumn{1}{l|}{}  & {} & \multicolumn{1}{l|}{$T_6 \gets T_6 + T_7$} & {$T_6 \gets T_6 + T_7$} \\ \cline{2-6} 
                        & \multirow{2}{*}{OP32:M} & \multicolumn{1}{l|}{\multirow{2}{*}{}} & {} & \multicolumn{1}{l|}{\multirow{2}{*}{$T_6 \gets T_6 \cdot $\hl{$\pmb{T_4}$}}} & {$T_6 \gets T_6 \cdot T_4 \cdot R^{-1}$}  \\
                        &                        & \multicolumn{1}{l|}{} & {} & \multicolumn{1}{l|}{} & {$T_6 \gets T_6 \cdot R^2 \cdot R^{-1}$} \\ \cline{2-6} 
                        & OP33:N                  & \multicolumn{1}{l|}{}  & {} & \multicolumn{1}{l|}{$T_9 \gets -T_8$} & {$T_9 \gets -T_8$} \\ \cline{2-6} 
                        & OP34:A                  & \multicolumn{1}{l|}{}  & {} & \multicolumn{1}{l|}{$T_2 \gets T_6 + T_9$} & {$T_2 \gets T_6 + T_9$} \\ \cline{2-6} 
                        & OP35:A                  & \multicolumn{1}{l|}{}  & {} & \multicolumn{1}{l|}{*} & {$T_0 \gets T_2 + T_5$} \\ \hline \hhline{|=|=|=|=|=|=|}

\multirow{9}{*}{$\Delta6$} & \multirow{2}{*}{OP36:M} & \multicolumn{1}{l|}{\multirow{2}{*}{}} & {} & \multicolumn{1}{l|}{\multirow{2}{*}{$T_3 \gets T_3 \cdot T_5$}} & {$T_3 \gets T_3 \cdot T_5 \cdot R^{-1}$} \\
                        &                        & \multicolumn{1}{l|}{} & {} & \multicolumn{1}{l|}{} & {$T_3 \gets T_3 \cdot R^2 \cdot R^{-1}$} \\ \cline{2-6} 
                        & OP37:N                  & \multicolumn{1}{l|}{}  & {} & \multicolumn{1}{l|}{$T_4 \gets -T_7$} & {$T_4 \gets -T_7$} \\ \cline{2-6} 
                        & OP38:A                  & \multicolumn{1}{l|}{}  & {} & \multicolumn{1}{l|}{$T_4 \gets T_1 + T_4$} & {$T_4 \gets T_1 + T_4$} \\ \cline{2-6} 
                        & \multirow{2}{*}{OP39:M} &  \multicolumn{1}{l|}{\multirow{2}{*}{}} & {} & \multicolumn{1}{l|}{\multirow{2}{*}{$T_5 \gets T_4 \cdot T_4$}} & {$T_5 \gets T_4 \cdot T_4 \cdot R^{-1}$} \\
                        &                        & \multicolumn{1}{l|}{} & {} & \multicolumn{1}{l|}{} & {$T_5 \gets T_5 \cdot R^2 \cdot R^{-1}$}  \\ \cline{2-6} 
                        & OP40:N                  &  \multicolumn{1}{l|}{}  & {} & \multicolumn{1}{l|}{$T_6 \gets -T_8$} & {$T_6 \gets -T_8$} \\ \cline{2-6} 
                        & OP41:A                  & \multicolumn{1}{l|}{}  & {} & \multicolumn{1}{l|}{$T_6 \gets T_2 + T_6$} & {$T_6 \gets T_2 + T_6$} \\ \cline{2-6} 
                        & OP42:A                  & \multicolumn{1}{l|}{}  & {} & \multicolumn{1}{l|}{*} & {$T_0 \gets T_2 + T_5$} \\ \hline
\caption{Implemented sequence of operations using constant-time FLECC functions for EC point doubling and EC point addition operations applying the atomic pattern MNAMNAA.}
\label{op-table}
\end{longtable}
\end{landscape}

Two columns denoted as "Our implementation using constant-time FLECC functions" show which operations we used as the dummy operations (instead of the operations indicated by * in Longa’s patterns). These dummy operations aim at making point doubling and point addition indistinguishable to each other. They do not change the computation results. Please note that we tried to apply identical operands and resultant registers on the arithmetic operations on both point doubling and point addition operations whenever any side has a dummy operation, to try to make both algorithms as indistinguishable as possible against address-bit SCA. For example, at point addition OP7 ($T_0 \gets T_1 +T_4$), we are not able to use the same register $T_4$ as resultant as in point doubling OP7 ($T_4 \gets T_1 + T_4$), since it would ruin other calculation steps, we use a dummy register $T_0$ instead. But we still try to match the operands $T_1$ and $T_4$ in the modular addition operation. 

Additionally, we inserted "no operations" (NOPs) between each atomic block within a point operation, and at the end of each EC point doubling and EC point addition operation. The NOPs are necessary for simplifying the identification of atomic blocks and point operations in the measured trace. 

\clearpage
\section{Measurements}\label{measurements}
\subsection{Experimental Setup}
\begin{figure}[H]
\centering
\includegraphics[width=0.4\textwidth]{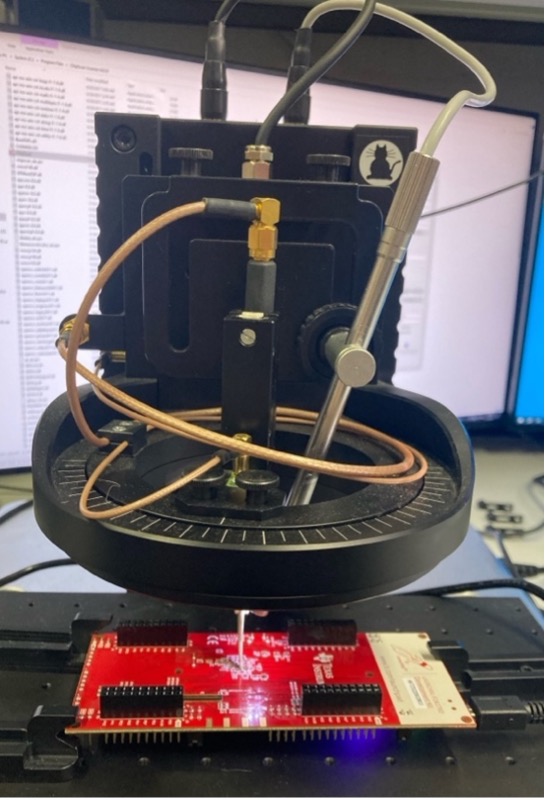}
\caption{The Integrated Circuit Scanner positioning the board and the EM-probe.}
\label{ics}
\end{figure}

\begin{figure}[H]
\centering
\includegraphics[width=0.45\textwidth]{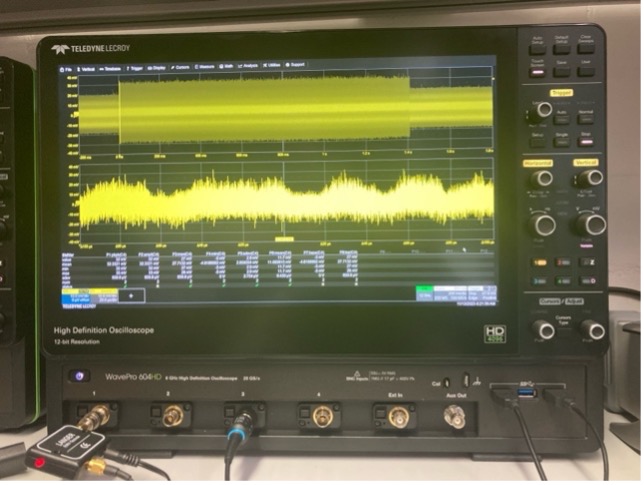}
\caption{The oscilloscope connected to the probe showing the captured trace.}
\label{osc}
\end{figure}

Measurements of electromagnetic (EM) emanation of cryptographic chips, unlike power measurements,  do not require any modification of the board of the attacked device. Thus, we decided to measure and analyse the EM emanation of our implementation captured during a single $k\bm{P}$ algorithm execution.
In our experimental setup, we used a near-field micro-probe to measure EM emanation, an Integrated Circuit Scanner (ICS) to precisely position the board and the EM-probe, and an oscilloscope to capture the EM emanation. Figures \ref{ics} and \ref{osc} show the equipment used. The list of equipment is as follows:
\begin{itemize}
    \item Langer ICS 105 \nocite{noauthor_ics_nodate}[99]
    \item Langer EM-probe MFA-R 0.2-75 \nocite{noauthor_mfa-r_nodate}[100]
    \item Teledyne LeCroy oscilloscope WavePro 604HD \nocite{noauthor_wavepro_nodate}[101]
\end{itemize}

\begin{figure}[H]
\centering
\includegraphics[width=0.45\textwidth]{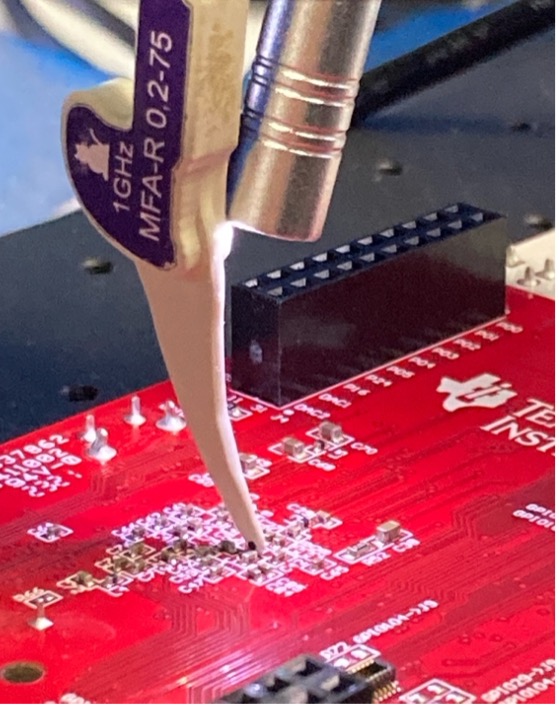}
\caption{Near-field micro probe placed near capacitor C78 on the board.}
\label{probe}
\end{figure}

To find the most optimal position to place the probe, we first looked at the schematic \nocite{texas_instruments_launchxl-f28379d_2019}[102] of the F28379D LaunchPad to find the list of capacitors in the power line for internal logic (1.2 V) as the power supply capacitors are the most suitable component for EM trace measurement. To find the best signal-to-noise ratio, we placed the probe near each of the capacitors C42, C46-49 and C75-78, while executing the $k\bm{P}$ algorithm code in RAM, to see which position received the highest amplitude of EM signals. We found that placing the probe at the side of C78, as shown in Figure \ref{probe}, gave us the strongest signals. Therefore, we used this position for the measurement of the EM trace.

For our $k\bm{P}$ algorithm implementation, without applying NOPs, for the board with a frequency of 100 MHz, the execution time of a point doubling operation run in RAM is around 290,944 clock cycles (2.91 $ms$ in our setup), while that of a point addition operation is about 436,416 (4.36 $ms$). The estimated processing time of a 256-bit scalar $k$ is between $290944 \cdot 255=74,190,720$ clock cycles (742 $ms$) for the 256-bit scalar $k=10...0$ and up to $(290944+436416)\cdot 255=185,476,800$ clock cycles (1855 $ms$) for the 256-bit scalar $k=11...1$. When using a sufficiently high sampling rate, e.g., 10 GS/$s$ which results into 100 samples per clock cycle, the average required memory to store the captured data is 391 GB, which takes tremendous time and resource to process and analyse, even if the oscilloscope could meet the memory requirement. 

We are aware that attackers need to deal with technical limits like the above mentioned when extracting a 256-bit long key. We assume that these demanding technical requirements are the reason why there is a lack of experimental investigation on the resistance of different atomic patterns.

Therefore, for the reasons explained above, we only used a 22-bit long scalar $k$ in our experiments. For the $k\bm{P}$ algorithm executed in RAM, we captured the trace with an oscilloscope sampling rate of 1 GS/$s$. By using a board with 100 MHz clock signal frequency, it results in only 10 samples per clock cycle. Taking into account that NOPs are inserted to the algorithm, the file size of the measured trace is about 6 GB. Similarly, for the $k\bm{P}$ algorithm executed in flash memory, we captured the trace with an oscilloscope sampling rate of 100 MS/$s$. By using a board with 20 MHz clock signal frequency, it results in only 5 samples per clock cycle and a trace file of 5.86 GB. 

As the inputs for the implemented $k\bm{P}$ algorithm in our experiments, we used a 22-bit binary scalar $k$\footnote{$k$=$k_{21}k_{20}k_{19}...k_{2}k_{1}k_0$=1001101101011111110111} and the base point $\bm{G}$ of the EC P-256 \nocite{chen_digital_2023}[7]\footnote{In hexadecimal: $\bm{G}$=(x,y)=(6b17d1f2e12c4247f8bce6e563a440f277037d812deb33a0f4a13945d898c296, 4fe342e2fe1a7f9b8ee7eb4a7c0f9e162bce33576b315ececbb6406837bf51f5)}. In total, 15 point additions and 21 point doublings are performed executing our $k\bm{P}$ algorithm with these inputs.

However, since the full length of the $k\bm{P}$ operation exceeds the limit we could capture with our measurement settings when being executed in RAM, the last point addition was not measured completely. We were only able to capture 14 point additions and 21 point doublings when the code was being executed in RAM. 

For the $k\bm{P}$ operation executed using the board's flash memory, we were able to measure the EM emanation during a whole $k\bm{P}$ operation execution (15 point additions and 21 point doublings). Table \ref{exp-settings} summarizes our measurement settings.


\begin{table}[H]
\centering
\begin{tabular}{|l|l|l|}
\hline
                                          & RAM                 & Flash \\ \hline
Frequency of microcontroller attacked & 100 MHz              & 20 MHz        \\ \hline
Sampling rate of the oscilloscope         & 1 GS/$s$                & 100 MS/$s$       \\ \hline
Samples per clock cycle                   & 10                  & 5            \\ \hline
Samples captured                          & 200 M                & 200 M         \\ \hline
$k\bm{P}$ execution time                         & \textgreater200 $ms$ & 1.35 $s$        \\ \hline
Size of raw data captured                 & 6.02 GB              & 5.86 GB       \\ \hline
\end{tabular}
\caption{Settings applied in our measurements.}\label{exp-settings}
\end{table}

Note that we chose to capture only 200 M samples using the oscilloscope, although our oscilloscope supports capturing a maximum of 500 M samples, due to memory limitations of the oscilloscope. This decision is made to compromise between the completeness of the whole $k\bm{P}$ operation trace and the number of sampling points per clock cycle for data quality of our analysis. Since the whole $k\bm{P}$ operation executed in RAM takes a bit more than 200 $ms$, it is much faster than that run in flash memory which takes about 1.35 $s$, we could use a higher rate on samples per second on the $k\bm{P}$ operation algorithm executed in RAM.

\subsection{Study of FLECC Constant-runtime Operations}\label{sec:study-of-flecc}
To determine whether the constant-runtime field operation functions from FLECC we used in our implementation are truly constant-runtime, we measured the execution time of those used in Doubling 2 of our atomic patterns algorithm. These measurements were performed using a hardware breakpoint in CCS with the code executed in RAM. We repeatedly measured the first Montgomery modular multiplication (X), the second Montgomery modular multiplication (X'), the first negation (N) and the first addition (A) in $\Delta1$ of Doubling 2, performing each measurement 10 times. We obtained the same results for each operation in all our measurements. The results are presented in Table \ref{flecc_exe_time}. The execution time of each operation was measured in number of clock cycles and then converted into time in millisecond according to our experimental equipment configurations. We observed a 1-clock cycle difference between the runtimes of X and X', as shown in Table \ref{flecc_exe_time}, which means the second Montgomery modular multiplication (X') required always 1 clock cycle less than the first one (X). 

\begin{table}[H]
\centering
\begin{tabular}{|l|c|c|}
\hline
               & No. of clock cycles & Time($ms$) \\ \hline
$1^{st}$ Multiplication (X) & 16570            &     0.166     \\ \hline
$2^{nd}$ Multiplication (X') & 16569            &     0.166     \\ \hline
Negation (N)       & 1197            &     0.0120     \\ \hline
Addition (A)       & 1353            &     0.0135     \\ \hline
\end{tabular}\caption{Execution time of FLECC constant-runtime operations in $\Delta1$ of Doubling 2, measured 10 times.}\label{flecc_exe_time}
\end{table}

Please note that we selected $\Delta1$ in Doubling 2 for our measurements to avoid any potential impact of the special operand $Z_1=1$ (from $\Delta1$ in Doubling 1) on the execution time of the multiplication operation. 
By analyzing the sub-traces of $\Delta1$ in Doubling 1 and Doubling 2 over time, we observed that Doubling 2 has a delay of 5 clock cycles compared to Doubling 1 at 20,000 indices after the synchronization anchor samples, which will be introduced in Chapter \ref{kp_analysis_ram}. This finding indicates that the duration of Doubling 1 is likely shorter than Doubling 2. A detailed description can be found in Chapter \ref{Appendix5}.

\begin{figure}[H]
\centering
\includegraphics[width=0.8\textwidth]{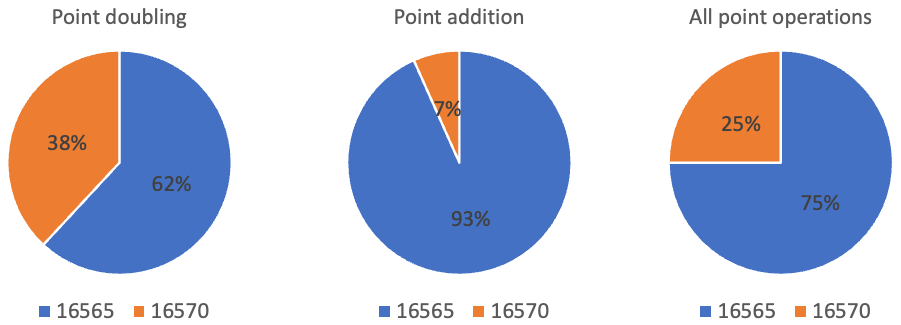}
\caption{Distribution of execution time of the first X in $\Delta1$ of all point doublings (left), all point additions (middle) and all EC point operations (right).}
\label{fig:x_exe_time_pie}
\end{figure}

Additionally, we measured the execution time of the first X in $\Delta1$ of each point operation, recorded the results in Table \ref{x_exec_time} and summarized them in Figure \ref{fig:x_exe_time_pie}. As illustrated in Figure \ref{fig:x_exe_time_pie}, the X operation consistently took either 16,565 or 16,570 clock cycles in our 36 measurements, with an overall 75\% likelihood of falling on the lower end.

However, these differences in the execution time between X and X', and among X itself are so negligible ($\leq5$ clock cycles) and consistent that we can classify the multiplication function as constant-runtime, with negation and addition functions also demonstrating constant-runtime behavior.

\begin{table}[H]
    \centering
    \begin{tabular}{|l|l|l|l|}
    \hline
        X & Start & End & Duration  \\ \hline
        Doubling 1 & 533427 & 549992 & 16565 \\ \hline
        Doubling 2 & 1052586 & 1069156 & 16570  \\ \hline
        Doubling 3 & 1571720 & 1588285 & 16565  \\ \hline
        Doubling 4 & 2673531 & 2690101 & 16570  \\ \hline
        Doubling 5 & 3775323 & 3791893 & 16570  \\ \hline
        Doubling 6 & 4294499 & 4311069 & 16570  \\ \hline
        Doubling 7 & 5396316 & 5412881 & 16565  \\ \hline
        Doubling 8 & 6498128 & 6514693 & 16565  \\ \hline
        Doubling 9 & 7017274 & 7033839 & 16565  \\ \hline
        Doubling 10 & 8119076 & 8135641 & 16565  \\ \hline
        Doubling 11 & 8638222 & 8654787 & 16565  \\ \hline
        Doubling 12 & 9740014 & 9756579 & 16565  \\ \hline
        Doubling 13 & 10841816 & 10858381 & 16565  \\ \hline
        Doubling 14 & 11943608 & 11960178 & 16570  \\ \hline
        Doubling 15 & 13045420 & 13061990 & 16570  \\ \hline
        Doubling 16 & 14147227 & 14163797 & 16570  \\ \hline
        Doubling 17 & 15249054 & 15265619 & 16565  \\ \hline
        Doubling 18 & 16350836 & 16367406 & 16570  \\ \hline
        Doubling 19 & 16870007 & 16886572 & 16565  \\ \hline
        Doubling 20 & 17971794 & 17988359 & 16565  \\ \hline
        Doubling 21 & 19073625 & 19090190 & 16565  \\ \Xhline{4\arrayrulewidth}
        Addition 1 & 2020871 & 2037436 & 16565  \\ \hline
        Addition 2 & 3122673 & 3139238 & 16565  \\ \hline
        Addition 3 & 4743655 & 4760225 & 16570  \\ \hline
        Addition 4 & 5845472 & 5862037 & 16565  \\ \hline
        Addition 5 & 7466429 & 7482994 & 16565  \\ \hline
        Addition 6 & 9087376 & 9103941 & 16565  \\ \hline
        Addition 7 & 10189168 & 10205733 & 16565  \\ \hline
        Addition 8 & 11290970 & 11307535 & 16565  \\ \hline
        Addition 9 & 12392772 & 12409337 & 16565  \\ \hline
        Addition 10 & 13494584 & 13511149 & 16565  \\ \hline
        Addition 11 & 14596411 & 14612976 & 16565  \\ \hline
        Addition 12 & 15698193 & 15714758 & 16565  \\ \hline
        Addition 13 & 17319150 & 17335715 & 16565  \\ \hline
        Addition 14 & 18420947 & 18437512 & 16565  \\ \hline
        Addition 15 & 19522734 & 19539299 & 16565 \\ \hline
    \end{tabular}
    \caption{Execution time of the first X operation in $\Delta1$ of each point operation, measured in clock cycles.}\label{x_exec_time}
\end{table}

\section{Simple Electromagnetic Analysis Attack}
\subsection{Analysis of \textit{k\textbf{P}} Operation Executed in RAM}\label{kp_analysis_ram}
\begin{figure}[H]
\centering
\includegraphics[width=1\textwidth]{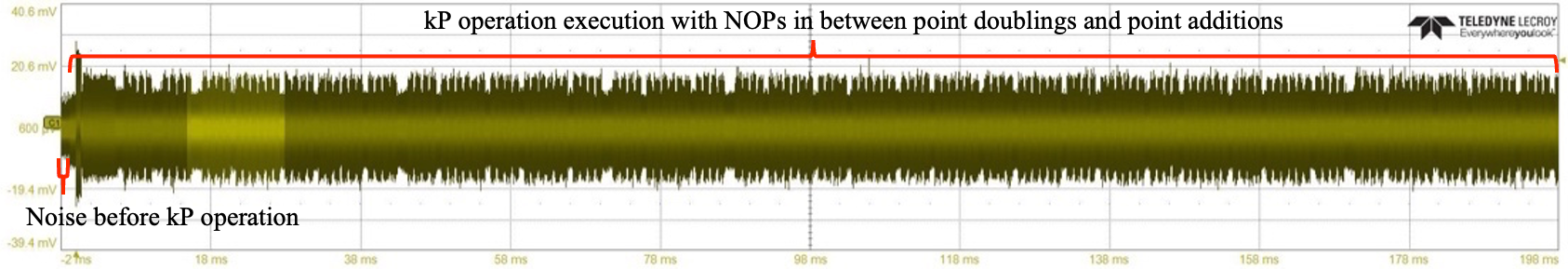}
\caption{Measurement of the $k\bm{P}$ execution: EM trace of the implemented atomic patterns $k\bm{P}$ algorithm executed in RAM of the board.}
\label{noise_kp}
\end{figure}

To ease the separation of EM traces for our analysis, in the atomic patterns $k\bm{P}$ algorithm, we separated each atomic block by placing a sequence of 2,000 redundant counter increment operations in between. We see these increment operations as no operations (NOPs) as they do very simple instructions that leak rather little EM emanation. Also, we placed a significantly longer sequence of 5,000 NOPs between each point doubling and addition operation to separate between different point operations, and 10,000 NOPs between two consecutive point doubling operations. 
Figure \ref{noise_kp} shows the entire EM trace we captured during the $k\bm{P}$ operation. By noticing the NOPs inserted between each point operation as well as each atomic block, we can clearly identify each point doubling operation and point addition operation, and each of their atomic blocks. 

By using a hardware breakpoint in CCS, we measured once the duration of each sequence of NOPs between the end of Doubling 2 and the start of Doubling 4, covering Doubling 3 and Addition 1. We found that the duration of the execution of 2,000 NOPs inserted between two atomic blocks has a median of 29,995 clock cycles (ranged from 27,996 to 30,006 clock cycles); while 5,000 NOPs between two EC point operation takes a median of 69,996 clock cycles (ranged from 69,996 to 70,012 clock cycles). The raw data of the measurements can be found in Table \ref{NOPs2000} and \ref{NOPs5000} in Chapter \ref{Appendix4}. We observed minor fluctuations in execution time for the same number of NOPs at different stages in the algorithm, which we attribute to the wait states of the memory causing the execution time to be slightly irregular. 

Microcontrollers usually require a non-zero wait state when the CPU's operating frequency is higher than the memory speed, which means the CPU has to wait for a number of clock cycles to complete data read/write operations at the memory address. Modern RAM uses sophisticated techniques to minimize wait states. 
Furthermore, for RAM, the wait time can differ depending on factors like the type of operation being performed (e.g., read, write) and the access patterns (e.g., sequential access, random access). 
Therefore, it is inevitable to have some varied amount of operational time added to the operations being performed, 
hindering the performance of SCA.

\begin{figure}[H]
\centering
\includegraphics[width=1\textwidth]{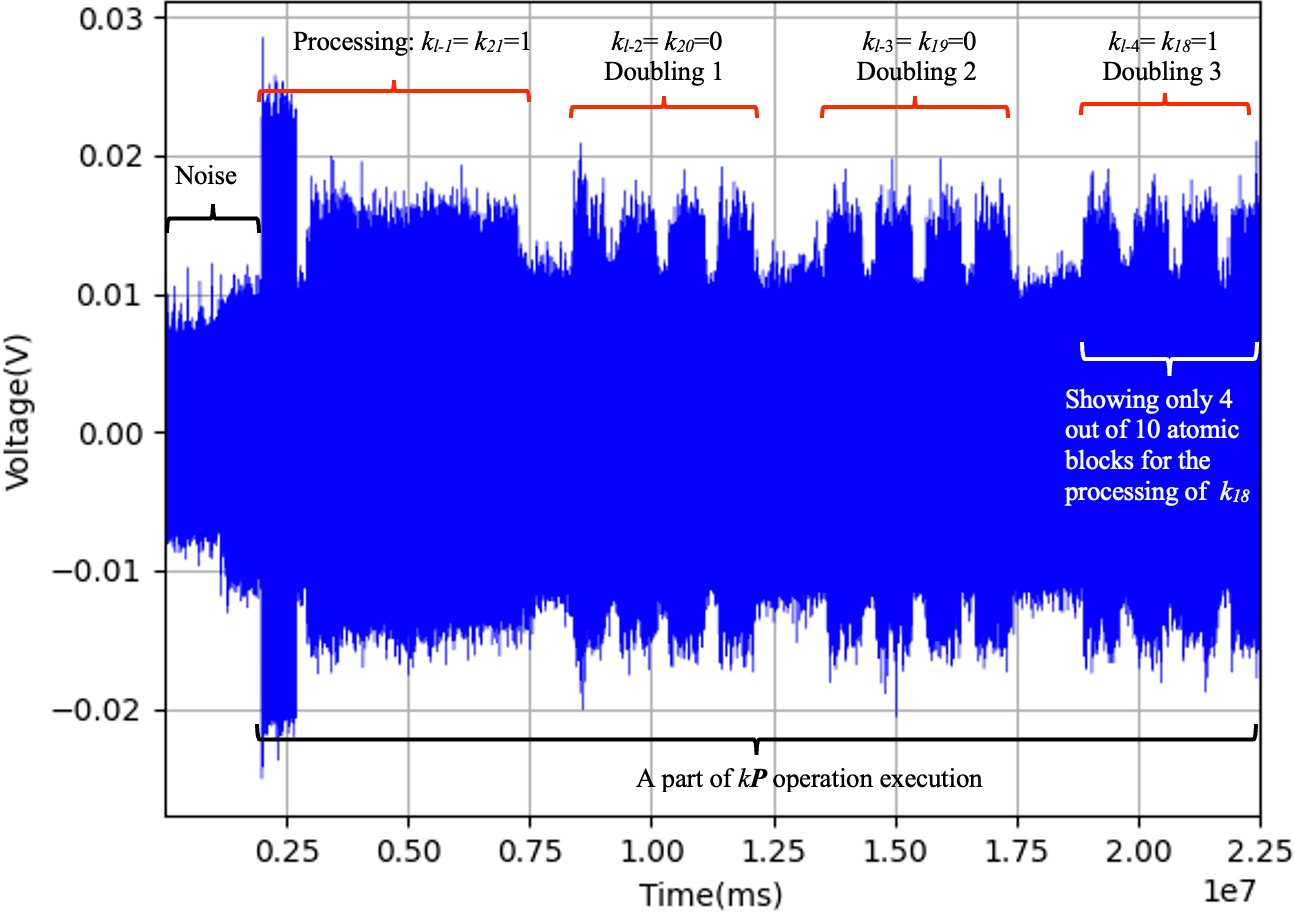}
\caption{A part of the captured EM trace: the noise before the start of the $k\bm{P}$ operation, the processing of the most significant bit $k_{l-1}$=1, and the atomic patterns of the first 3 point doublings are shown; $k\bm{P}$ executed in RAM of the board.}
\label{emt_part}
\end{figure}

Figure \ref{emt_part} magnifies the beginning of the trace in Figure \ref{noise_kp}, showing the shapes of the noise before the start of the $k\bm{P}$ operation, the processing of the most significant bit of the scalar $k$, and the shapes of some atomic patterns in the main loop of the $k\bm{P}$ algorithm.

Before the atomic patterns in the $k\bm{P}$ operation begin, there is a significant peak and some continuous peaks without drops in amplitudes. This part of the $k\bm{P}$ operation refers to the initialization phase of the $k\bm{P}$ operation before the main loop iterations. In the main loop, 21 of 22 bits of the scalar $k$ are processed, executing 21 point doublings and 15 point additions as atomic patterns. We denote in our analysis all atomic patterns as “Doubling $i$” with $1\leq i \leq 21$ or “Addition $j$” with $1\leq j\leq 15$. In Figure \ref{emt_part}, the first 3 EC point doubling operations are shown, namely Doubling 1, Doubling 2 and Doubling 3.

\begin{figure}[H]
\centering
\includegraphics[width=1\textwidth]{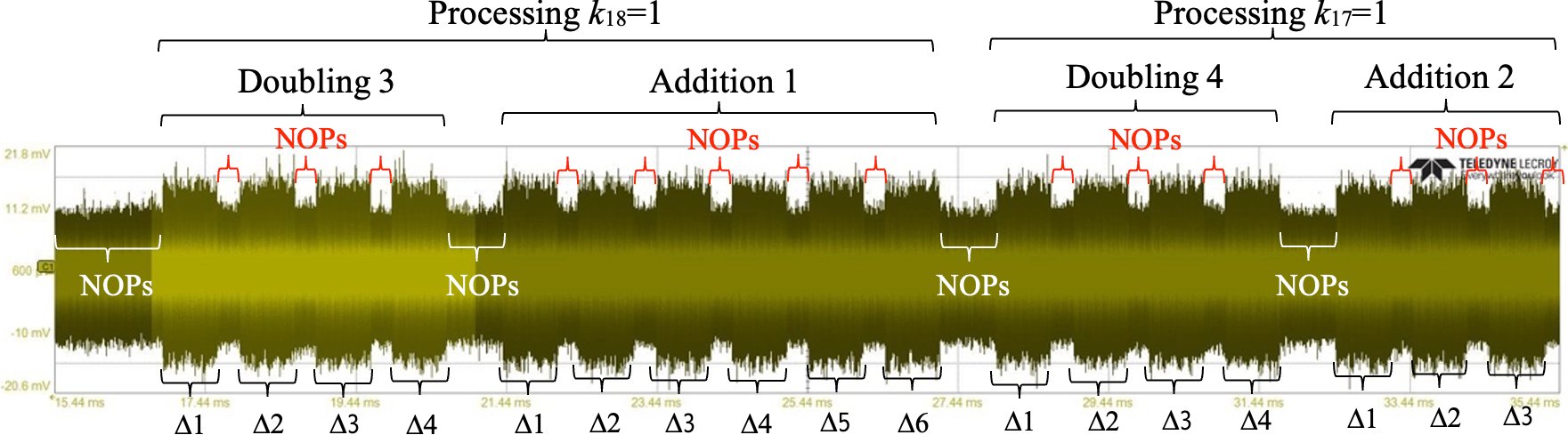}
\caption{EM trace representing a part of the atomic pattern $k\bm{P}$ algorithm executed in RAM of the board.}
\label{dbl_add_zoomed}
\end{figure}

Figure \ref{dbl_add_zoomed} shows the part of the captured EM trace corresponding to the processing of the bits $k_{l-4}=k_{18}=1$ and $k_{l-5}=k_{17}=1$ of the key. Processing a key bit value ‘1’ requires the execution of a point doubling and a point addition. Thus, Figure \ref{dbl_add_zoomed} shows the atomic patterns Doubling 3, Addition 1, Doubling 4 and Addition 2, corresponding to our numbering of the atomic patterns. 
Each doubling consists of 4 atomic blocks and each addition consists of 6 atomic blocks. 
The short NOP sequences inserted between each atomic block help us separate the atomic blocks $\Delta$1 to $\Delta$4 or $\Delta$1 to $\Delta$6. The long NOP sequences inserted between atomic patterns help us separate the EC point operations executed in the main loop iterations, so the doubling operations and the addition operations are visibly distinguishable by observing the presence of long troughs of NOPs in between and counting the number of atomic blocks they have. Each atomic block of the corresponding atomic pattern, the short NOP blocks between the atomic blocks and the long NOP blocks between different operations are shown in Figure \ref{dbl_add_zoomed} too.

We are aware that in a real-world scenario there would not be such long breaks between the atomic blocks as well as between atomic patterns. So, it would not be easy for an adversary to separate the atomic blocks or distinguish the operations. Nevertheless, our work concentrates on the investigation of the distinguishability of the atomic patterns. Thus, it is reasonable to ignore the difficulty in separating the atomic blocks and only focus on analysing the differences between the atomic blocks of point doubling and point addition operations.

Additionally, in order to analyse the shapes of point doubling operations and point addition operations, breakpoints were added in the CCS program of our $k\bm{P}$ algorithm implementation to help us measure the clock cycles needed until the first point operation starts and the clock cycles each point operation then takes. We used these measured clock cycles, as shown in Table \ref{exec-time}, to calculate the actual time each point operation approximately takes, and then split the whole captured trace into sub-traces of single point operations according to the time calculated, resulting in 21 separate point doubling sub-traces and 14 separate point addition sub-traces. 

\begin{table}[H]
\centering
\scalebox{0.95}{
\begin{tabular}{|l|lll|lll|}
\hline
           & \multicolumn{3}{c|}{RAM}                                         & \multicolumn{3}{c|}{Flash memory}                                       \\ \hline
           & \multicolumn{1}{l|}{Start} & \multicolumn{1}{l|}{End} & Duration & \multicolumn{1}{l|}{Start} & \multicolumn{1}{l|}{End} & Duration \\ \hline
Doubling 1 & \multicolumn{1}{l|}{642930}  &  \multicolumn{1}{l|}{3364774} & 376912 & \multicolumn{1}{l|}{3364774} & \multicolumn{1}{l|}{5830998} & 2466224\\ \hline
Doubling 2 & \multicolumn{1}{l|}{1169884} & \multicolumn{1}{l|}{1546776} & 376892 & \multicolumn{1}{l|}{6631158}  & \multicolumn{1}{l|}{9097318} & 2466160\\ \hline
Doubling 3 & \multicolumn{1}{l|}{1696818} & \multicolumn{1}{l|}{2073735} & 376917 & \multicolumn{1}{l|}{9897478}  & \multicolumn{1}{l|}{12363718} & 2466240\\ \hline
Doubling 4 & \multicolumn{1}{l|}{2804121} & \multicolumn{1}{l|}{3181028} & 376907 & \multicolumn{1}{l|}{16943334}  & \multicolumn{1}{l|}{19409542} & 2466208\\ \hline
Doubling 5 & \multicolumn{1}{l|}{3911407} & \multicolumn{1}{l|}{4288339} & 376932 & \multicolumn{1}{l|}{23989126}  & \multicolumn{1}{l|}{26455414} & 2466288 \\ \hline
Doubling 6 & \multicolumn{1}{l|}{4438382} & \multicolumn{1}{l|}{4815299} & 376917 & \multicolumn{1}{l|}{27255574}  & \multicolumn{1}{l|}{29721814} & 2466240 \\ \hline
Doubling 7 & \multicolumn{1}{l|}{5545693} & \multicolumn{1}{l|}{5922610} & 376917 & \multicolumn{1}{l|}{34301446}  & \multicolumn{1}{l|}{36767686} & 2466240\\ \hline
Doubling 8 & \multicolumn{1}{l|}{6652999} & \multicolumn{1}{l|}{7029901} & 376902 & \multicolumn{1}{l|}{41347302}  & \multicolumn{1}{l|}{43813494} & 2466192 \\ \hline
Doubling 9 & \multicolumn{1}{l|}{7179943} & \multicolumn{1}{l|}{7556855} & 376912 & \multicolumn{1}{l|}{44613654}  & \multicolumn{1}{l|}{47079878} & 2466224 \\ \hline
Doubling 10 & \multicolumn{1}{l|}{8287239} & \multicolumn{1}{l|}{8664141} & 376902 & \multicolumn{1}{l|}{51659478}  & \multicolumn{1}{l|}{54125670} & 2466192 \\ \hline
Doubling 11 & \multicolumn{1}{l|}{8814183} & \multicolumn{1}{l|}{9191090} & 376907 & \multicolumn{1}{l|}{54925830}  & \multicolumn{1}{l|}{57392038} & 2466208 \\ \hline
Doubling 12 & \multicolumn{1}{l|}{9921469} & \multicolumn{1}{l|}{10298376} & 376907 & \multicolumn{1}{l|}{61971622}  & \multicolumn{1}{l|}{64437830} & 2466208 \\ \hline
Doubling 13 & \multicolumn{1}{l|}{11028765} & \multicolumn{1}{l|}{11405672} & 376907 & \multicolumn{1}{l|}{69017446}  & \multicolumn{1}{l|}{71483654} & 2466208 \\ \hline
Doubling 14 & \multicolumn{1}{l|}{12136051} & \multicolumn{1}{l|}{12512968} & 376917 & \multicolumn{1}{l|}{76063238}  & \multicolumn{1}{l|}{78529478} & 2466208 \\ \hline
Doubling 15 & \multicolumn{1}{l|}{13243357} & \multicolumn{1}{l|}{13620274} & 376917 & \multicolumn{1}{l|}{83109094}  & \multicolumn{1}{l|}{85575334} & 2466240 \\ \hline
Doubling 16 & \multicolumn{1}{l|}{14350658} & \multicolumn{1}{l|}{14727595} & 376937 & \multicolumn{1}{l|}{90154934}  & \multicolumn{1}{l|}{92621238} & 2466304 \\ \hline
Doubling 17 & \multicolumn{1}{l|}{15457979} & \multicolumn{1}{l|}{15834871} & 376892 & \multicolumn{1}{l|}{97200838}  & \multicolumn{1}{l|}{99666998} & 2466160\\ \hline
Doubling 18 & \multicolumn{1}{l|}{16565255} & \multicolumn{1}{l|}{16942182} & 376927 & \multicolumn{1}{l|}{104246598}  & \multicolumn{1}{l|}{106712870} & 2466272\\ \hline
Doubling 19 & \multicolumn{1}{l|}{17092224} & \multicolumn{1}{l|}{17469116} & 376892 & \multicolumn{1}{l|}{107513030}  & \multicolumn{1}{l|}{109979190} & 2466160 \\ \hline
Doubling 20 & \multicolumn{1}{l|}{18199505} & \multicolumn{1}{l|}{18576407} & 376902 & \multicolumn{1}{l|}{114558806}  & \multicolumn{1}{l|}{117024998} & 2466192\\ \hline
Doubling 21 & \multicolumn{1}{l|}{19306776} & \multicolumn{1}{l|}{19683688} & 376912 & \multicolumn{1}{l|}{121604550}  & \multicolumn{1}{l|}{124070774} & 2466224\\ \Xhline{4\arrayrulewidth}
Addition 1 & \multicolumn{1}{l|}{2148757} & \multicolumn{1}{l|}{2729118} & 580361 & \multicolumn{1}{l|}{12763798}  & \multicolumn{1}{l|}{16543366} & 3779568 \\ \hline
Addition 2 & \multicolumn{1}{l|}{3256050} & \multicolumn{1}{l|}{3836401} & 580351 & \multicolumn{1}{l|}{19809622}  & \multicolumn{1}{l|}{23589158} & 3779536 \\ \hline
Addition 3 & \multicolumn{1}{l|}{4890321} & \multicolumn{1}{l|}{5470687} & 580366 & \multicolumn{1}{l|}{30121894}  & \multicolumn{1}{l|}{33901478} & 3779584 \\ \hline
Addition 4 & \multicolumn{1}{l|}{5997632} & \multicolumn{1}{l|}{6577993} & 580361 & \multicolumn{1}{l|}{37167766}  & \multicolumn{1}{l|}{40947334} & 3779568 \\ \hline
Addition 5 & \multicolumn{1}{l|}{7631877} & \multicolumn{1}{l|}{8212233} & 580356 & \multicolumn{1}{l|}{47479958}  & \multicolumn{1}{l|}{51259510} & 3779552 \\ \hline
Addition 6 & \multicolumn{1}{l|}{9266112} & \multicolumn{1}{l|}{9846463} & 580351 & \multicolumn{1}{l|}{57792118}  & \multicolumn{1}{l|}{61571654} & 3779536\\ \hline
Addition 7 & \multicolumn{1}{l|}{10373398} & \multicolumn{1}{l|}{10953759} & 580361 & \multicolumn{1}{l|}{64837910}  & \multicolumn{1}{l|}{68617478} & 3779568\\ \hline
Addition 8 & \multicolumn{1}{l|}{11480694} & \multicolumn{1}{l|}{12061045} & 580351 & \multicolumn{1}{l|}{71883734}  & \multicolumn{1}{l|}{75663270} & 3779536 \\ \hline
Addition 9 & \multicolumn{1}{l|}{12587990} & \multicolumn{1}{l|}{13168351} & 580361 & \multicolumn{1}{l|}{78929558}  & \multicolumn{1}{l|}{82709126} & 3779568 \\ \hline
Addition 10 & \multicolumn{1}{l|}{13695296} & \multicolumn{1}{l|}{14275652} & 580356 & \multicolumn{1}{l|}{85975414}  & \multicolumn{1}{l|}{89754966} & 3779552 \\ \hline
Addition 11 & \multicolumn{1}{l|}{14802617} & \multicolumn{1}{l|}{15382973} & 580356 & \multicolumn{1}{l|}{93021318}  & \multicolumn{1}{l|}{96800870} & 3779552\\ \hline
Addition 12 & \multicolumn{1}{l|}{15909893} & \multicolumn{1}{l|}{16490249} & 580356 & \multicolumn{1}{l|}{100067078}  & \multicolumn{1}{l|}{103846630} & 3779552 \\ \hline
Addition 13 & \multicolumn{1}{l|}{17544138} & \multicolumn{1}{l|}{18124499} & 580361 & \multicolumn{1}{l|}{110379270}  & \multicolumn{1}{l|}{114158838} & 3779568 \\ \hline
Addition 14 & \multicolumn{1}{l|}{18651429} & \multicolumn{1}{l|}{19231770} & 580341 & \multicolumn{1}{l|}{117425078}  & \multicolumn{1}{l|}{121204582} & 3779504\\ \hline
Addition 15 & \multicolumn{1}{l|}{19758710} & \multicolumn{1}{l|}{20339063} & 580353 & \multicolumn{1}{l|}{124470854}  & \multicolumn{1}{l|}{128250406} & 3779552\\ \hline
\end{tabular}}
\caption{Atomic patterns point operations execution time measured in clock cycles, using breakpoints.}\label{exec-time}
\end{table}

\begin{figure}[H]
\centering
\includegraphics[width=1\textwidth]{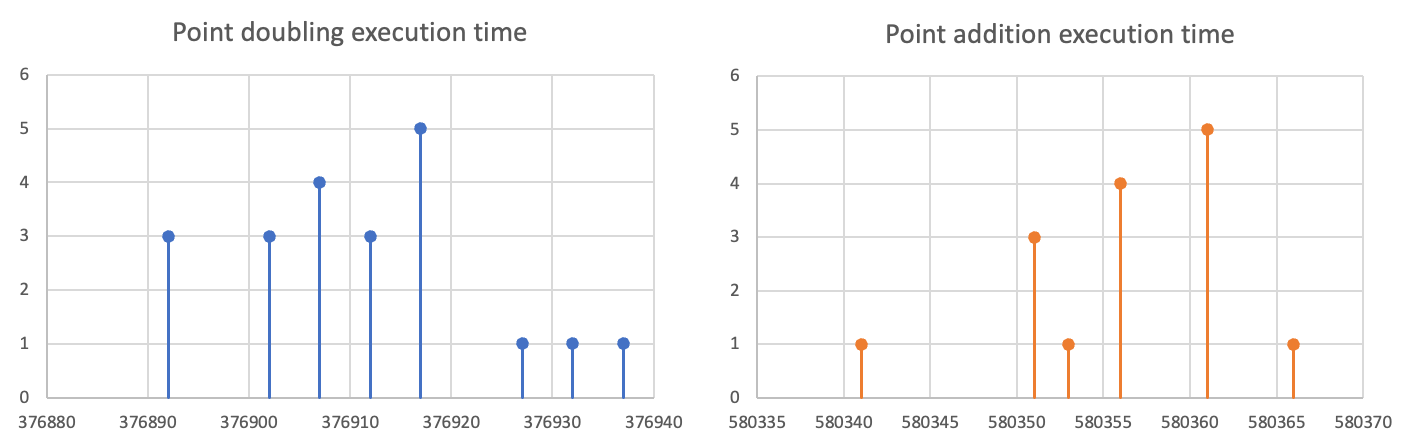}
\caption{The distribution of measured execution time of point doubling operations (left) and point addition operations (right) in a $k\bm{P}$ algorithm executed in RAM.}
\label{ram_po_exec_time}
\end{figure}

\begin{figure}[H]
\centering
\includegraphics[width=1\textwidth]{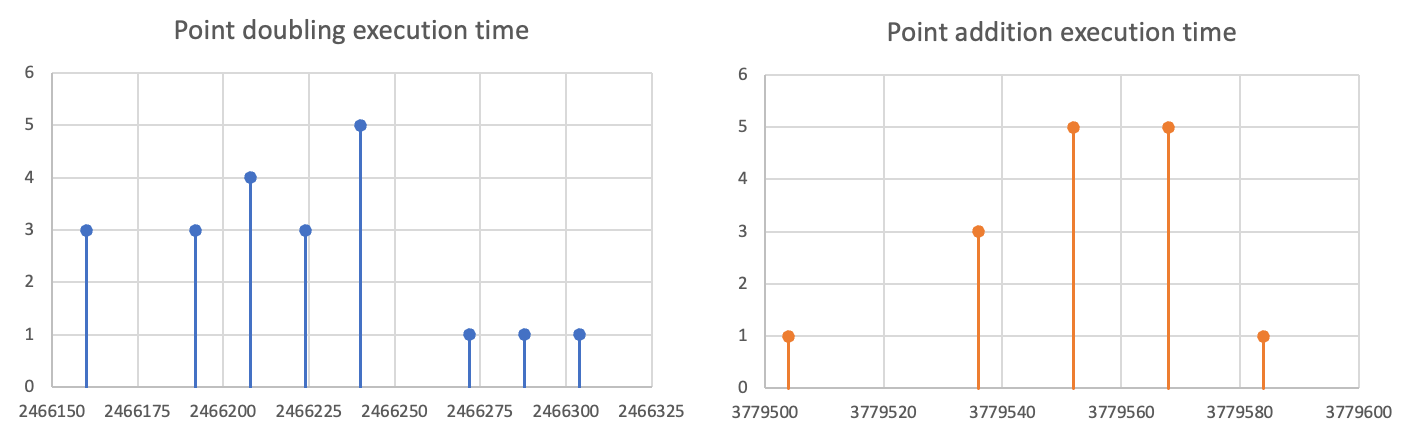}
\caption{The distribution of measured execution time of point doubling operations (left) and point addition operations (right) in a $k\bm{P}$ algorithm executed in flash memory.}
\label{flash_po_exec_time}
\end{figure}

Figures \ref{ram_po_exec_time} and \ref{flash_po_exec_time} represent the duration of EC point operations given in Table \ref{exec-time} graphically. We can see that the execution time of point operations in RAM fluctuates within 50 clock cycles, while those executed in flash memory has a bigger range of 150 clock cycles for the fluctuation.

\subsubsection{Synchronization Alignment}\label{sync_ram}
In all the following analysis, we excluded the sub-trace of Doubling 1 from the dataset because Doubling 1 consists of operations with operands using initialization values, i.e., OP1 in Table \ref{op-table}, which computes essentially $1\times1\times R^{-1}$ and $R^{-1}\times R^2 \times R^{-1}$ that take much less computing power than any other multiplication operations with more complicated operands. Removing it from the dataset facilitates an unbiased analysis on typical doubling operations.

All the sub-traces were then finely synchronized manually using Microsoft Excel charts to have the shapes of their atomic blocks and NOPs aligned to precision for further analysis. We used three methods to synchronize the sub-traces, i.e., Synchronization A, B and C, see Table \ref{sync_methods}. 

\begin{table}[H]
\begin{tabular}{|l|l|l|}
\hline
Sync.            & Addition operations                                                                                          & Doubling operations                                                                                                   \\ \hline
A & \begin{tabular}[c]{@{}l@{}}Align local maximums and\\ minimums using bare eyes\end{tabular} & \begin{tabular}[c]{@{}l@{}}Align local maximums and minimums\\ using bare eyes\end{tabular}          \\ \hline
B & \begin{tabular}[c]{@{}l@{}}Based on Sync. A, align the rise\\ of the lines between some samples\end{tabular} & \begin{tabular}[c]{@{}l@{}}Based on Sync. A, align the rise of the\\ lines between some samples\end{tabular}          \\ \hline
C  & \begin{tabular}[c]{@{}l@{}}Same as Sync. A\end{tabular} & \begin{tabular}[c]{@{}l@{}}Based on Sync. A, align according to\\ the fixed clock period of the samples\end{tabular} \\ \hline
\end{tabular}
\caption{Methods used for the fine synchronization of EC point operations.}
\label{sync_methods}
\end{table}

\begin{figure}[H]
\centering
\includegraphics[width=1\textwidth]{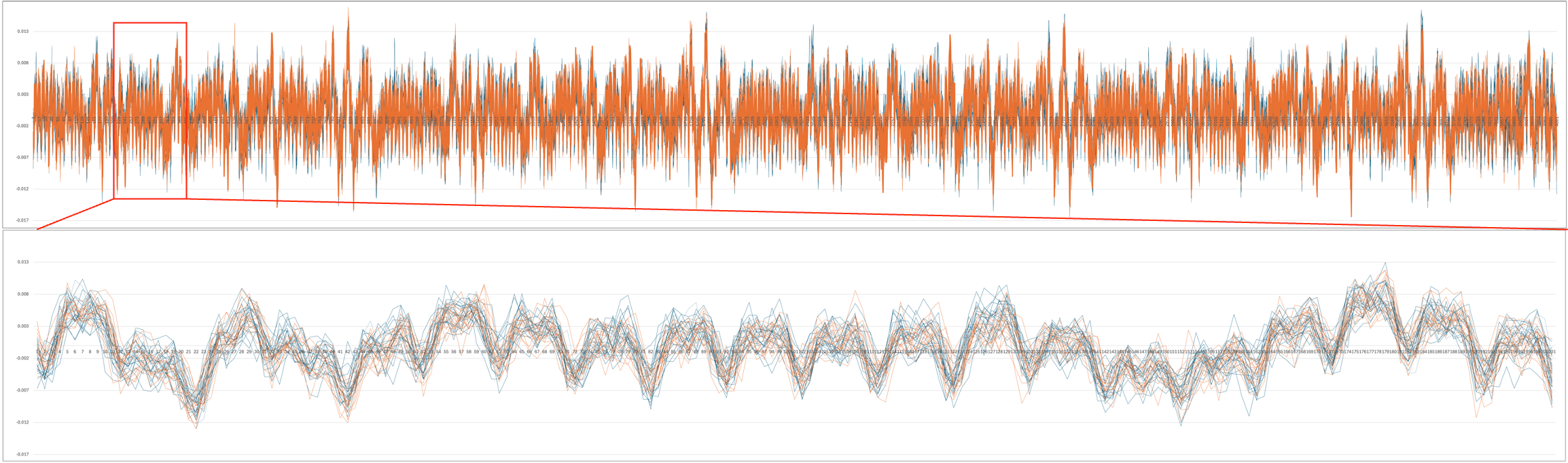}
\caption{Above: alignment of 4,000 anchor samples in Synchronization A. Below: zoomed in at samples 200-400.}
\label{syncA_4000}
\end{figure}

In Synchronization A, we started the alignment of sub-traces by fixating a close observation on a range of 4,000 anchor samples in one of the point operation sub-traces, which is at around sample 198,000 to sample 202,000 from the start of the sub-trace, so it is well assured that they are not noise but samples within the point operation in $\Delta$1. Then, we overlaid other sub-traces on top of it to synchronize them one by one, by shifting each of them to the left or right to fit the shapes of our anchoring sub-trace. Figure \ref{syncA_4000} shows parts of the Excel chart after the synchronization, with point addition traces in orange and point doubling traces in blue.

\begin{figure}[H]
\centering
\includegraphics[width=1\textwidth]{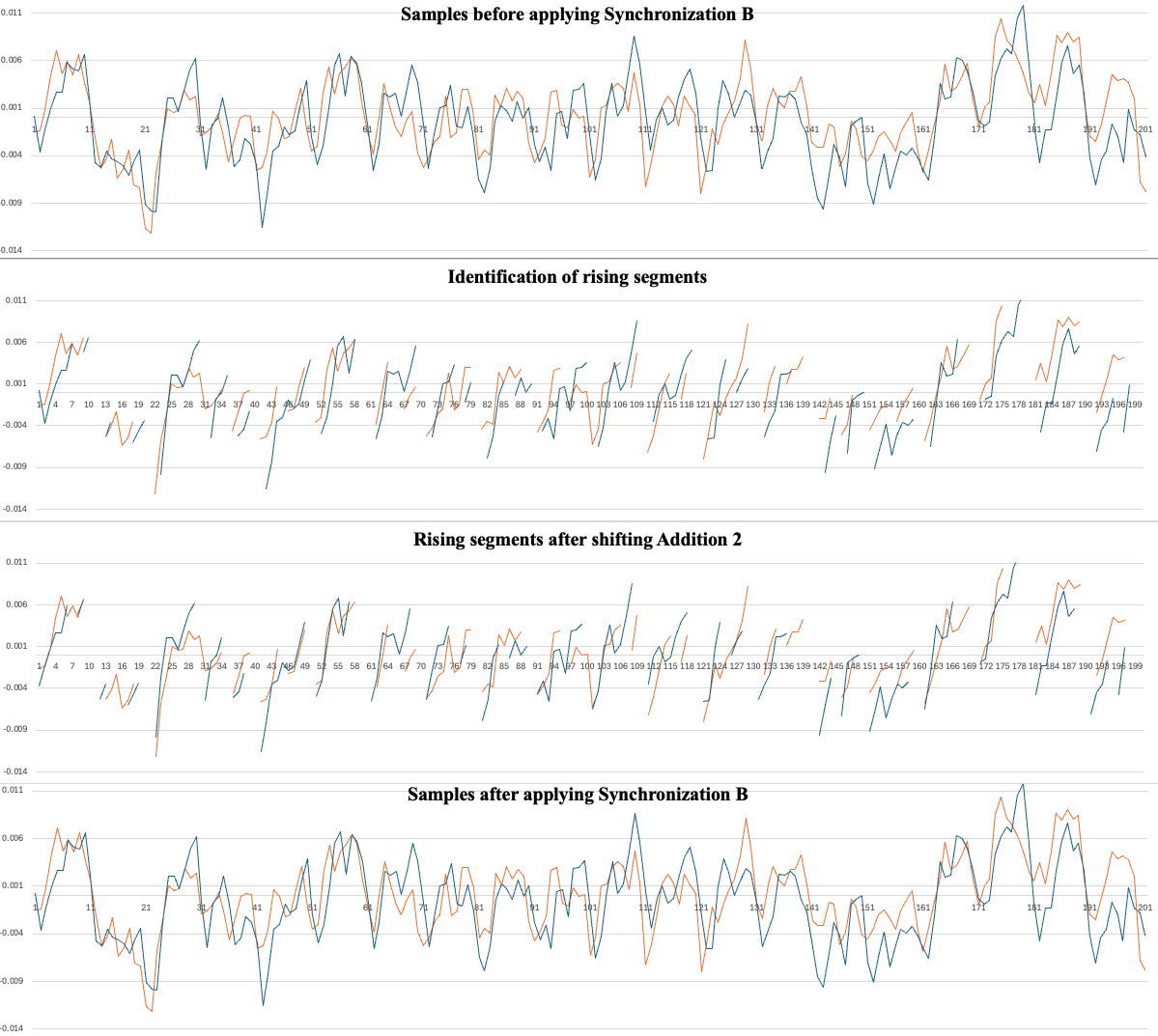}
\caption{Method used in Synchronization B: Identification and alignment of rising segments.}
\label{syncB_method}
\end{figure}

\begin{figure}[H]
\centering
\includegraphics[width=1\textwidth]{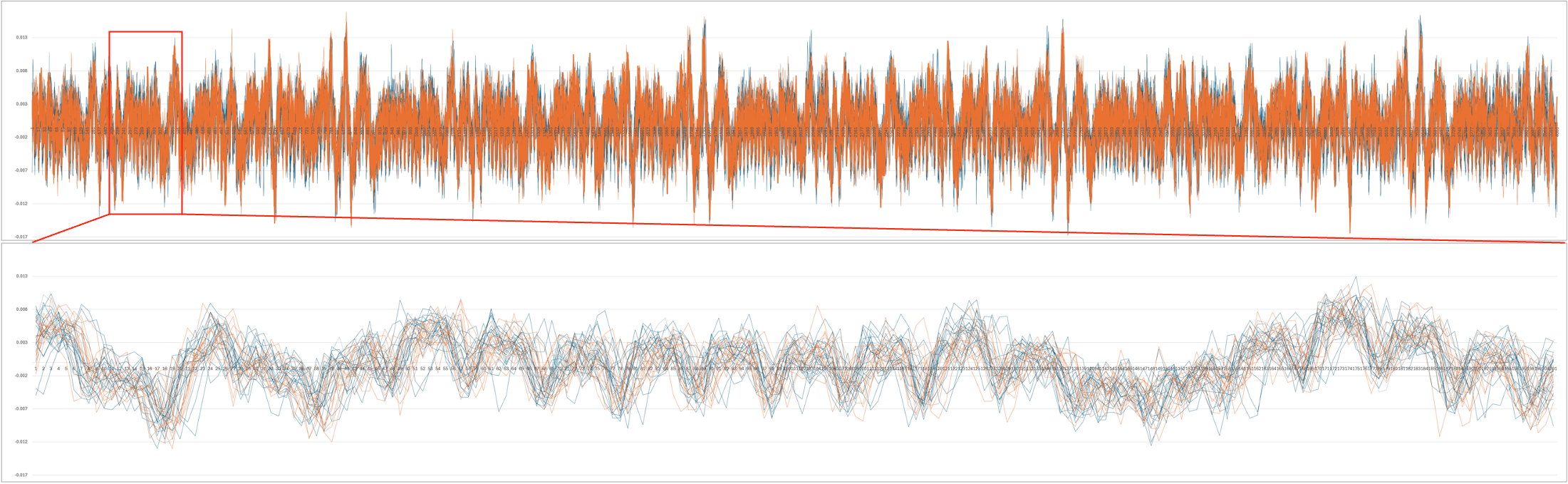}
\caption{Above: alignment of 4,000 anchor samples in Synchronization B. Below: zoomed in at samples 200-400.}
\label{syncB_4000}
\end{figure}

In Synchronization B, we first took the results from Synchronization A, then started aligning the rising segments of specific anchor samples in a smaller range (2,000 samples) within the same anchor samples range as in Synchronization A. Figure \ref{syncB_method} depicts the essential steps involved in Synchronization B of Addition 1 (orange) and Addition 2 (blue). First, the rising segments of the data are identified, keeping all consecutive samples that exhibit increasing values in the plot. Subsequently, we observe the difference and decide to shift Addition 2 by one unit to the left in order to align the slopes and lengths of the majority of the rising segments in Addition 1. The overall synchronization alignment results are shown in Figure \ref{syncB_4000}.

\begin{figure}[H]
\centering
\includegraphics[width=1\textwidth]{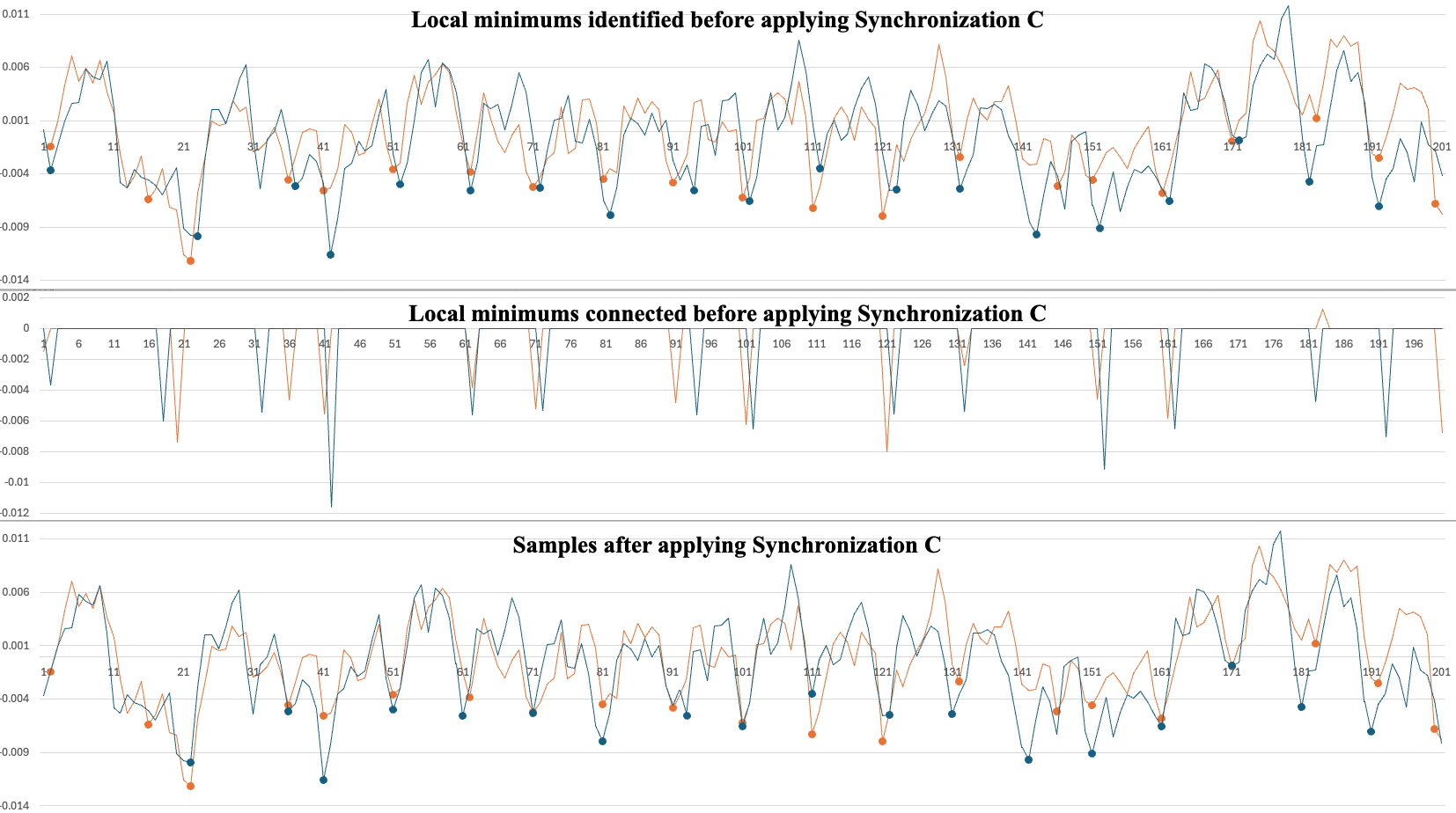}
\caption{Method used in Synchronization C: Identification of local minimum in each clock cycle (Above). Connection of local minimums (Middle). Synchronization done by shifting Addition 2 one unit to the left. (Below)}
\label{syncC_method}
\end{figure}

\begin{figure}[H]
\centering
\includegraphics[width=1\textwidth]{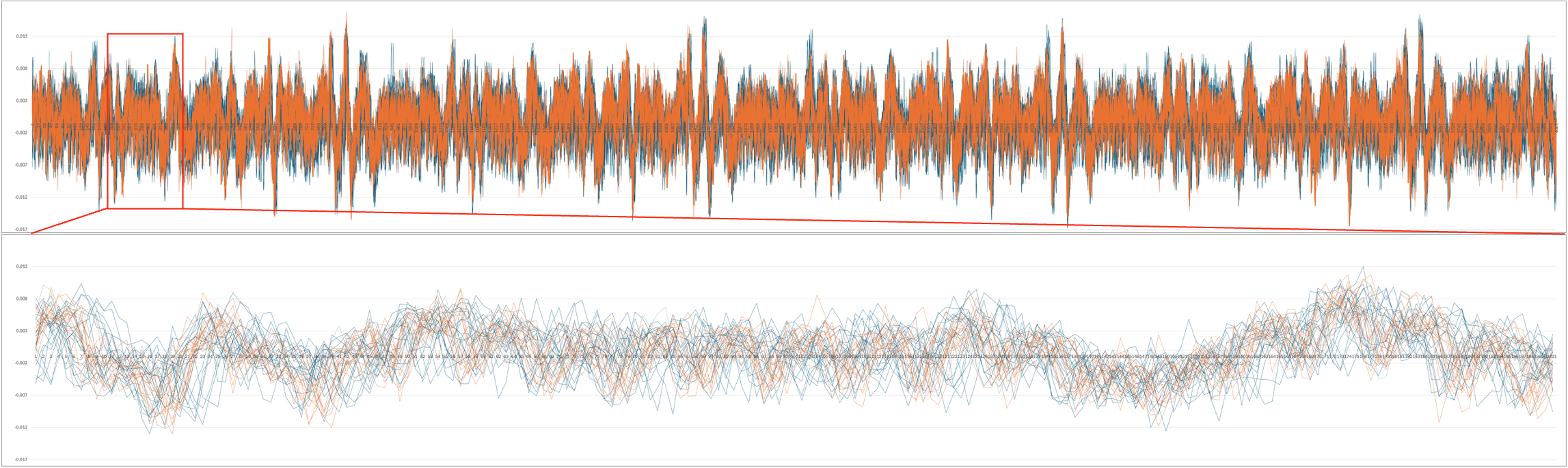}
\caption{Above: alignment of 4,000 anchor samples in Synchronization C. Below: zoomed in at samples 200-400.}
\label{syncC_4000}
\end{figure}

In Synchronization C, we employed the same synchronization method used in Synchronization A for point addition sub-traces. While for point doubling, based on Synchronization A, we aligned the sub-traces according to the fixed clock period (10 samples/period) of the samples. Similar to Synchronization B, we focused on the synchronization of specific samples within 2,000 samples in the chosen anchor samples. We first find out the first local minimum of each plot within the chosen range, then indicate the next local minimums at an interval of 10 samples, which corresponds to the period of the clock signal. Finally, we connected the local minimums obtained in each plot as a new plot, compared these plots and synchronized them. An example of the alignment of Addition 1 and Addition 2 using this method is illustrated in Figure \ref{syncC_method}, first indicating the identified local minimums in every 10 samples with dots, then moving these dots into another plot for better visualization and finally shifting one of the sub-traces to finely align the majority of its minimums to that of the other sub-trace. The overall synchronization alignment results are shown in Figure \ref{syncC_4000}.

\begin{figure}[H]
\centering
\includegraphics[width=0.9\textwidth]{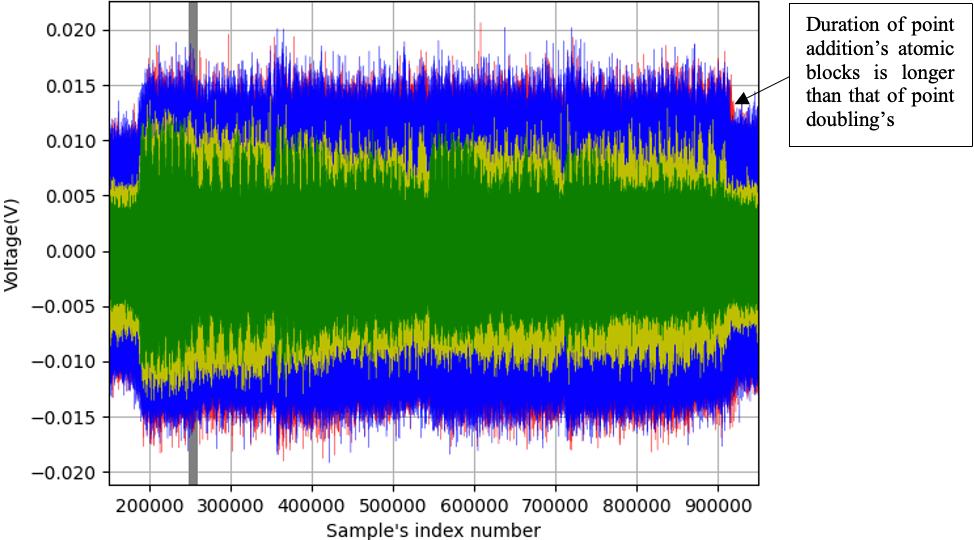}
\caption{Overlay of $\Delta$1 of all sub-traces synchronized using Synchronization A, with the algorithm executed in RAM.}
\label{atom1_syncA}
\end{figure}

After that, we calculated for the case "Synchronization A" the average atomic patterns for point doublings as well as for point additions to examine the synchronization of the overlaid sub-traces of the first atom $\Delta$1 of each atomic pattern. In Figure \ref{atom1_syncA}, the red lines represent the sub-traces of all point addition operations; the blue lines represent that of all point doubling operations; the yellow line marks the mean trace of all the point addition sub-traces; the green line marks the mean trace of all the point doubling sub-traces; the grey vertical area indicates the anchor samples we used for the fine synchronization (the rough synchronization was done at the start of the atomic patterns by counting clock cycles, as explained right before this sub-section). The values on the $x$-axis represent the sample's index numbers, counted from the beginning of each point operation sub-trace we separated for $\Delta1$.

\begin{figure}[H]
\centering
\includegraphics[width=0.8\textwidth]{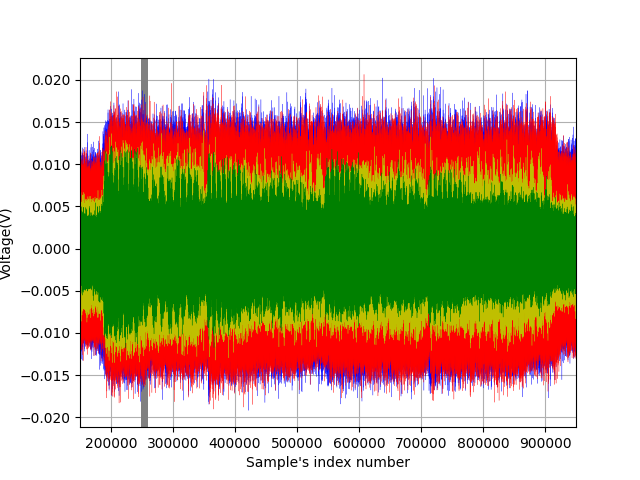}
\caption{Overlay of $\Delta$1 of all sub-traces synchronized using Synchronization A, with point addition traces laying on top of point doubling traces, and the algorithm executed in RAM.}
\label{atom1_syncA_inverted}
\end{figure}

The part of the red traces at the end of the atomic block for point additions demonstrates that the duration of $\Delta$1 in point addition patterns is slightly longer than (and not equal to) that of point doubling patterns. 
In Figure \ref{atom1_syncA_inverted}, we inverted the order of lines overlay regarding point addition and point doubling in Figure \ref{atom1_syncA}.

From Figure \ref{atom1_syncA} and \ref{atom1_syncA_inverted}, we noticed via visual observation in a fine scale that the shapes of the red and blue lines are adequately aligned. However, the amplitude of the green line (i.e., the mean trace of EC point doubling sub-traces) in Figure \ref{atom1_syncA_inverted} diminishes gradually and gets closer to zero as it is further away from the synchronization anchor samples. Having a mean value close to zero potentially means that the sub-traces are not well synchronized at the end of the atomic blocks, as the samples in the sub-traces possibly have more distinct values, resulting in cancelling each other when a mean value is being calculated. Thus, we examined other methods, i.e., “Synchronization B” and “Synchronization C” (see Table \ref{sync_methods}) in hope of improving the synchronization. 

\begin{figure}[H]
\centering
\includegraphics[width=0.8\textwidth]{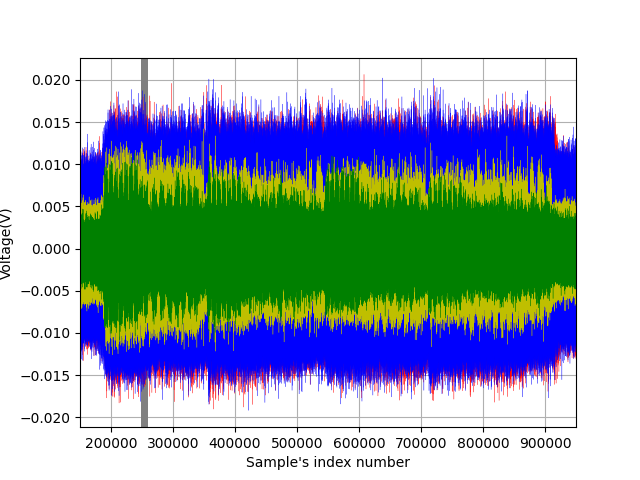}
\caption{Overlay of $\Delta$1 of all sub-traces synchronized using Synchronization B, with the algorithm executed in RAM.}
\label{atom1_syncB}
\end{figure}

Figure \ref{atom1_syncB} shows the results of Synchronization B in $\Delta$1, colored in the same way as in Figure \ref{atom1_syncA} for Synchronization A. All figures in this sub-section adhere to the same format.

\begin{figure}[H]
\centering
\includegraphics[width=1\textwidth]{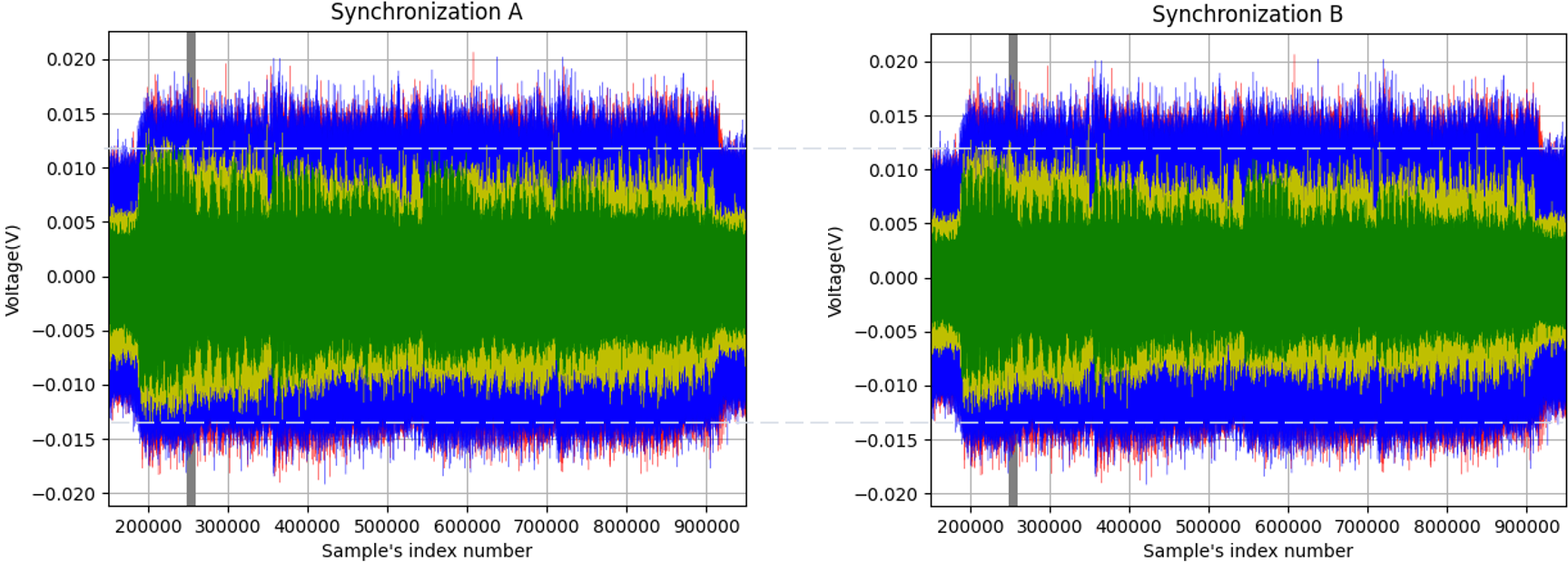}
\caption{Comparison of amplitudes of mean of sub-traces between Synchronization A and Synchronization B.}
\label{compare_syncAB}
\end{figure}

We noticed that Synchronization B does not show any improvements when compared to Synchronization A. In Figure \ref{compare_syncAB}, the minimum value of the mean trace of point addition sub-traces (yellow) and the maximum value of the mean trace of point addition sub-traces (green) from Synchronization A are marked by the horizontal grey dashed lines. The amplitudes of the mean values of both the point addition and point doubling sub-traces from Synchronization B are visibly much smaller than those from Synchronization A, especially at the beginning of the atomic blocks. Therefore, we consider that Synchronization A is better than Synchronization B. However, the mean trace of point doubling sub-traces has always smaller amplitudes than that of point addition’s. One of the reasons of this phenomenon can be a larger number of sub-traces being used in point doubling for calculating the means than in point addition.

\begin{figure}[H]
\centering
\includegraphics[width=0.8\textwidth]{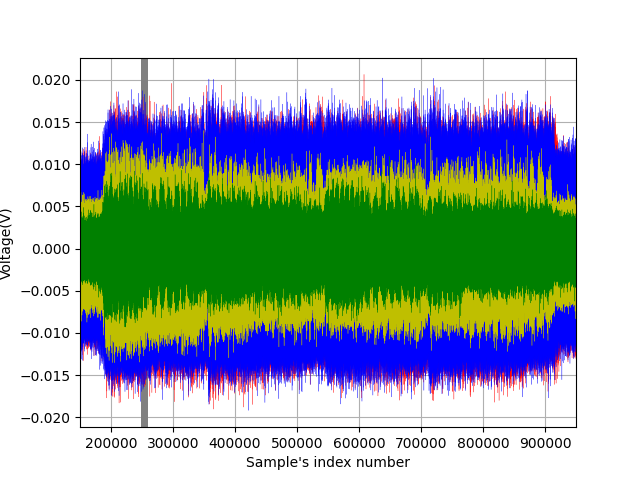}
\caption{Overlay of $\Delta$1 of all sub-traces synchronized using Synchronization C, with the algorithm executed in RAM.}
\label{atom1_syncC}
\end{figure}

Figure \ref{atom1_syncC} shows the results of Synchronization C. This synchronization relying on the fixed clock period on the point doubling sub-traces is not better than the method used in Synchronization A, showing even smaller amplitudes in the mean trace of point doublings.

Comparing the mean trace of $\Delta1$ sub-traces of all three synchronizations, we assert that Synchronization A is our best-effort synchronization, which has decent alignment of point addition and point doubling sub-traces. Therefore, we would use Synchronization A for our plots and analysis going forward. It is important to note that point addition's atomic blocks are visibly slightly longer than point doubling's atomic blocks for all three synchronizations, i.e., the shapes of $\Delta1$ in point doublings and in point additions are distinguishable. Exploring the factors contributing to this distinguishability could be a topic for future research.

\begin{figure}[H]
\centering
\includegraphics[width=0.8\textwidth]{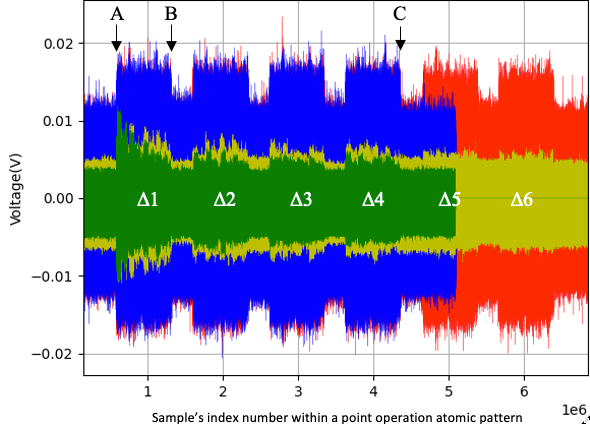}
\caption{Atomic blocks of the synchronized point addition and point doubling sub-traces overlaid (additions: red; doublings: blue) and the mean trace of their sub-traces (addition: yellow; doubling: green).}
\label{6atom_syncB}
\end{figure}

Next, we extended the plot of synchronized sub-traces in $\Delta$1 to all the atomic blocks $\Delta$1 to $\Delta$6, as shown in Figure \ref{6atom_syncB}. We observed that the mean trace of point addition sub-traces and mean trace of point doubling sub-traces have the most distinct peaks in $\Delta$1. From $\Delta$2 onwards, the synchronization does not work as good as in $\Delta$1 anymore, as seen in diminishing peaks in both mean traces, possibly due to different execution time of the NOPs between atomic blocks. Therefore, we considered focusing our distinguishability analysis solely on samples in $\Delta$1. We will next zoom in Figure \ref{6atom_syncB} to gain more insights from the sub-traces at locations A, B and C.

\begin{figure}[H]
\centering
\includegraphics[width=1\textwidth]{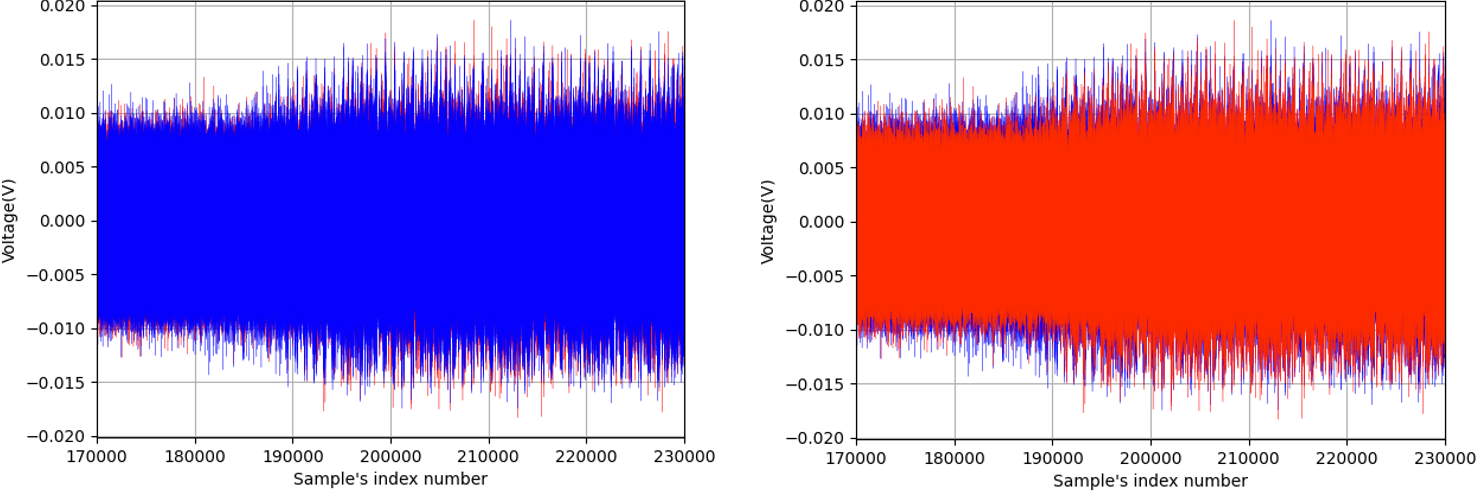}
\caption{The samples at the beginning of $\Delta$1 at location A in Figure \ref{6atom_syncB}, lines overlaid in different orders.}
\label{syncA_atom_start}
\end{figure}

\begin{figure}[H]
\centering
\includegraphics[width=1\textwidth]{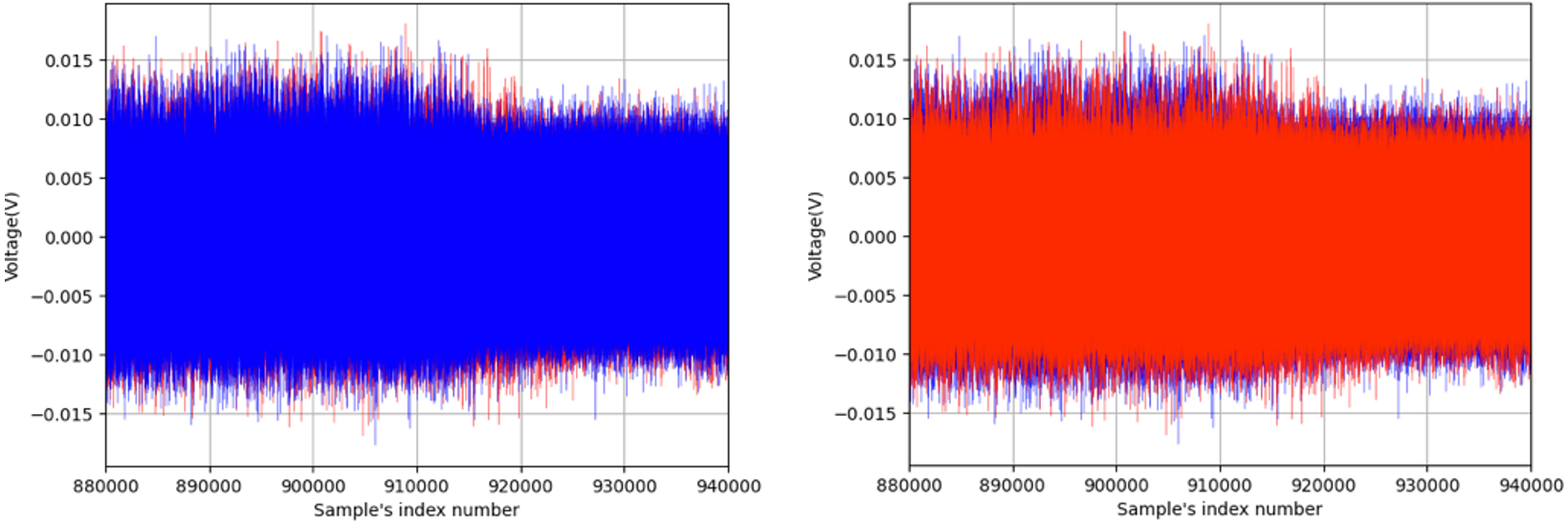}
\caption{The samples at the end of $\Delta$1 at location B in Figure \ref{6atom_syncB}, lines overlaid in different orders.}
\label{syncA_1atom_end}
\end{figure}

The start and the end of $\Delta$1 are particularly interesting to us. In Figure \ref{syncA_atom_start}, there are the first 60,000 samples of $\Delta$1 (from location A in Figure \ref{6atom_syncB}) showing how well the synchronization works in the beginning. We can see that the peaks of both the voltage and the average voltage of both operations are generally well aligned, despite some blue peaks with slightly greater amplitudes from point doubling at around sample index 186,000-191,000 (5,000 samples) in the beginning. In Figure \ref{syncA_1atom_end}, oppositely, we can see some red peaks with greater amplitudes from point addition at around sample index 918,000-923,000 (5,000 samples) at the end of $\Delta$1 (from location B in Figure \ref{6atom_syncB}). Notably, these outstanding peaks have values quite similar to the noise, making it challenging to determine whether these observations could contribute to distinguishing between point addition and point doubling operations. If these outstanding peaks can be indicators for different duration of $\Delta1$ in point doublings in comparison to point additions, an attacker can use them to reveal the key. A possible attack scenario is described in Chapter \ref{Appendix6}.

\begin{figure}[H]
\centering
\includegraphics[width=1\textwidth]{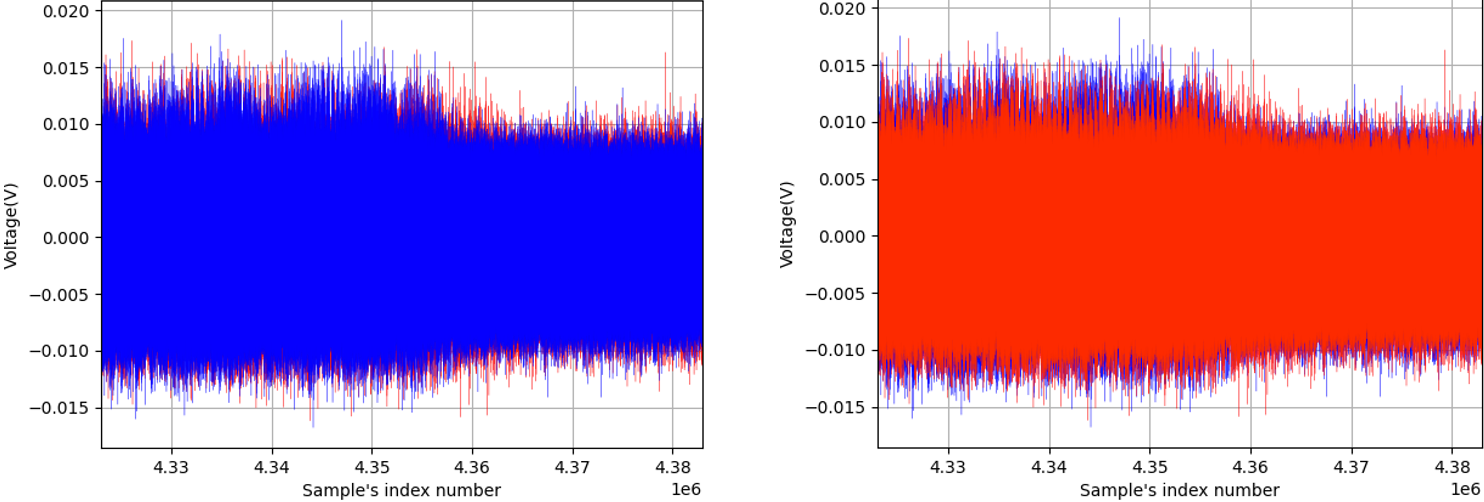}
\caption{The samples at the end of $\Delta$4 at location C in Figure \ref{6atom_syncB}, lines overlaid in different orders.}
\label{syncA_4atom_end}
\end{figure}

Similarly, we used Figure \ref{syncA_4atom_end} to examine the synchronization at the end of $\Delta$4 to see if both point addition and point doubling sub-traces remain aligned. We observed that the sub-traces are still reasonably synchronized. Furthermore, similar to Figure \ref{syncA_1atom_end}, red peaks from point addition stand out at the last 5,000 samples of the atomic block (from location C in Figure \ref{6atom_syncB}).

Through visual comparison of the execution time in the sub-traces, we observed that the atomic patterns of point doubling begin approximately 5,000 samples earlier in $\Delta1$ compared to point addition. i.e., red lines are delayed in comparison to blue lines.
Both EC point operations take the same execution time, but point additions have a delay of 500 clock cycles in comparison to the point doublings. 
Additionally, $\Delta1$ of point additions show slightly more obvious peaks before their shape is fully aligned with that of point doubling. This discrepancy (or delay) could be due to possible alternative synchronization alignments between the two sets, suggesting that multiple locations within the sub-traces in $\Delta1$ could exhibit very similar characteristic shapes and our current alignment may not be optimal. 

We cannot explain these discrepancies (delays of point additions) observed at the beginning and the end of the atomic blocks clearly.
The red and blue lines were perfectly aligned at multiple locations within an atomic block. Despite our best efforts to synchronize the sub-traces and minimize discrepancies, we were unable to identify a better alignment option. The reasons behind these discrepancies require further investigation in future research.

In summary, the point doubling and point addition sub-traces are generally well synchronized throughout $\Delta1$ using Synchronization A. Therefore, we decided to use these synchronized sub-traces in $\Delta1$ for further SSCA.

\subsubsection{Identification of Operations in Atomic Blocks}
\begin{figure}[H]
\centering
\includegraphics[width=1\textwidth]{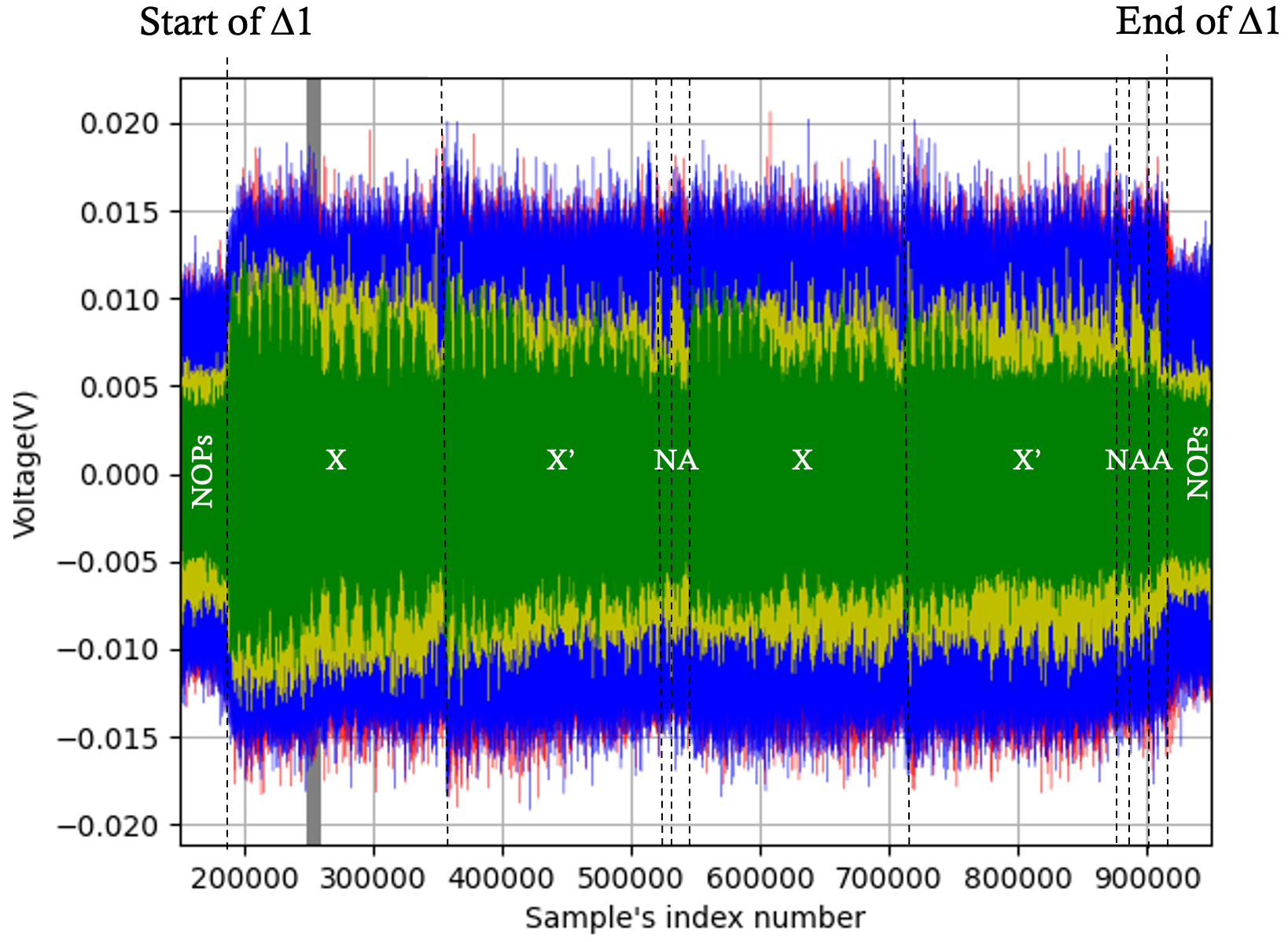}
\caption{Identification of field operations based on the measured execution time of the operations.}
\label{fig:1st_atom_pattern}
\end{figure}

The shape of $\Delta1$ does not allow us to identify any single field operation performed. The mean trace of point doubling sub-traces and the mean trace of point addition sub-traces indicating the synchronization are helpful to identify the multiplication operations: during the multiplication, all sub-traces are better synchronized than during other operations. Figure \ref{fig:1st_atom_pattern} represents our assumption regarding the start and end of each operation within $\Delta1$. 
We used the execution time measured using breakpoints to estimate the duration of each field operation, see Table \ref{flecc_exe_time}. In Figure \ref{fig:1st_atom_pattern}, we marked the areas of the atomic block's sub-traces in white fonts and dashed lines to indicate the execution time of each field operation according to their execution time measured using breakpoints. 

\subsubsection{Automated Simple SCA}
\begin{figure}[H]
\centering
\includegraphics[width=0.75\textwidth]{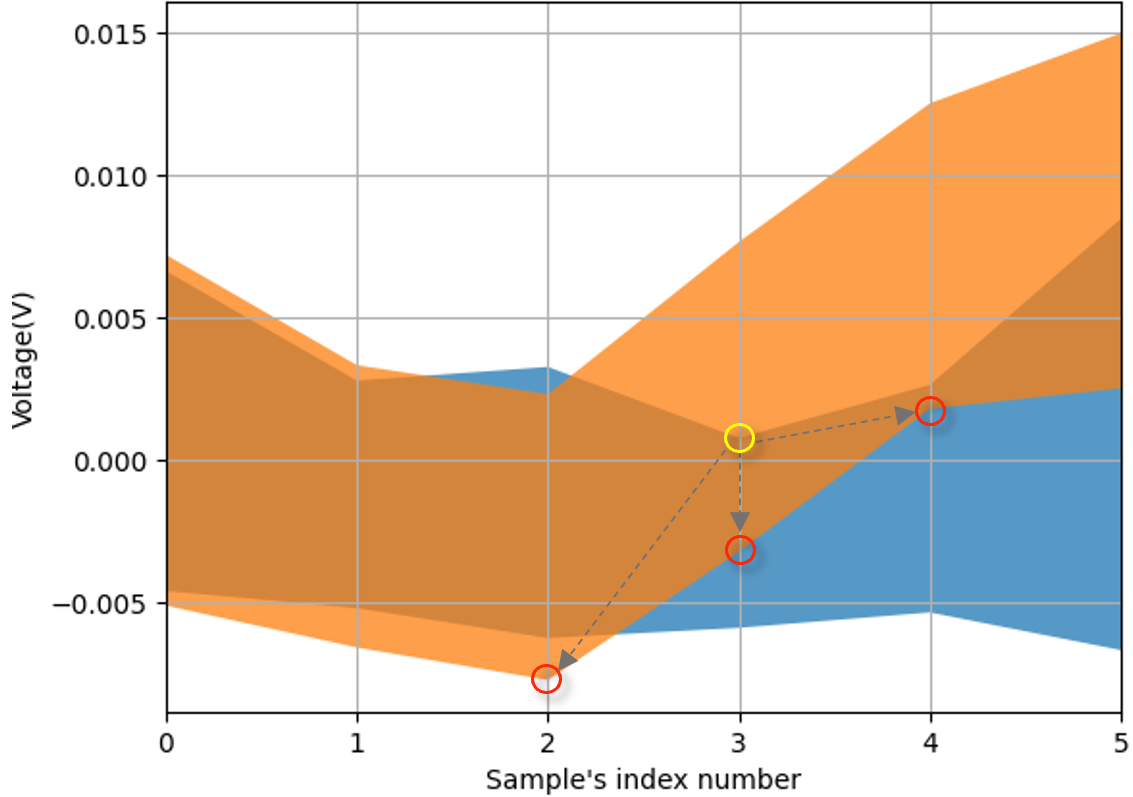}
\caption{An example of identifying a distinguishable sample using a 3-sample window.}
\label{minmax_example}
\end{figure}

After applying Synchronization A on all the sub-traces, we adapted the Automated Simple SCA described in \nocite{kabin_horizontal_2023}[3] to analyse our synchronized and uncompressed sub-traces, focusing on the first atomic block only. Similar to \nocite{kabin_horizontal_2023}[3], we search for the samples' index numbers, which the amplitudes of the set of point addition sub-traces will be completely separated from the set of point doubling sub-traces'. The notation "completely separated" means:
\begin{itemize}
    \item at sample's index number $i$, the maximum value of all point doubling samples is less than the minimum value of all point addition samples, i.e., $max_d(i) < min_a(i)$ or
    \item the minimum value of all point doubling samples is greater than the maximum value of all point addition samples, i.e., $max_a(i) < min_d(i)$
\end{itemize}
By doing so, we can identify potentially distinguishable samples in the measured sub-traces of the atomic pattern point operations. We developed a Python program to look for the occurrence of the above two cases. However, given that perfect synchronization for every sample is unlikely, slight alignment discrepancies may arise between sub-traces over time due to unknown factors. Our program was unable to detect any occurrences of the two cases described above, when comparing the maximum and minimum values of each sample in point doubling sub-traces directly against the corresponding values in point addition sub-traces using Synchronization A. To account for these potential synchronization issues, we extended our analysis to compare a window of sample values instead of individual samples. Figure \ref{minmax_example} illustrates how a 3-sample window works. The orange area represents the range between the minimum and maximum values of the point addition sub-traces, and the blue area represents the corresponding range for the point doubling sub-traces. In this example, we compare the maximum value of point doubling at sample number 3 (indicated by a yellow circle) to the minimum values of point addition at sample numbers 2 to 4 (indicated by red circles). We can see that the maximum value of point doubling at sample number 3 is lower than the minimum value of point addition at sample number 4. Please note that comparing to the traditional method difference-of-means \nocite{heyszl_impact_2013}[103] (Section 5.3.2), our proposed and implemented method allows to determine areas with a trend of capability of being distinguished. More investigations and comparison of the quality of the distinguishability of the point operation sets were not done in this work.

We performed this comparison using windows ranging from 1 to 6 samples and recorded the number of occurrences and the corresponding differences between the maximum and minimum values.

\begin{figure}[H]
\centering
\includegraphics[width=0.7\textwidth]{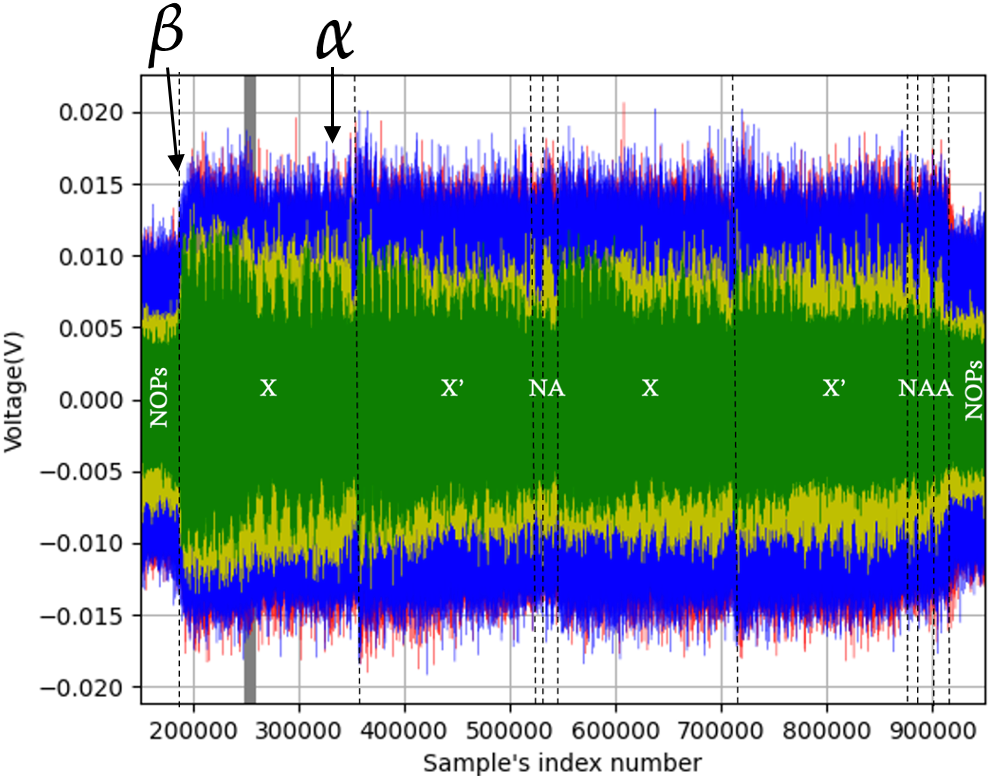}
\caption{Indication of the locations of samples in Figures \ref{minmax_2} ($\beta$) and \ref{minmax_1} ($\alpha$) in $\Delta1$, with each atomic pattern's duration estimated from the execution time measured in Chapter \ref{sec:study-of-flecc}.}
\label{minmax_0}
\end{figure}

With the help of the program, a lot of the target samples were identified, where the set of point doubling sub-traces can be distinguished from the point addition sub-traces. We selected an area in $\Delta1$ of all synchronized sub-traces, which assumingly refers to the first field multiplication in Montgomery space (between sample index 180,000 and 360,000, see Figure \ref{minmax_0}, locations $\alpha$ and $\beta$) to demonstrate some examples of the automatically determined separations of the atomic pattern sub-traces.

\begin{table}[H]
    \centering
    \begin{tabular}{|l|l|l|l|l|}
    \hline
        ~ & $v\leq0.002$ & $0.002<v\leq0.003$ & $0.003<v\leq0.004$ & $v>0.004$  \\ \hline
        1-sample window & 0 & 0 & 0 & 0  \\ \hline
        2-sample window & 0 & 0 & 0 & 0  \\ \hline
        3-sample window & 6 & 0 & 0 & 0  \\ \hline
        4-sample window & 22 & 0 & 0 & 0  \\ \hline
        5-sample window & 84 & 2 & 0 & 1  \\ \hline
        6-sample window & 296 & 16 & 3 & 1 \\ \hline
    \end{tabular}
    \caption{Number of distinguishable samples found using different sample window values, with each sample’s distinguishability segmented by the voltage difference $v$ between the two separated sets of voltages (measured in Volts).}\label{minmax_dist}
\end{table}

Table \ref{minmax_dist} shows the number of distinguishable samples found when using the sample window values 1 to 6. Each column shows the magnitude of the distinguishability, i.e., the distance between the minimum voltage of the point addition sub-traces and the maximum voltage of the point doubling sub-traces at sample index $i$, or oppositely, between the minimum voltage of the point doubling sub-traces and the maximum voltage of the point addition sub-traces at sample index $i$. This helps us determine how distinguishable these samples are. Then, we further looked into the samples with a high degree of distinguishability.

\begin{figure}[H]
\centering
\includegraphics[width=0.7\textwidth]{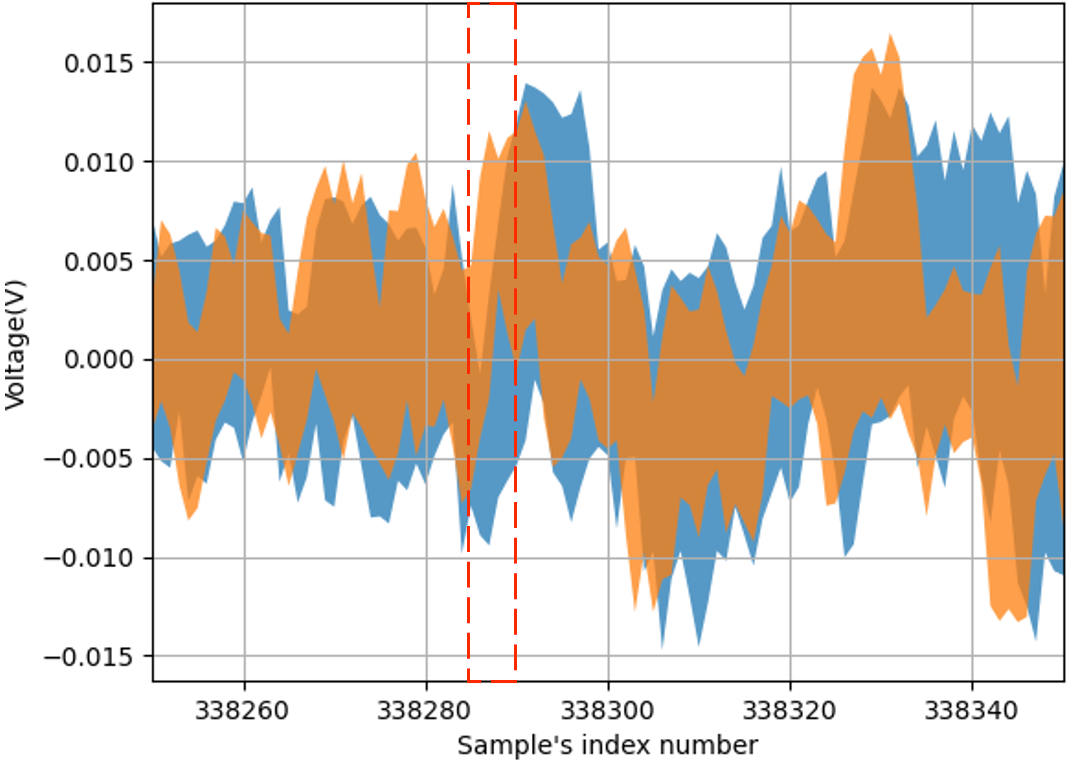}
\caption{Minimum and maximum voltages range for point addition (orange) and point doubling (blue) operations showing a distinguishable sample at index 338,287 with $v\geq0.004$ V in a 5-sample window.}
\label{minmax_2}
\end{figure}

Using a 5-sample window, we identified a notable case at sample index 338,287, where the minimum voltage of point addition sub-trace exceeded the maximum voltage of point doubling sub-trace by more than 0.004 V. This difference serves as a fairly significant indicator of distinguishability. See Figure \ref{minmax_2}, which is located at indicator $\alpha$ in Figure \ref{minmax_0}, approximately at the end of the first Montgomery modular multiplication operation in $\Delta1$. We outlined the distinguishable area of the graph in red.

Having identified many other potential distinguishability indicators in Table \ref{minmax_dist} by analyzing the distances between maximum and minimum voltages of the samples from two point operation sets, we used another technique to narrow our focus on identifying the most effective indicators. In addition to comparing these voltage differences, we also investigated consecutive distinguishable samples using various sample windows to find long sequences of distinguishable samples that could signal success in SCA. The results are shown in Table \ref{streak_count}. Our analysis shows that no consecutive distinguishable samples are found when using a sample window smaller than 5. With a 5-sample window, we identified 3 instances of two consecutive distinguishable samples; this number increased to 15 occurrences with a 6-sample window. However, no sequences longer than two consecutive distinguishable samples were identified. 

\begin{table}[H]
    \centering
    \begin{tabular}{|l|l|l|}
    \hline
        ~ & $s = 2$ & $s \geq 2$  \\ \hline
        1-sample window & 0 & 0  \\ \hline
        2-sample window & 0 & 0  \\ \hline
        3-sample window & 0 & 0  \\ \hline
        4-sample window & 0 & 0  \\ \hline
        5-sample window & 3 & 0  \\ \hline
        6-sample window & 15 & 0 \\ \hline
    \end{tabular}
    \caption{Count of occurrences of consecutive distinguishable samples ($s$) across different sample windows.}\label{streak_count}
\end{table}

\begin{figure}[H]
\centering
\includegraphics[width=0.7\textwidth]{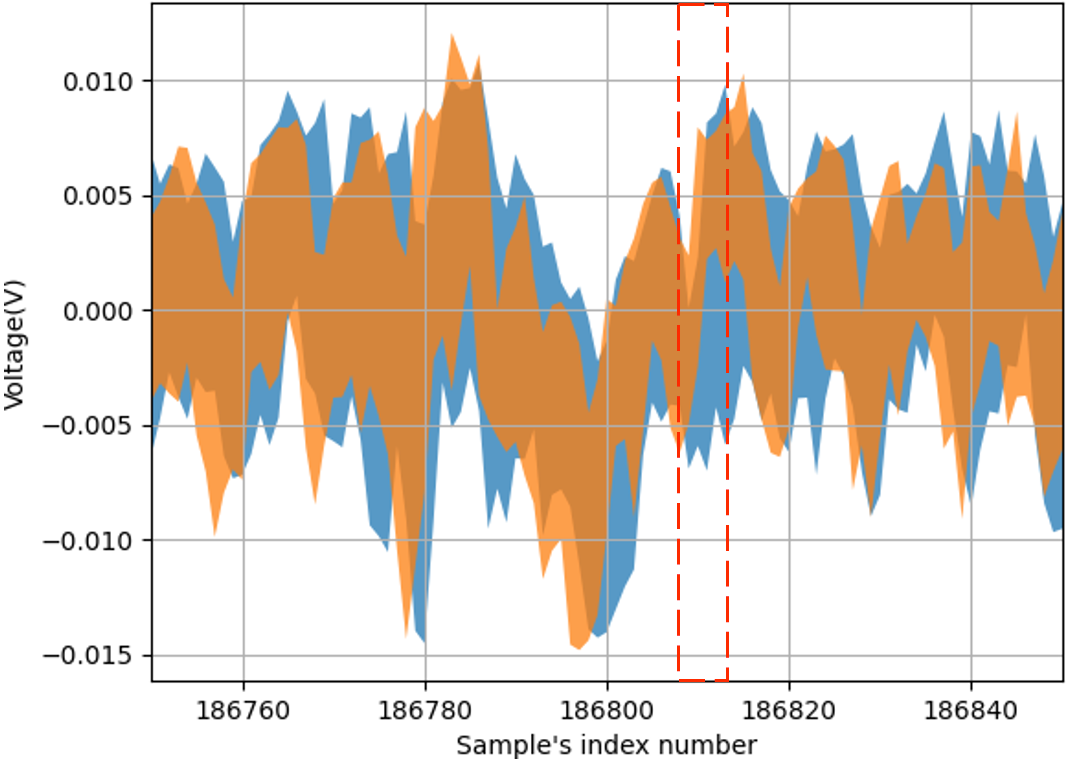}
\caption{Minimum and maximum voltages range for point addition (orange) and point doubling (blue) operations showing distinguishable samples with indices 186,810 and 186,811, in voltage range $0.002$ V$<v\leq0.003$ V in a 5-sample window.}
\label{minmax_1}
\end{figure}

By combining the techniques of distance comparison and the detection of consecutive distinguishable samples, we identified the only case that exhibits both a relatively large distance between the minimum and maximum voltages of different operations and two consecutive distinguishable samples, as shown in Figure \ref{minmax_1}. This case is located at the beginning of $\Delta1$, marked as $\beta$ in Figure \ref{minmax_0}. At sample numbers 186,810 and 186,811, the minimum voltages of point addition sub-traces are higher than the maximum voltages of point doubling sub-traces in a 5-sample window, by a distinguishable distance of 0.002 V to 0.003 V. 

Although no longer streaks were found to indicate more significant distinguishability, it is noteworthy that in these two cases, which represent our best distinguishability indicators, the adjacent samples from point doublings and point additions appear well synchronized, with their maximum and minimum ranges closely aligned.

\clearpage
\subsection{Analysis of \textit{k\textbf{P}} Operation Executed in Flash Memory}
\subsubsection{Synchronization Alignment}

Since we used 100 Mega samples (MS) per second in the oscilloscope and 20 MHz clock rate in the CCS program run in flash memory of the board, the trace was captured with only 5 samples per clock cycle, which is very low and makes it difficult to obtain meaningful insights from any analysis. Nonetheless, we split the trace into sub-traces using similar methods we used on RAM traces as in Chapter \ref{kp_analysis_ram}.

\begin{figure}[H]
\centering
\includegraphics[width=1\textwidth]{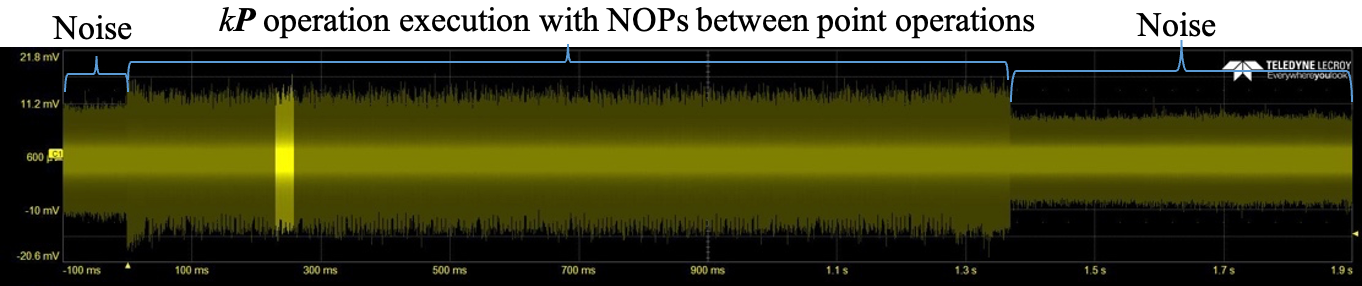}
\caption{EM trace of noise and the implemented atomic pattern $k\bm{P}$ algorithm executed in flash memory of the board attacked.}
\label{full_trace_flash}
\end{figure}

\begin{figure}[H]
\centering
\includegraphics[width=0.7\textwidth]{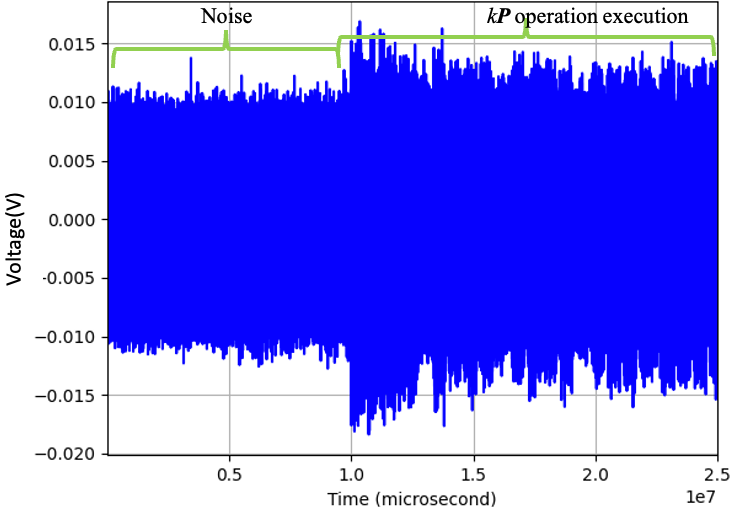}
\caption{EM trace of noise and the beginning of the $k\bm{P}$ algorithm executed in flash memory of the board attacked.}
\label{kp_start_flash}
\end{figure}

\begin{figure}[H]
\centering
\includegraphics[width=1\textwidth]{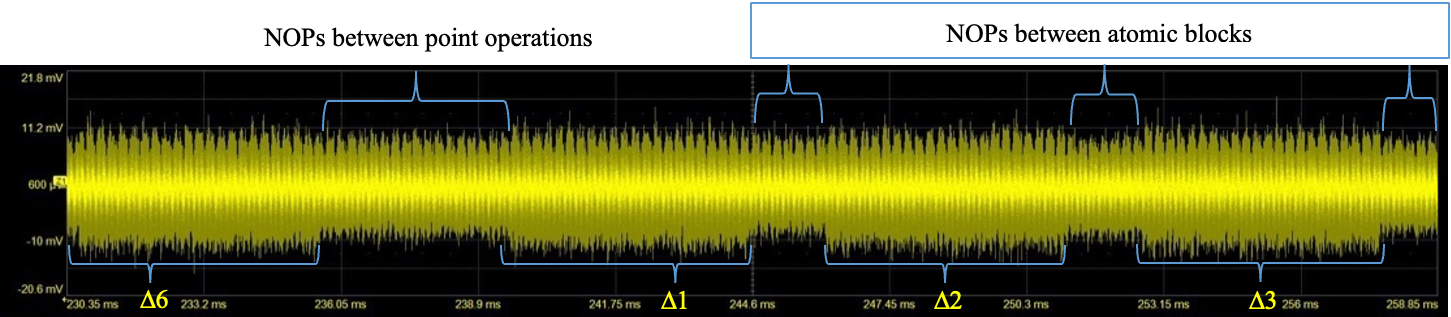}
\caption{EM trace of atomic blocks of point operations of the $k\bm{P}$ algorithm executed in flash memory of the board attacked.}
\label{kp_flash_zoomed}
\end{figure}

First, we identified the noise and the $k\bm{P}$ operation in the full trace, as shown in Figure \ref{full_trace_flash} and a magnified version in Figure \ref{kp_start_flash}. Then, with the help of NOPs inserted as described in Chapter \ref{kp_analysis_ram}, we were able to separate each atomic block in point doubling and point addition operations into sub-traces, as shown in Figure \ref{kp_flash_zoomed}. 

 We tried to synchronize the sub-traces by using a simple averaging method, which is to first convert each sample in each sub-trace into the average voltage value of its closest 10 samples, and then try to align the samples with another sub-trace's manually using Excel charts. Despite our efforts in manual synchronization, the processed sub-traces cannot be synchronized. It is important to recognize that both the low number of samples per clock cycle and the high noise level are significant problems hindering the success of simple SCA. Therefore, unfortunately we cannot further carry out simple electromagnetic analysis attack on the $k\bm{P}$ operation executed in flash memory like what we did for the $k\bm{P}$ operation executed in RAM.

\subsection{Comparison of the Results to Sigourou et al.'s Work [27]}

As mentioned in Chapter \ref{Chapter3}, no investigation of the SCA-resistance of Longa's atomic patterns was performed experimentally prior to this work.
Our objectives, code implementation and experimental setup closely resemble those used in the SCA study by Sigourou et al. \nocite{sigourou_successful_2023}[27]. In this section, we compare our work to \nocite{sigourou_successful_2023}[27] to identify any overarching findings. A short summary is given in Table \ref{compare-sigourou} below.

\begin{table}[!ht]
    \centering
    \begin{tabular}{|l|l|l|}
    \hline
        ~ & Sigourou et al. \nocite{sigourou_successful_2023}[27] & This work \\ \hline
        $k\bm{P}$ algorithm & Double-and-add L-to-R & Double-and-add L-to-R \\ \hline
        Atomic patterns & Rondepierre \nocite{francillon_revisiting_2014}[23] & Longa [2] \\ \hline
        Cryptographic library & FLECC & FLECC \\ \hline
        Embedded device & LAUNCHXL-F280025C & LAUNCHXL-F28379D \\ \hline
        \begin{tabular}[c]{@{}l@{}}Samples per clock \\cycle\end{tabular} & 50 & \begin{tabular}[c]{@{}l@{}}RAM: 10 \\ Flash: 5\end{tabular} \\ \hline
        Signal-to-noise ratio & $\approx 2.69$ & $\approx 1.36$ \\ \hline
        \begin{tabular}[c]{@{}l@{}}No. of samples \\per point operation\end{tabular} & \begin{tabular}[c]{@{}l@{}}PD: 1,970,000 \\ PA: 1,970,000\end{tabular} & \begin{tabular}[c]{@{}l@{}}PD: 3,200,000 \\ PA: 4,800,000 \end{tabular} \\ \hline
        File size & 2.7 GB &  \begin{tabular}[c]{@{}l@{}}RAM: 6.02 GB \\ Flash: 5.86 GB\end{tabular}\\ \hline
        SCA type & Horizontal & Horizontal \\ \hline
        SCA method & Simple Power Analysis & Automated Simple SCA \\ \hline
        Major findings & \begin{tabular}[c]{@{}l@{}}
            - SCA leakage in $1^{st}$ and $4^{th}$ OPs \\- Desynchronization found \\
            - FLECC is not constant-time
        \end{tabular} & 
        \begin{tabular}[c]{@{}l@{}} 
            - Negative SSCA results \\ - Desynchronization found \\
            - FLECC is constant-time \end{tabular} \\  \hline
    \end{tabular}
    \caption{Summary of comparison between [27] by Sigourou et al. and our work.}\label{compare-sigourou}
\end{table}

 Both works use FLECC for the implementation of the binary double-and-add left-to-right (L-to-R) $k\bm{P}$ algorithm (Algorithm \ref{modified-kp}). Sigourou et al. used Rondepierre's atomic patterns \nocite{francillon_revisiting_2014}[23] while we used Longa's [2], both with the insertion of NOPs in the algorithm to ease the separation of sub-traces. 

Similar to the finding of Sigourou et al., we noticed some desynchronization in our aligned sub-traces, which led us to investigate the execution time of the supposed constant-runtime functions in FLECC. Our experimental board enables us to use one hardware breakpoint within the implemented code to measure the duration of different operations, offering an alternative to visually analysing the trace.
We repeatedly measured the execution time of the "constant-runtime" Montgomery modular multiplication, negation and addition functions in FLECC and confirmed that they are indeed constant-runtime, with only a minor fluctuation of 1 clock cycle in the runtime of Montgomery modular multiplication. This evidence supports that FLECC is truly constant-runtime, contrary to the claim made in \nocite{sigourou_successful_2023}[27].

Sigourou et al. performed a Simple Power Analysis in their work, while we applied Automated Simple SCA. They identified the part of the $k\bm{P}$ trace, which allows to clearly differentiate between PD and PA operations. We were not able to show a high degree of distinguishability using the distinguishability indicators we found. Thus, our atomic patterns $k\bm{P}$ implementation executed in our target embedded device is resistant against our applied SSCA techniques. 
They assumed the distinguishability is caused by the different addresses of the registers in memory and the Hamming distance between these addresses. Further research is required to validate these assumptions. In our work, due to a small number of samples per clock cycle, the vulnerability to horizontal address-bit SCA cannot be confirmed nor ruled out.

\chapter{Conclusion} 

\label{Chapter8} 
Atomic patterns were introduced as a countermeasure against simple SCA attacks targeting EC $k\bm{P}$ algorithms. This work primarily aims to experimentally investigate the resistance of Longa's atomic patterns [2] to simple SCA attacks, with the goal of applying these atomic patterns in EC-based cryptographic protocols for embedded devices using an open-source cryptographic library. In Longa's work, atomic pattern algorithms with 4, 6 and 8 atomic blocks were proposed to realize EC point doubling, point addition and point tripling operations respectively. Given the frequent use of binary $k\bm{P}$ algorithms in the implementation of cryptographic operations for embedded devices, our work focused specifically on Longa's atomic patterns for point doubling and point addition which are essential operations for double-and-add $k\bm{P}$ algorithms, selecting the atomic pattern MNAMNAA for our investigation.

As the first step of our investigation, we performed an exhaustive search of literature citing Longa's work [2]. Our goal was to determine whether the resistance of Longa's atomic patterns to SSCA or other horizontal SCA attacks had been experimentally tested. As a result, we identified 63 papers citing [2] across various contexts.
Some papers referred [2] as state-of-the-art SSCA countermeasure, some others focused on improving the efficiency of $k\bm{P}$ algorithms. Only a few papers discussed the potential vulnerabilities of the atomic patterns to horizontal SCA attacks. However, none of these papers experimentally examined the SCA-resistance of Longa's atomic patterns.

To address this gap, we implemented a binary left-to-right double-and-add $k\bm{P}$ algorithm using P-256, which is a recommended EC for ECDSA and EC key establishment by NIST since 2024. Our implementation was tested on a LAUNCHXL-F28379D LaunchPad board with a MS320F28379D C2000 32-bit dual-core microcontroller. We evaluated three open-source cryptographic libraries for their suitability to meet our objectives and selected the library FLECC in C, which is proven to offer constant-runtime functions.

During the implementation, we identified several mistakes in Longa's atomic patterns published in [2]. The atomic patterns were corrected corresponding to the point doubling and mixed point addition formulae proposed by Longa for the following cases:
    \begin{itemize}
        \item MNAMNAA-based atomic point doubling ([2], Appendix B2)
        \item MANA-based atomic mixed addition ([2], Appendix B3)
        \item MNAMNAA-based atomic mixed addition ([2], Appendix B4)
        \item MANA-based atomic special addition with identical Z-coordinate ([2], Appendix B5)
        \item MNAMNAA-based atomic special addition ([2], Appendix B6)
        \item MANA-based atomic point tripling ([2], Appendix B7)
        \item MNAMNAA-based atomic point tripling ([2], Appendix B8)
    \end{itemize}

To evaluate the resistance of the implemented $k\bm{P}$ algorithm against SSCA, we performed a horizontal electromagnetic analysis attack, measuring the EM emanations of the target board during a $k\bm{P}$ execution. Due to technical limitations, we used a 22-bit scalar $k$ and focused on analysing only the first atomic blocks in EC point doublings and additions. Automated Simple SCA was conducted to identify sample indices where the sub-traces of point doubling significantly differ from those of point addition. However, no significant difference was found simply using this technique.
We further enhanced this technique by developing two additional methods - distance comparison and detection of consecutive distinguishable samples - to identify trends. Specifically, we used Automated Simple SCA to compare the amplitudes of the synchronized samples in both sets, as well as the amplitudes of the neighboring samples within a small window of samples.

Our analysis revealed no significant differences in execution time or the shape of the atomic block sub-traces, likely due to high noise levels, an insufficient number of samples per clock cycle, long execution time of the $k\bm{P}$ implementation using FLECC library and the large file size of the EM trace. Based on the knowledge of the limitations described in this thesis, further investigation is required to conclusively determine whether Longa’s atomic patterns in the $k\bm{P}$ algorithm are vulnerable to single-trace SCA attacks, particularly horizontal address-bit SCA.

\chapter{Appendix} 

\label{Chapter9}


\section{Results of Literature Overview} 

\label{Appendix1} 

The 63 papers referenced in the column "Reference" in Table \ref{tab:lit-overview} below all truly cited Longa’s master thesis [2]. 

The column "Theme" represents the topic of discussion of the reference paper.

The column "State-of-the-art" represents whether the reference paper considers Longa’s MNAMNAA atomic pattern as an example of atomic patterns in the context of state-of-the-art countermeasure against simple SCA attacks. 

"Basis for improvement" column shows whether the author(s) of the reference paper used Longa’s MNAMNAA-pattern(s) for their theoretical investigations or in practical experiments.

The column "Analysed" shows whether the reference paper analysed the efficiency or SCA-resistance of Longa's MNAMNAA-pattern(s) theoretically, or made claims to Longa's pattern(s) based on analysis done on other similar atomic patterns. 

The column "Implemented" represents whether the author(s) actually implemented Longa's MNAMNAA atomic pattern(s) on hardware in their work.

The column "Used for $k\bm{P}$" represents whether the reference paper has made an improvement on $k\bm{P}$ algorithm without any change on atomic patterns.

The column "Remarks" describes in which context [2] was referred to in the reference paper.

\clearpage
\begin{landscape}
\begin{longtable}[H]{|l|l|l|l|l|l|l|l|}
\hline
\begin{tabular}[c]{@{}l@{}}Refe-\\ rence\end{tabular} & Theme & \begin{tabular}[c]{@{}l@{}}State-\\ of-the\\-art\end{tabular} & \begin{tabular}[c]{@{}l@{}}Basis\\for\\ improv-\\ement\end{tabular} & \begin{tabular}[c]{@{}l@{}}Anal-\\ysed\end{tabular} & \begin{tabular}[c]{@{}l@{}}Imple-\\ mented\end{tabular} & \begin{tabular}[c]{@{}l@{}}Used\\for\\ $k\bm{P}$\end{tabular} & Remarks \\ \hline
                                        
\endfirsthead

\hline
\begin{tabular}[c]{@{}l@{}}Refe-\\ rence\end{tabular} & Theme & \begin{tabular}[c]{@{}l@{}}State-\\ of-the\\-art\end{tabular} & \begin{tabular}[c]{@{}l@{}}Basis\\for\\ improv-\\ement\end{tabular} & \begin{tabular}[c]{@{}l@{}}Anal-\\ysed\end{tabular} & \begin{tabular}[c]{@{}l@{}}Imple-\\ mented\end{tabular} & \begin{tabular}[c]{@{}l@{}}Used\\for\\ $k\bm{P}$\end{tabular} & Remarks \\ \hline
\endhead

\nocite{hutchison_atomicity_2010}[28] & SCA & Y                & Y                     & Y        & N & Y & \begin{tabular}[c]{@{}l@{}}Introduced general doubling and readdition algorithms using Longa's \\atomic patterns.\end{tabular} \\ \hline

\nocite{verneuil_elliptic_2012}[31]& \begin{tabular}[c]{@{}l@{}}SCA \end{tabular} & Y                & Y                     & Y        & N & Y & \begin{tabular}[c]{@{}l@{}}Used Longa's MNAMNAA pattern to build readdition and general\\ doubling algorithms\end{tabular}         \\ \hline

\nocite{lu_general_2013}[30] & SCA & Y                & Y                     & Y        & N & Y & \begin{tabular}[c]{@{}l@{}}Provided a general framework that can accommodate Longa's base-\{2,3\} \\number systems, $wmb$NAF representations and atomic patterns.\end{tabular}         \\ \hline

\nocite{francillon_revisiting_2014}[23] & SCA & Y                & Y                     & Y        & N & N & \begin{tabular}[c]{@{}l@{}}Compared the performance of his new atomic pattern algorithm\\ with Longa's.\end{tabular}         \\ \hline

\nocite{bauer_horizontal_2015}[12] & SCA  & Y                & Y                     & Y        & N & N & \begin{tabular}[c]{@{}l@{}}Theoretically applied their novel attack with input randomization\\ against Longa's atomic implementation.\end{tabular}        \\ \hline

\nocite{das_improved_2016}[32] & SCA  & Y                & N                     & Y        & N & N & \begin{tabular}[c]{@{}l@{}}Showed vulnerabilities in Longa's atomic pattern against HCCA.\end{tabular}         \\ \hline

\nocite{kabin_ec_2021-1}[33]& \begin{tabular}[c]{@{}l@{}}SCA\end{tabular} & Y                & N                     & N        & N & N & \begin{tabular}[c]{@{}l@{}}Mentioned Longa's atomic pattern as state-of-the-art scalar\\ multiplication algorithm.\end{tabular}         \\ \hline

\begin{tabular}[c]{@{}l@{}}\nocite{kabin_randomized_2023, kabin_horizontal_2023}[3, 21,\\ \nocite{kabin_ec_2021, sigourou_successful_2023}27, 34]\end{tabular} & SCA & Y                & N                     & N        & N & N & \begin{tabular}[c]{@{}l@{}}Mentioned Longa's atomic pattern as state-of-the-art SCA \\countermeasure.\end{tabular}         \\ \hline

\nocite{longa_high-speed_2011}[35]& \begin{tabular}[c]{@{}l@{}}EC kP\\efficiency\end{tabular} & Y                & N                     & N        & N & Y & \begin{tabular}[c]{@{}l@{}}Evaluated costs of point doubling and addition in doubling-addition\\ $2\bm{P}+\bm{Q}$ formula from [2].\end{tabular}         \\ \hline

\nocite{venelli_contribution_2011}[37]& \begin{tabular}[c]{@{}l@{}}SCA \end{tabular} & Y                & N                     & N        & N & Y & \begin{tabular}[c]{@{}l@{}}Mentioned Longa's atomic pattern as state-of-the-art scalar\\ multiplication algorithm.\end{tabular}         \\ \hline

\begin{tabular}[c]{@{}l@{}}\nocite{houssain_elliptic_2012, houssain_power_2012, tawalbeh_towards_2016}[38-40]\end{tabular} & SCA & Y                & N                     & N        & N & N & \begin{tabular}[c]{@{}l@{}}Mentioned Longa's atomic patterns as state-of-the-art SPA \\countermeasure. \end{tabular}         \\ \hline

\nocite{faye_algorithmes_2014}[36]& \begin{tabular}[c]{@{}l@{}}EC kP\\efficiency\end{tabular} & Y                & N                     & N        & N & Y & \begin{tabular}[c]{@{}l@{}}Compared the costs of Longa's point tripling and special point\\ addition formulae with others.\end{tabular}         \\ \hline

\nocite{cramer_new_2008}[84]& \begin{tabular}[c]{@{}l@{}}EC kP\\efficiency\end{tabular} & N                & N                     & N        & N & Y & \begin{tabular}[c]{@{}l@{}}Mentioned Longa's $mb$NAF method.\end{tabular}          \\ \hline

\nocite{longa_setting_2008, chung_fast_2011}[73, 75]& \begin{tabular}[c]{@{}l@{}}EC kP\\efficiency\end{tabular} & N                & N                     & N        & N & Y & \begin{tabular}[c]{@{}l@{}}Mentioned Longa's generic multibase scalar representations.\end{tabular}          \\ \hline


\nocite{jarecki_fast_2009}[85]& \begin{tabular}[c]{@{}l@{}}EC kP\\efficiency\end{tabular} & N                & N                     & N        & N & Y & \begin{tabular}[c]{@{}l@{}}Mentioned Longa's $mb$NAF and $wmb$NAF representations and\\ compared them with the proposed representation.\end{tabular}          \\ \hline

\nocite{hutchison_efficient_2010}[86]& \begin{tabular}[c]{@{}l@{}}EC kP\\efficiency\end{tabular} & N                & N                     & N        & N & Y & \begin{tabular}[c]{@{}l@{}}Mentioned Longa's doubling-addition as state-of-the-art formula\\ and implemented it to compare point multiplication costs.\end{tabular}          \\ \hline

\nocite{longa_analysis_2010}[74]& \begin{tabular}[c]{@{}l@{}}EC kP\\efficiency\end{tabular} & N                & N                     & N        & N & Y & \begin{tabular}[c]{@{}l@{}}Implemented Longa's doubling-addition algorithm.\end{tabular}          \\ \hline

\nocite{adikari_hybrid_2011}[76] & \begin{tabular}[c]{@{}l@{}}EC kP\\efficiency\end{tabular} & N                & N                     & N        & N & Y & \begin{tabular}[c]{@{}l@{}}Mentioned Longa's multibase single scalar multiplication, as it is \\very similar to the authors' work.      \end{tabular}   \\ \hline

\begin{tabular}[c]{@{}l@{}}\nocite{almohaimeed_increasing_2013, meier_side-channel_2014, das_exploiting_2015, sako_side-channel_2016, das_inner_2016, ryan_safe-errors_2016, das_automatic_2019}[41-72],\\ \nocite{gebotys_elliptic_2010, purohit_fast_2011, hutchison_recoding_2012,purohit_elliptic_2012, chabrier_arithmetic_2013, chabrier_--fly_2013, reyes_performance_2013, dygin_efficient_2013, hutchison_joint_2013, hutchison_expansion_2013, li_fast_2014, al_saffar_improved_2014, ajeena_point_2014, ahmad_x-tract_2015, lopez_analysis_2015, dou_fast_2015, _jacobian_2015, __2015,  das_secure_2016, romaniuk_usage_2016, quetny__2016, guerrini_randomized_2018, khleborodov_fast_2018, ho_kim_speeding_2020, cai_handshake_2018, wojcik_partially_2018, khleborodov_fast_2018-1, yq_dou_revisiting_2018, ilyenko_perspectives_2019, andres_lara-nino_comparison_2021, shuang-gen_liu_fast_2021, xie_secure_2022}[77-83]\end{tabular}& Misc. & N                & N                     & N        & N & Misc. & Not relevant to our goal of study in this thesis.        \\ \hline

\caption{Literature overview: aspects relevant for this thesis were taken as the basis for classification of papers referring to [2]}\label{tab:lit-overview}
\end{longtable}
\end{landscape}

\section{Revision of Longa's Atomic Pattern Algorithms} 

\label{Appendix2} 
We reviewed all of Longa’s atomic pattern and SSCA-protected algorithms (Appendices B1-B10, C1-C3 and E1-E3 in [2]). Below are the atomic patterns in which we have found errors (Appendices B2, B3, B4, B5, B6, B7 and B8 in [2]). The corrections are shown in bold text with yellow highlights in the following.

\subsection{Longa’s Appendix B2 algorithm revised}
    \label{appendix-mnamnaa-point-doubling}
The MNAMNAA-based atomic point doubling algorithm taken from [2](Appendix B2) is revised as follows:
\\ \\ \noindent Input: $P=(X_1,Y_1,Z_1)$ \\
Output: $2P=(X_3,Y_3,Z_3)$ \\
$T_1\gets X_1, T_2\gets Y_1, T_3\gets Z_1$

\begin{table}[hbt!]
\centering
\begin{tabular}{|lr|lr|}
\hline
\multicolumn{2}{|c|}{$\Delta1$} & \multicolumn{2}{c|}{$\Delta2$} \\ \hline
$T_4=T_3^2$ & $(Z_1^2)$ & $T_5=T_4\times T_5$ & $(A.B)$ \\ 
* && * &\\
$T_5=T_1+T_4$ & $(A=X_1+Z_1^2)$ & $T_4=T_5+T_5$ & $(2A.B)$ \\
$T_6=T_2^2$ & $(Y_1^2)$ & $T_3=T_2\times T_3$ & $(Z_3)$ \\
$T_4=-T_4$ & $(-Z_1^2)$ & * &\\
$T_2=T_2+T_2$ & $(2Y_1)$&  $T_4=T_4+T_5$ & $(\alpha)$ \\
$T_4=T_1+T_4$ & $(B=X_1-Z_1^2)$ &  $T_2=T_6+T_6$ & $(2Y_1^2)$ \\ \hline

\multicolumn{2}{|c|}{$\Delta3$} & \multicolumn{2}{c|}{$\Delta4$} \\ \hline
$T_5=T_4^2$ & $(\alpha^2)$ & $T_2=T_2^2$ & $(4Y_1^4)$ \\
* && $T_5=-T_1$ & $(-X_3)$ \\
$T_6=T_2+T_2$ & $(4Y_1^2)$ & $T_5=T_5+T_6$ & $(\beta-X_3)$ \\
$T_6=T_1\times T_6$ & $(\beta)$ & $T_5=T_4\times T_5$ & $(\alpha(\beta-X_3))$ \\
$T_1=-T_6$ & $(-\beta)$ & $T_2=-T_2$ & $(-4Y_1^4)$ \\
$T_1=T_1+T_1$ & $(-2\beta)$ & $T_2=T_2+T_2$ & $(-8Y_1^4)$ \\
$T_1=T_1+T_5$ & $(X_3)$ & $T_2=T_2+$\hl{$\pmb{T_5}$} & $(Y_3)$ \\ \hline
\end{tabular}
\caption{\label{mnamnaa-point-doubling}Revised MNAMNAA-based atomic point doubling. The newly revised register is shown in bold with yellow highlight.}
\end{table}

\clearpage
\subsection{Longa’s Appendix B3 algorithm revised}\label{appendix-mana-mixed-addition}
The MANA-based atomic mixed addition algorithm taken from [2](Appendix B3) is revised as follows:
\\ \\ \noindent Input: $P=(X_1,Y_1,Z_1)$ and $Q=(X_2,Y_2)$ \\
Output: $P+Q=(X_3,Y_3,Z_3,X_1',Y_1')$ \\
$T_1\gets X_1, T_2\gets Y_1, T_3\gets Z_1, T_x\gets X_2, T_y\gets Y_2$

\begin{table}[hbt!]
\centering
\begin{tabular}{|lr|lr|}
\hline
\multicolumn{2}{|c|}{$\Delta1$} & \multicolumn{2}{c|}{$\Delta2$}  \\ \hline
$T_4=T_3^2$ & $(Z_1^2)$ & 
$T_5=T_x\times T_4$ & $(Z_1^2X_2)$ \\

* && * & \\
* && $T_6=-T_1$ & $(-X_1)$ \\
* && $T_5=T_5+T_6$ & $(A=Z_1^2X_2-X_1)$\\
 \hline
 
\multicolumn{2}{|c|}{$\Delta3$} & \multicolumn{2}{|c|}{$\Delta4$} \\ \hline
$T_6=T_5^2$ & $(A^2)$ & $T_7=T_1\times T_6$ & $(X_1')$ \\
* & & $T_8=T_1+T_1$ & $(2$\hl{$\pmb{X_1}$}$)$ \\
* & & * & \\
* & & * & \\
\hline

\multicolumn{2}{|c|}{$\Delta5$} & \multicolumn{2}{c|}{$\Delta6$} \\ \hline
 $T_9=T_5\times T_6$ & $(A^3)$ & $T_4=T_3\times T_4$ & $(Z_1^3)$ \\
 $T_8=T_8+T_9$ & $(A^3+2$\hl{$\pmb{X_1}$}$)$ & * &  \\
 * && * & \\
 * && * & \\ 

\hline

\multicolumn{2}{|c|}{$\Delta7$} & \multicolumn{2}{c|}{$\Delta8$} \\ \hline

$T_4=T_y\times T_4$ & $(Z_1^3Y_2)$ & $T_{10}=T_4^2$ & $(B^2)$ \\
* && * &\\ 

$T_{10}=-T_2$ & $(-Y_1)$ & $T_8=-T_8$ & $(-A^3-2$\hl{$\pmb{X_1}$}$)$ \\

$T_4=T_4+T_{10}$ & $(B=Z_1^3Y_2-Y_1)$ & $T_1=$\hl{$\pmb{T_4}$}$+T_8$ & $(X_3)$ \\ \hline

\multicolumn{2}{|c|}{$\Delta9$} & \multicolumn{2}{|c|}{$\Delta10$} \\ \hline
$T_8=T_2\times T_9$ & $(Y_1')$ & $T_6=T_6\times $\hl{$\pmb{T_4}$} & $(B(X_1'-X_3))$ \\
* & & * & \\
$T_6=-T_1$ & $(-X_3)$ & $T_9=-T_8$ & $(-Y_1')$ \\
$T_6=T_6+T_7$ & $(X_1'-X_3)$ & $T_2=T_6+T_9$ & $(Y_3)$ \\
\hline

 \multicolumn{2}{|c|}{$\Delta11$}  \\ \cline{1-2}
$T_3=T_3\times T_5$ & $(Z_3)$ \\
* & \\ 

$T_4=-T_7$ & $(-X_1')$  \\

$T_4=T_1+T_4$ & $(X_3-X_1')$\\ \cline{1-2}

\end{tabular}

\caption{\label{mana-mixed-addition}Revised MANA-based atomic mixed addition. The newly revised registers are shown in bold with yellow highlight.}
\end{table}

\clearpage
\subsection{Longa’s Appendix B4 algorithm revised}\label{appendix-mnamnaa-mixed-addition}
The MNAMNAA-based atomic mixed addition algorithm taken from [2](Appendix B4) is revised as follows:
\\ \\ \noindent Input: $P=(X_1,Y_1,Z_1)$ and $Q=(X_2,Y_2)$ \\
Output: $P+Q=(X_3,Y_3,Z_3,X_1',Y_1')$ \\
$T_1\gets X_1, T_2\gets Y_1, T_3\gets Z_1, T_x\gets X_2, T_y\gets Y_2$

\begin{table}[hbt!] 
\centering
\begin{tabular}{|lr|lr|}
\hline
\multicolumn{2}{|c|}{$\Delta1$} & \multicolumn{2}{c|}{$\Delta2$}\\ \hline
$T_4=T_3^2$ & $(Z_1^2)$ & 
$T_6=T_5^2$ & $(A^2)$ \\

* && * &\\
* && * &\\
$T_5=T_x\times T_4$ & $(Z_1^2X_2)$ & $T_7=T_1\times T_6$ & $(X_1')$ \\
$T_6=-T_1$ & $(-X_1)$ & * & \\
$T_5=T_5+T_6$ & $(A=Z_1^2X_2-X_1)$ & $T_8=$\hl{$\pmb{T_7}$}$+$\hl{$\pmb{T_7}$} & $(2X_1')$ \\
* && * &\\
 \hline

\multicolumn{2}{|c|}{$\Delta3$} & \multicolumn{2}{c|}{$\Delta4$}  \\ \hline
$T_9=T_5\times T_6$ & $(A^3)$ & $T_4=T_y\times T_4$ & $(Z_1^3Y_2)$ \\
* & & $T_{10}=-T_2$ & $(-Y_1)$  \\
$T_8=T_8+T_9$ & $(A^3+2X_1')$ & $T_4=T_4+T_{10}$ & $(B=Z_1^3Y_2-Y_1)$ \\
$T_4=T_3\times T_4$ & $(Z_1^3)$ & $T_{10}=T_4^2$ & $(B^2)$ \\
* & & $T_8=-T_8$ & $(-A^3-2X_1')$ \\
* & & $T_1=$\hl{$\pmb{T_{10}}$}$+T_8$ & $(X_3)$ \\
* & & * &\\
\hline

 \multicolumn{2}{|c|}{$\Delta5$} & \multicolumn{2}{c|}{$\Delta6$} \\ \hline
$T_8=T_2\times T_9$ & $(Y_1')$ & $T_3=T_3\times T_5$ & $(Z_3)$ \\
$T_6=-T_1$ & $(-X_3)$ & $T_4=-T_7$ & $(-X_1')^{(a)}$ \\
$T_6=T_6+T_7$ & $(X_1'-X_3)$ & $T_4=T_1+T_4$ & $(X_3-X_1')^{(a)}$\\
$T_6=T_6\times $\hl{$\pmb{T_4}$} & $(B(X_1'-X_3))$ & $T_5=T_4^2$ & $(A^2)^{(a)}$ \\
$T_9=-T_8$ & $(-Y_1')$ & $T_6=-T_8$ & $(-Y_1')^{(a)}$\\
$T_2=T_6+T_9$ & $(Y_3)$ & $T_6=T_2+T_6$ & $(B)^{(a)}$\\
* && * &\\
\hline
\end{tabular}
\parbox{15cm}{\caption{\label{mnamnaa-mixed-addition}Revised MNAMNAA-based atomic mixed addition. The newly revised registers are shown in bold with yellow highlight.}}
\end{table}
(a) Field operations in $\Delta$6 if a special addition follows.

\clearpage
\subsection{Longa’s Appendix B5 algorithm revised}\label{appendix-mana-special-addition}
The MANA-based atomic special addition with identical Z-coordinate algorithm taken from [2](Appendix B5) is revised as follows. The variables $X_1’$ and $Y_1’$ are outputs from the MANA-based atomic mixed addition algorithm.
\\ \\ \noindent Input: $P=(X_1,Y_1,Z)$ and $Q=(X_2,Y_2,Z)$ \\
Output: $P+Q=(X_3,Y_3,Z_3)$ \\
$T_1\gets $\hl{$\pmb{X_2}$}$, T_2\gets $\hl{$\pmb{Y_2}$}$, T_3\gets Z, T_7\gets $\hl{$\pmb{X_1'}$}$, T_8\gets $\hl{$\pmb{Y_1'}$}

\begin{table}[hbt!]
\centering
\begin{tabular}{|lr|lr|}
\hline
\multicolumn{2}{|c|}{$\Delta1$} & \multicolumn{2}{c|}{$\Delta2$} \\ \hline
* &  & 
$T_5=T_4\times T_4$ & $(A^2)$ \\

* && * & \\
$T_4=-T_7$ & $(-X_1')$ & $T_6=-T_8$ & $(-Y_1')$ \\
$T_4=T_1+T_4$ & $(A=X_2-X_1')$& $T_6=T_2+T_6$ & $(B=Y_2-Y_1')$ \\

 \hline

\multicolumn{2}{|c|}{$\Delta3$} & \multicolumn{2}{|c|}{$\Delta4$} \\ \hline
$T_7=T_5\times T_7$ & $(X_1'A^2)$ & $T_5=T_4\times T_5$ & $(A^3)$ \\
$T_9=T_7+T_7$ & $(2X_1'A^2)$ & $T_9=T_5+T_9$ & $(A^3+2X_1'A^2)$ \\
* & &* & \\
* & &* & \\

\hline
 
\multicolumn{2}{|c|}{$\Delta5$} & \multicolumn{2}{c|}{$\Delta6$} \\ \hline
$T_1=T_6\times T_6$ & $(B^2)$ & $T_2=T_5\times T_8$ & $(Y_1'A^3)$ \\
* & & * &  \\
$T_9=-T_9$ & $(-A^3-2X_1'A^2)$ & $T_5=-T_1$ & $(-X_3)$ \\
$T_1=T_1+T_9$ & $(X_3)$ & $T_5=T_5+T_7$ & $(C=X_1'A^2-X_3)$ \\ 

\hline

\multicolumn{2}{|c|}{$\Delta7$} & \multicolumn{2}{c|}{$\Delta8$} \\ \hline

$T_5=T_5\times T_6$ & $(B.C)$ & $T_3=T_3\times T_4$ & $(Z_3)$\\
* && * & \\ 
$T_2=-T_2$ & $(-Y_1'A^3)$ & * & \\ 
$T_2=T_2+T_5$ & $(Y_3)$ & * &  \\ \cline{1-4}

\end{tabular}
\caption{\label{mana-special-addition}Revised MANA-based atomic special addition with identical Z-coordinate. The newly revised registers are shown in bold with yellow highlight.}
\end{table}
\clearpage

\subsection{Longa’s Appendix B6 algorithm revised}\label{appendix-mnamnaa-special-addition}
The MNAMNAA-based atomic special addition algorithm with identical Z-coordinate taken from [2](Appendix B6) is revised as follows. The variables $X_1’$ and $Y_1’$ are outputs from the MNAMNAA-based atomic mixed addition algorithm.
\\ \\ \noindent Input: $P=(X_1,Y_1,Z)$ and $Q=(X_2,Y_2,Z)$ \\
Output: $P+Q=(X_3,Y_3,Z_3)$ \\
$T_1\gets $\hl{$\pmb{X_2}$}$, T_2\gets $\hl{$\pmb{Y_2}$}$, T_3\gets Z, T_7\gets $\hl{$\pmb{X_1'}$}$, T_8\gets $\hl{$\pmb{Y_1'}$}

\begin{table}[hbt!]
\centering
\begin{tabular}{|lr|lr|}
\hline
\multicolumn{2}{|c|}{$\Delta1$} & \multicolumn{2}{c|}{$\Delta2$} \\ \hline
*& & $T_7=T_5\times T_7$ & $(X_1'A^2)$ \\ 
$T_4=-T_7$ & $(-X_1')$& * &\\
$T_4=T_1+T_4$ & $(A=X_2-X_1')$ & $T_9=T_7+T_7$ & $(2X_1'A^2)$ \\
$T_5=T_4^2$ & $(A^2)$ & $T_5=T_4\times T_5$ & $(A^3)$ \\
$T_6=-T_8$ & $(-Y_1')$ & * &\\
$T_6=T_2+T_6$ & $(B=Y_2-Y_1')$&  $T_9=T_5+T_9$ & $(A^3+2X_1'A^2)$ \\
*&&*& \\ \hline

\multicolumn{2}{|c|}{$\Delta3$} & \multicolumn{2}{c|}{$\Delta4$} \\ \hline
$T_1=T_6^2$ & $(B^2)$ & $T_5=T_5\times T_6$ & $(B.C)$ \\
$T_9=-T_9$ & $(-A^3-2X_1'A^2)$& $T_2=-T_2$ & $(-Y_1'A^3)$ \\
$T_1=T_1+T_9$ & $(X_3)$ & $T_2=T_2+T_5$ & $(Y_3)$ \\
$T_2=T_5\times T_8$ & $(Y_1'A^3)$ & $T_3=T_3\times T_4$ & $(Z_3)$ \\
$T_5=-T_1$ & $(-X_3)$ & * & \\
$T_5=T_5+T_7$ & $(C=X_1'A^2-X_3)$ & *& \\
*&&*& \\ \hline
\end{tabular}
\caption{\label{mnamnaa-special-addition}Revised MNAMNAA-based atomic special addition with identical Z-coordinate. The newly revised registers are shown in bold with yellow highlight.}
\end{table}

\clearpage
\subsection{Longa’s Appendix B7 algorithm revised}\label{appendix-mana-tripling}
The MANA-based atomic point tripling algorithm taken from [2](Appendix B7) is revised as follows.
\\ \\ \noindent Input: $P=(X_1,Y_1,Z_1)$\\
Output: $3P=(X_3,Y_3,Z_3,X_2,Y_2)$ \\
$T_1\gets X_1, T_2\gets Y_1, T_3\gets Z_1$

\begin{table}[hbt!]
\centering
\scalebox{0.9}{
\begin{tabular}{|lr|lr|}
\hline
\multicolumn{2}{|c|}{$\Delta1$} & \multicolumn{2}{c|}{$\Delta2$} \\ \hline
$T_4=T_3^2$ & $(Z_1^2)$ & 
$T_3=T_2\times T_3$ & $(Z_1')$ \\

$T_3=T_3+T_3$ & $(2Z_1)$& $T_4=T_1+T_4$ & $(A=X_1+Z_1^2)$ \\

$T_5=-T_4$ & $(-Z_1^2)$ & * &\\

* & & $T_5=T_1+T_5$ & $(B=X_1-Z_1^2)$ \\
 \hline

\multicolumn{2}{|c|}{$\Delta3$} & \multicolumn{2}{|c|}{$\Delta4$} \\ \hline
$T_4=T_4\times T_5$ & $(A.B)$ & $T_2=T_2^2$ & $(Y_1^2)$ \\
$T_5=T_4+T_4$ & $(2A.B)$ & $T_4=T_4+T_5$ & $(C=3A.B)$ \\
* & & *& \\
* & & $T_2=T_2+T_2$& $(2Y_1^2)$ \\
\hline

\multicolumn{2}{|c|}{$\Delta5$} & \multicolumn{2}{c|}{$\Delta6$} \\ \hline
$T_5=T_4^2$ & $(C^2)$ & 
$T_6=T_1\times T_2$ & $(X_1')$\\
$T_1=T_1+T_1$ & $(2X_1)$& * & \\
* && $T_1=-T_6$ & $(-X_1')$\\
* & & $T_1=T_1+T_1$ & $(-2X_1')$ \\ \hline

\multicolumn{2}{|c|}{$\Delta7$} & \multicolumn{2}{c|}{$\Delta8$} \\ \hline

$T_2=T_2^2$ & $(4Y_1^4)$ & 
$T_4=T_1\times T_4$ & $(C.D)$ \\

$T_7=T_1+T_5$ & $(X_2)$& * &  \\

$T_1=-T_7$ & $(-X_2)$ & $T_2=-T_2$ & $(-4Y_1^4)$\\

$T_1=T_1+T_6$ & $(D=X_1'-X_2)$& $T_2=T_2+T_2$ & $(-Y_1')$ \\ \hline

\multicolumn{2}{|c|}{$\Delta9$} & \multicolumn{2}{c|}{$\Delta10$} \\ \hline
$T_5=T_1^2$ & $(D^2)$ &$T_6=T_5\times T_6$ & $(F=X_1'D^2)$ \\
$T_8=T_2+T_4$ & $(Y_2)$ & *& \\
* & & $T_9=-T_1$ & $(-D)$ \\
$T_4=T_2+T_8$ & $(E=Y_2-Y_1')$ &* & \\

\hline
\multicolumn{2}{|c|}{$\Delta11$} & \multicolumn{2}{c|}{$\Delta12$} \\ \hline
$T_5=$\hl{$\pmb{T_9}$}$\times T_5$ & $(-D^3)$ & 
$T_{10}=T_4^2$ & $(E^2)$\\

$T_1=T_6+T_6$ & $(2F)$& $T_1=T_1+T_{10}$ & $(X_3)$ \\

$T_1=-T_1$ & $(-2F)$& $T_{10}=-T_1$ & $(-X_3)$ \\

$T_1=T_1+T_5$ & $(-D^3-2F)$ & * & \\ \hline

\multicolumn{2}{|c|}{$\Delta13$} & \multicolumn{2}{c|}{$\Delta14$} \\ \hline
$T_5=T_2\times T_5$ & $(Y_1'D^3)$ & 
$T_2=T_2\times T_4$ & $(E(F-X_3))$ \\

$T_2=T_6+T_{10}$ & $(F-X_3)$& * & \\

* & &$T_5=-T_5$ & $(-Y_1'D^3)$  \\

* & & $T_2=T_2+T_5$ & $(Y_3)$ \\
\hline

\multicolumn{2}{|c|}{$\Delta15$} \\ \cline{1-2}
$T_3=T_3\times T_9$ & $(Z_3)$\\
*& \\
$T_4=-T_7$ & $(-X_2)^{(a)}$ \\
$T_4=T_1+T_4$& $(X_3-X_2)^{(a)}$\\

\cline{1-2}
\end{tabular}}
\caption{\label{mana-point-tripling}Revised MANA-based atomic point tripling. The newly revised register is shown in bold with yellow highlight.}
\end{table}
(a) Field operations in $\Delta$15 if a special addition follows.

\clearpage
\subsection{Longa’s Appendix B8 algorithm revised}\label{appendix-mnamnaa-tripling}
The MNAMNAA-based atomic point tripling algorithm taken from [2](Appendix B8) is revised as follows.
\\ \\ \noindent Input: $P=(X_1,Y_1,Z_1)$\\
Output: $3P=(X_3,Y_3,Z_3,X_2,Y_2)$ \\
$T_1\gets X_1, T_2\gets Y_1, T_3\gets Z_1$

\begin{table}[hbt!]
\centering
\begin{tabular}{|lr|lr|}
\hline
\multicolumn{2}{|c|}{$\Delta1$} & \multicolumn{2}{c|}{$\Delta2$} \\ \hline
$T_4=T_3^2$ & $(Z_1^2)$ & 
$T_4=T_4\times T_5$ & $(A.B)$ \\

* && * &\\
$T_3=T_3+T_3$ & $(2Z_1)$& $T_5=T_4+T_4$ & $(2A.B)$ \\

$T_3=T_2\times T_3$ & $(Z_1')$ & $T_2=T_2^2$ & $(Y_1^2)$ \\

$T_5=-T_4$ & $(-Z_1^2)$ & * &  \\
$T_4=T_1+T_4$ & $(A=X_1+Z_1^2)$& $T_4=T_4+T_5$ & $(C=3A.B)$ \\

$T_5=T_1+T_5$ & $(B=X_1-Z_1^2)$& $T_2=T_2+T_2$ & $(2Y_1^2)$  \\
 \hline

\multicolumn{2}{|c|}{$\Delta3$} & \multicolumn{2}{c|}{$\Delta4$} \\ \hline
$T_5=T_4^2$ & $(C^2)$ & $T_2=T_2^2$ & $(4Y_1^4)$ \\
* &  & $T_1=-T_7$ & $(-X_2)$ \\
$T_1=T_1+T_1$ & $(2X_1)$ & $T_1=T_1+T_6$ & $(D=X_1'-X_2)$ \\
$T_6=T_1\times T_2$ & $(X_1')$ & $T_4=T_1\times T_4$ & $(C.D)$ \\
$T_1=-T_6$ & $(-X_1')$ & $T_2=-T_2$ & $(-4Y_1^4)$ \\
 $T_1=T_1+T_1$ & $(-2X_1')$ & $T_2=T_2+T_2$ & $(-Y_1')$ \\
 $T_7=T_1+T_5$ & $(X_2)$ & $T_8=T_2+T_4$ & $(Y_2)$\\

\hline 
\multicolumn{2}{|c|}{$\Delta5$} & \multicolumn{2}{c|}{$\Delta6$} \\ \hline
 $T_5=T_1^2$ & $(D^2)$ & $T_5=$\hl{$\pmb{T_9}$}$\times T_5$ & $(-D^3)$ \\
* && * &  \\
$T_4=T_2+T_8$ & $(E=Y_2-Y_1')$ & $T_1=T_6+T_6$ & $(2F)$\\
$T_6=T_5\times T_6$ & $(F=X_1'D^2)$ & $T_{10}=T_4^2$ & $(E^2)$\\
$T_9=-T_1$ & $(-D)$ & $T_1=-T_1$ & $(-2F)$ \\
* & & $T_1=T_1+T_5$ & $(-D^3-2F)$ \\
* &  & $T_1=T_1+T_{10}$ & $(X_3)$ \\

\hline

\multicolumn{2}{|c|}{$\Delta7$} & \multicolumn{2}{c|}{$\Delta8$} \\ \hline
$T_5=T_2\times T_5$ & $(Y_1'D^3)$ & 
$T_3=T_3\times T_9$ & $(Z_3)$ \\

$T_{10}=-T_1$ & $(-X_3)$& $T_4=-T_7$ & $(-X_2)^{(a)}$ \\
$T_2=T_6+T_{10}$ & $(F-X_3)$& $T_4=T_1+T_4$ & $(X_3-X_2)^{(a)}$ \\

$T_2=T_2\times T_4$ & $(E(F-X_3))$ & $T_5=T_4^2$ & $(A^2)^{(a)}$ \\

$T_5=-T_5$ & $(-Y_1'D^3)$ & $T_6=-T_8$ & $(-Y_2)^{(a)}$  \\
$T_2=T_2+T_5$ & $(Y_3)$& $T_6=T_2+T_6$ & $(B)^{(a)}$  \\

* &&*&
\\ \hline

\end{tabular}
\caption{\label{mnamnaa-point-tripling}Revised MNAMNAA-based atomic point tripling. The newly revised register is shown in bold with yellow highlight.}
\end{table}

(a) Field operations in $\Delta$8 if a special addition follows.
\clearpage
\section{Distinguishability of Doubling 1 and Doubling 2}

\label{Appendix5}



\begin{figure}[H]
\centering
\includegraphics[width=0.9\textwidth]{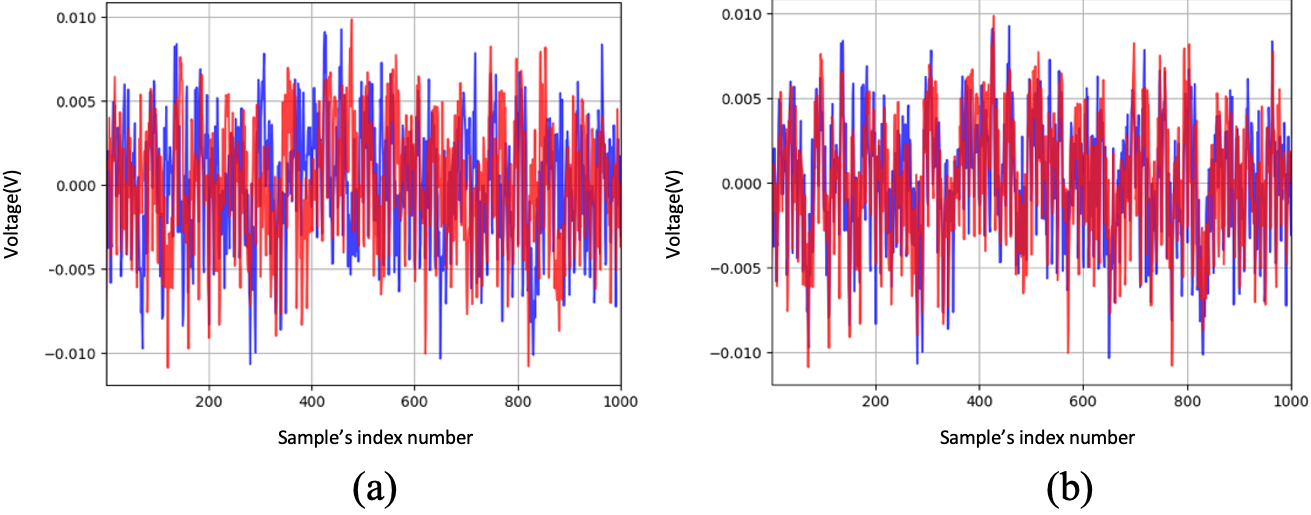}
\caption{(a) Showing Doubling 1 (blue) and Doubling 2 (red) are non-synchronous at 20,000 samples after the previous synchronization, (b) synchronous again after shifting Doubling 2 by 50 samples left.}
\label{unsynced_dbl1dbl2}
\end{figure}

By focusing on the first atomic block (with about 800,000 samples) from the sub-traces of Doubling 1 and Doubling 2, we overlay the synchronized sub-traces, so they have matching patterns in the beginning of $\Delta1$. At 20 µ$s$ (20,000 samples) after where we have initially synchronized the patterns, the traces become non-synchronous, as shown in Figure \ref{unsynced_dbl1dbl2} (a). The red trace (Doubling 2) has a delay of about 50 $ns$ (50 samples, i.e., 5 clock cycles) when compared to the blue trace (Doubling 1). The discrepancy is then corrected as shown in Figure \ref{unsynced_dbl1dbl2} (b). 

Since we are unable to identify and separate each arithmetic field operation within an atomic block in the sub-traces exactly, as we did not place any NOPs in between, we studied the execution time of the FLECC constant-time field operations using breakpoints. It was found that the constant-time Montgomery modular multiplication, negation and addition operations from FLECC are quite constant-time, according to our extensive study in Chapter \ref{sec:study-of-flecc}. 
Despite of that, the 5-clock cycle difference causing the delay in the red line in Figure \ref{unsynced_dbl1dbl2} (above) aligns with the duration difference between Doubling 1 and Doubling 2 measured using breakpoints (see Table \ref{exec-time} in Chapter \ref{kp_analysis_ram}). Further investigation is necessary to determine the reasons for this delay.



\clearpage
\section{Execution Time of No Operations} 

\label{Appendix4}

\begin{table}[H]
\centering
\begin{tabular}{|cc|}
\hline
\multicolumn{2}{|c|}{2000 NOPs}                              \\ \hline
\multicolumn{1}{|c|}{Measured between}          & No. of clock cycles \\ \hline
\multicolumn{1}{|l|}{Doubling 3 $\Delta1$ \& $\Delta2$} & 27996               \\ \hline
\multicolumn{1}{|l|}{Doubling 3 $\Delta2$ \& $\Delta3$} & 29995               \\ \hline
\multicolumn{1}{|l|}{Doubling 3 $\Delta3$ \& $\Delta4$} & 29995               \\ \hline
\multicolumn{1}{|l|}{Addition 1 $\Delta1$ \& $\Delta2$} & 30006               \\ \hline
\multicolumn{1}{|l|}{Addition 1 $\Delta2$ \& $\Delta3$} & 28007               \\ \hline
\multicolumn{1}{|l|}{Addition 1 $\Delta3$ \& $\Delta4$} & 30006               \\ \hline
\multicolumn{1}{|l|}{Addition 1 $\Delta4$ \& $\Delta5$} & 28007               \\ \hline
\multicolumn{1}{|l|}{Addition 1 $\Delta5$ \& $\Delta6$} & 30006               \\ \hline
\end{tabular}
\caption{Measured execution time of 2,000 NOPs.}\label{NOPs2000}
\end{table}

\begin{table}[H]
\centering
\begin{tabular}{|cc|}
\hline
\multicolumn{2}{|c|}{5000 NOPs}                                      \\ \hline
\multicolumn{1}{|c|}{Measured between}                  & No. of clock cycles \\ \hline
\multicolumn{1}{|l|}{Doubling 2 \& Doubling 3} & 69996               \\ \hline
\multicolumn{1}{|l|}{Doubling 3 \& Addition 1} & 70012               \\ \hline
\multicolumn{1}{|l|}{Addition 1 \& Doubling 4} & 69996               \\ \hline
\end{tabular}
\caption{Measured execution time of 5,000 NOPs.}\label{NOPs5000}
\end{table}
\section{Attack Scenario} 

\label{Appendix6}




Assume an attacker can differentiate atomic blocks in the $k\bm{P}$ trace measured, whereby the duration of the first atomic block $\Delta1$ of EC point addition (PA) is (slightly) longer than that of EC point doubling (PD).

Consider a left-to-right $k\bm{P}$ algorithm implemented using Longa's MNAMNAA atomic patterns, the first point operation is a point doubling. 
Since a point doubling operation consists of 4 atomic blocks, the next $\Delta1$ in the $k\bm{P}$ trace will be at atomic block number $i=1+4=5$, with duration $t_{\Delta1}(x)$, where $x$ is either PD or PA at the $5^{th}$ atomic block, the attacker does not know yet. An attacker can compare his observed $t_{\Delta1}(x)$ with $t_{\Delta1}(PD)$ to reveal the operation executed:
\begin{itemize}
    \item If $t_{\Delta1}(x)\approx t_{\Delta1}(PD)$, $x$ is the atomic pattern of point doubling. The attacker will then take the $(i=i+4)^{th}$ atomic block in the $k\bm{P}$ trace as the next $\Delta1$ to analyse.
    \item Otherwise, if $t_{\Delta1}(x) > t_{\Delta1}(PD)$, $x$ is the atomic pattern of point addition. The attacker will then take the $(i=i+6)^{th}$ atomic block in the $k\bm{P}$ trace as the next $\Delta1$ to analyse.
\end{itemize}

Therefore, the attacker can step-by-step reveal all point doublings and point additions from the measured $k\bm{P}$ trace, based on the assumption that the attacker can effectively separate each atomic block within the $k\bm{P}$ trace.

Similarly, this algorithm applies when $t_{\Delta1}(PA) < t_{\Delta1}(PD)$. The key condition is that $t_{\Delta1}(PA) \neq t_{\Delta1}(PD)$.
However, further analysis of the $k\bm{P}$ trace is required to identify additional indicators for distinguishing the atomic patterns. Applying multiple indicators enhances the likelihood of a successful attack. 





\sloppy
\printbibliography[heading=bibintoc]

\end{CJK*}
\end{document}